\documentclass[11pt]{article}
\usepackage{epsfig}
\usepackage{amsmath,amsfonts}

\begin{document}

\title{\textbf{Introduction to the Gribov Ambiguities In Euclidean Yang-Mills Theories }}
\author{R.F. Sobreiro and S.P. Sorella\footnote{Lectures given by S.P. Sorella at the 13th Jorge Andre Swieca Summer School on
Particles and Fields, Campos de Jord{\~a}o, Brazil, 9-22 January
2005.}{\ }\footnote{Work supported by FAPERJ, Funda{\c c}{\~a}o de
Amparo {\`a} Pesquisa do Estado do Rio de Janeiro, under the program {\it Cientista do Nosso Estado}, E-26/151.947/2004.}\\
\textit{\ UERJ, Universidade do Estado do Rio de Janeiro}\\
\textit{\ Instituto de F{\'\i}sica}\\
\textit{\ Departamento de F{\'\i}sica Te\'{o}rica}\\
\textit{\ Rua S\~{a}o\ Francisco Xavier, 524, } \\
\textit{\ 20550-013 Maracan\~{a}, Rio de Janeiro, Brazil}}
\date{}
\maketitle

\begin{abstract}
An introduction to the Gribov ambiguities and their consequences
on the infrared behavior of Euclidean Yang-Mills theories is
presented.
\end{abstract}

\vspace{2cm}

\newpage

\tableofcontents

\newpage

\section{Introduction}

Nowadays, that the Gribov ambiguities play an important role in
the quantization of Yang-Mills theories is becoming more and more
evident. These ambiguities, affecting the Faddeev-Popov
quantization formula, deeply modify the infrared behavior of the
theory and might have a crucial role in the understanding of color
confinement. \\\\The aim of these notes is to give a simple
introduction to Gribov's work and to its consequences on the
infrared regime of nonabelian Euclidean gauge theories. The
discussion will be focused mainly on the original paper by Gribov
\cite{g}, whose content will be reproduced in a somewhat detailed
way.\\\\This work is divided in two parts. Part I is devoted to
the study of the Gribov ambiguities. We begin by reviewing the
examples provided by Gribov \cite{g}. Further, a class of Gribov
copies proposed by Henyey \cite{hen} will be considered.\\\\In
Part II we introduce the so called Gribov horizons and we analyze
their properties. Following Gribov's suggestion, the restriction
of the domain of integration in the Feynman path integral to the
first horizon will be discussed. The ensuing modifications on the
infrared behavior of the gluon and ghost propagators in the Landau
gauge will be worked out.\\\\We conclude this short introduction
by mentioning a result due to Singer \cite{is}. Although the
examples of the Gribov copies which we shall consider in the
following will be worked out in the transverse gauge, $\partial
A=0$, the existence of the Gribov ambiguities is not related to a
specific gauge condition. As pointed out in \cite{is}, the
presence of Gribov copies is in fact a general feature of
nonabelian gauge theories.

\section{Part I: The Gribov Pendulum}

\subsection{The Gribov ambiguities}

\subsubsection{Quantization of Euclidean Yang-Mills theories.
Non-uniqueness of the gauge condition}

The Euclidean Yang-Mills action\footnote{%
See Appendix A for the notation.}
\begin{equation}
S_{YM}=\frac{1}{4g^{2}}\int {d^{4}x}F_{\mu \nu }^{a}F_{\mu \nu }^{a}\;,
\label{eq1}
\end{equation}
is left invariant if one replaces $A_{\mu }$ by the gauge transformed field $%
\widetilde{A}_{\mu }$
\begin{equation}
A_{\mu }\rightarrow \widetilde{A}_{\mu }={S}^{\dagger }\partial _{\mu }{S}%
+S^{\dagger }A_{\mu }{S}\;.  \label{eq2}
\end{equation}
The fields $A_{\mu }$ and $\widetilde{A}_{\mu }$ define equivalent
configurations, leading to the same value for the expression
$\left( \ref {eq1}\right) $. In order to properly quantize the
theory, only inequivalent configurations should be taken into
account in the Feynman path integral. The first step towards the
implementation of this procedure is provided by the Faddeev-Popov
quantization formula. One integrates over configurations which
have a certain divergence, {\it i.e.}
\begin{equation}
\partial _{\mu }A_{\mu }=f\;.  \label{eq3}
\end{equation}
Therefore, for the partition function one gets
\begin{equation}
\mathcal{Z=N}\int DA_{\mu }\delta (\partial A-f) \mathrm{\det
}\left(\mathcal{M}^{ab}(A)\right)e^{-\frac{1}{4g^{2}}\int
{d^{4}x}F_{\mu \nu }^{a}F_{\mu \nu }^{a}\;}\;, \label{eq4}
\end{equation}
where
\begin{equation}
\mathcal{M}^{ab}(A)=-\partial _{\mu }(\partial _{\mu }\delta
^{ab}-f^{abc}A_{\mu }^{c})\;,  \label{eq5}
\end{equation}
is the Faddeev-Popov operator and $\mathcal{N}$ a normalization factor. In
the following we shall limit ourselves to the choice $f=0$, which amounts to
impose the following transversality condition
\begin{equation}
\partial A=0\;.  \label{eq6}
\end{equation}
This condition is known as the Landau gauge condition. Thus
\begin{equation}
\mathcal{Z=N}\int DA_{\mu }\delta (\partial A)\mathrm{\det }\left(\mathcal{M}%
^{ab}\right)e^{-\frac{1}{4g^{2}}\int {d^{4}x}F_{\mu \nu
}^{a}F_{\mu \nu }^{a}\;}\;. \label{eq7}
\end{equation}
However, as pointed out by Gribov \cite{g}, the condition
$\partial A=0$ does not fix uniquely the gauge configurations.
This means that, for a given $A_{\mu }$ satisfying the gauge
condition $\partial A=0$, there exist equivalent fields
$\widetilde{A}_{\mu }$ obeying the same condition, \textit{i.e.}
\begin{eqnarray}
\widetilde{A}_{\mu } &=&{S}^{\dagger }\partial _{\mu }{S}+S^{\dagger }A_{\mu
}{S}= A_{\mu}+ S^{\dagger}(\partial_{\mu}S+[A_{\mu},S]) \;,  \nonumber  \\
\partial \widetilde{A} &=&\partial A=0\;.  \label{eq8}
\end{eqnarray}
Field configurations $\widetilde{A}_{\mu }$ satisfying the
conditions $\left( \ref{eq8}\right) $ are copies of $A_{\mu}$. In
terms of the gauge transformation $S$, the condition $\partial
\widetilde{A}=\partial A =0 $ reads
\begin{equation}
\partial_{\mu} (S^{\dagger}(\partial_{\mu}S+[A_{\mu},S]))=0 \;, \label{eq9}
\end{equation}
which, due to $\partial A=0$, becomes
\begin{equation}
\partial _{\mu }S^{\dagger }\partial _{\mu }{S+S}^{\dagger }\partial _{\mu
}\partial _{\mu }{S+\partial }_{\mu }S^{\dagger }A_{\mu }{S+}S^{\dagger
}A_{\mu }{\partial }_{\mu }{S=0\;.}  \label{eq10}
\end{equation}
\begin{itemize}
\item  \textbf{Remark}

\textit{At the infinitesimal level, $S=1+\alpha ,$ $\alpha \ll 1$,
expression $\left( \ref{eq10}\right) $ reduces to
\begin{equation}
\partial ^{2}\alpha -(\partial _{\mu }\alpha) A_{\mu }+A_{\mu }(\partial_{\mu}\alpha) =0\;,
\label{eq11}
\end{equation}
\textit{i.e.}
\begin{equation}
\partial _{\mu }\left( \partial _{\mu }\alpha +\left[ A_{\mu },\alpha
\right] \right) =0\;.  \label{eq12}
\end{equation}
We see that, in the infinitesimal case, the condition for the existence of
Gribov's copies is equivalent to state that the operator $-\partial _{\mu
}\left( \partial _{\mu }\cdot +\left[ A_{\mu },\cdot \right] \right) $,
whose determinant enters the Faddeev-Popov quantization formula $\left( \ref
{eq7}\right) $, has zero eigenvalues. Notice that the eigenvalues equation
for the Faddeev-Popov operator $-\partial _{\mu }\left( \partial _{\mu
}\cdot +\left[ A_{\mu },\cdot \right] \right) $,
\begin{equation}
-\partial _{\mu }\left( \partial _{\mu }\psi +\left[ A_{\mu },\psi \right]
\right) =\epsilon (A)\psi \;,  \label{eq13}
\end{equation}
can be seen as a kind of Schr{\"{o}}dinger equation, with $A_{\mu
}$ playing
the role of the potential. Therefore, for large enough values of the field $%
A_{\mu }$, we might expect that zero energy solutions, $\epsilon
=0$, of eq.$\left( \ref{eq13}\right) $ do in fact exist. }
\end{itemize}

\subsubsection{The Gribov pendulum}

In order to analyze the condition for the existence of Gribov's
copies, we shall consider the three-dimensional case,
$A_{i}$,$\;i=1,2,3$, assuming that the gauge group is $SU(2)$ and
that the gauge field is spherically symmetric, \textit{i.e.} $A_i$
depends on the unit vector $n_{i}=x_{i}/r$, $r=\sqrt{x_{i}x_{i}}$,
$n_{i}n_{i}=1$. Let $\sigma _{i}$ denote the $2\times 2$ Pauli
matrices
\begin{equation}
\sigma _{i}\sigma _{j}=\delta _{ij}+i\varepsilon _{ijk}\sigma _{k}\;.
\label{eq14}
\end{equation}
\begin{equation}
\sigma _{1}=\left(
\begin{array}{ll}
0 & 1 \\
1 & 0
\end{array}
\right) ,\;\;\;\;\sigma _{2}=\left(
\begin{array}{ll}
0 & i \\
-i & 0
\end{array}
\right) ,\;\;\;\;\sigma _{3}=\left(
\begin{array}{ll}
1 & 0 \\
0 & -1
\end{array}
\right) \;.  \label{pm}
\end{equation}
It follows that the quantity $\widehat{n}$
\begin{equation}
\widehat{n}=in_{i}\sigma _{i}\;,  \label{eq15}
\end{equation}
obeys
\begin{equation}
\widehat{n}^{2}=-n_{i}n_{j}\sigma _{i}\sigma _{j}=-n_{i}n_{i}=-1\;.
\label{eq16}
\end{equation}
Since the gauge field $A_{i}$ is Lie algebra valued, it can be parametrized
as
\begin{equation}
A_{i}=a_{1}(r)In_{i}+a_{2}(r)\sigma _{i}+a_{3}(r)n_{i}(\overrightarrow{n}%
\cdot \overrightarrow{\sigma })+a_{4}(r)\sigma _{i}(\overrightarrow{n}\cdot
\overrightarrow{\sigma })\;,  \label{eq17}
\end{equation}
where $I$ stands for the $2\times 2$ unit matrix\footnote{%
The set $(I,\sigma _{i})$ is a basis for the $2\times 2$ matrices of $SU(2)$.%
} and where we have taken into account that $A_{i}$ is spherically
symmetric. Notice that, due to $n_{i}n_{i}=1$, and
$(\overrightarrow{n}\cdot \overrightarrow{\sigma })^{2}=1$, higher
powers of $\left( n_{i}n_{i}\right) ^{n}$ and
$(\overrightarrow{n}\cdot \overrightarrow{\sigma })^{m}$ are
absent in the expression $\left( \ref{eq17}\right) $. Also, we
observe that
\begin{equation}
i\varepsilon _{ijk}n_{j}\sigma _{k}=n_{j}\left( \sigma _{i}\sigma
_{j}-\delta _{ij}\right) =\sigma _{i}(\overrightarrow{n}\cdot
\overrightarrow{\sigma })-n_{i}\;,  \label{eq18}
\end{equation}
meaning that this term is not independent. In addition, from eq.$\left( \ref
{eq14}\right) $, it follows that
\begin{equation}
(\overrightarrow{n}\cdot \overrightarrow{\sigma })\sigma _{i}=-\sigma _{i}(%
\overrightarrow{n}\cdot \overrightarrow{\sigma })+2n_{i}\;.  \label{eq19}
\end{equation}
Thus, expression $\left( \ref{eq17}\right) $ yields the most
general form for a $SU(2)$ field which is spherically symmetric.
Moreover, due to the traceless condition
\begin{equation}
 Tr\sigma ^{a}A_{i}^{a}=0 \;, \nonumber
\end{equation}
we get the relationship
\begin{equation}
a_{4}=-a_{1}\;,  \label{eq20}
\end{equation}
so that $A_{i}$ turns out to be parametrized by three independent quantities
\begin{equation}
A_{i}=a_{1}(r)In_{i}+a_{2}(r)\sigma _{i}+a_{3}(r)n_{i}(\overrightarrow{n}%
\cdot \overrightarrow{\sigma })-a_{1}(r)\sigma _{i}(\overrightarrow{n}\cdot
\overrightarrow{\sigma })\;.  \label{eq21}
\end{equation}
Of course, we can adopt the same parametrization employed by
Gribov \cite{g}, namely
\begin{equation}
A_{i}=f_{1}(r)\frac{\partial \widehat{n}}{\partial x_{i}}+f_{2}(r)\widehat{n}%
\frac{\partial \widehat{n}}{\partial x_{i}}+f_{3}(r)\widehat{n}n_{i}\;.
\label{eq22}
\end{equation}
Indeed, from
\begin{equation}
\frac{\partial \widehat{n}}{\partial x_{i}}=\frac{i}{r}\sigma _{k}(\delta
_{ik}-n_{i}n_{k})=\frac{i}{r}\left( \sigma _{i}-(\overrightarrow{n}\cdot
\overrightarrow{\sigma })n_{i}\right) \;,  \label{eq23}
\end{equation}
it follows
\begin{equation}
A_{i}=\frac{i}{r}f_{1}(r)\sigma _{i}-\frac{i}{r}f_{1}(r)(\overrightarrow{n}%
\cdot \overrightarrow{\sigma })n_{i}-\frac{1}{r}f_{2}(r)(\overrightarrow{n}%
\cdot \overrightarrow{\sigma })\sigma _{i}+\frac{1}{r}%
f_{2}(r)n_{i}+if_{3}(r)(\overrightarrow{n}\cdot \overrightarrow{\sigma }%
)n_{i}\;  \label{eq24}
\end{equation}
which has precisely the same form of expression $\left( \ref{eq21}\right) $.
\begin{itemize}
\item  \textit{Observe that for $f_{1}=f_{3}=0$, expression $\left( \ref
{eq22}\right) $ becomes
\begin{eqnarray}
A_{i} &=&-\frac{1}{r}f_{2}(r)(\overrightarrow{n}\cdot \overrightarrow{\sigma
})\sigma _{i}+\frac{1}{r}f_{2}(r)n_{i}\;  \label{eq25} \\
&=&-\frac{1}{r}f_{2}(r)\left( \delta _{ij}+i\varepsilon _{jik}\sigma
_{k}\right) n_{j}+\frac{1}{r}f_{2}(r)n_{i}  \nonumber \\
&=&\frac{i}{r^{2}}\varepsilon _{ijk}x_{j}\sigma _{k}f_{2}(r)\;,  \nonumber
\end{eqnarray}
which is purely transverse,\textit{\ i.e.}
\begin{equation}
\partial _{i}A_{i}=0\;.  \label{eq26}
\end{equation}
}
\end{itemize}
\noindent Having found the most general parametrization for the
$SU(2)$ gauge field, eq.$\left( \ref{eq22}\right) $, let us work
out the condition for the existence of copies, {\it i.e.}
\begin{eqnarray}
\widetilde{A}_{i} &=&{S}^{\dagger }\partial _{i}{S}+S^{\dagger }A_{i}{S\;,}
\label{eq27} \\
\partial _{i}\widetilde{A}_{i} &=&\partial _{i}A_{i}\;,  \nonumber
\end{eqnarray}
where we shall consider the class of gauge transformations $S$
parametrized by
\begin{equation}
S=e^{\frac{i}{2}\alpha (r)\overrightarrow{n}\cdot \overrightarrow{\sigma }%
}=\cos \frac{\alpha(r) }{2}+i\overrightarrow{n}\cdot \overrightarrow{\sigma }%
\sin \frac{\alpha(r) }{2}\;.  \label{eq28}
\end{equation}
From eqs.$\left( \ref{eq27}\right) $, $\left( \ref{eq28}\right) $
it follows
\begin{eqnarray}
\widetilde{A}_{i} &=&\left( \cos \frac{\alpha }{2}-\widehat{n}\sin \frac{%
\alpha }{2}\right) \left[ \left( -\sin \frac{\alpha }{2}+\widehat{n}\cos
\frac{\alpha }{2}\right) \frac{1}{2}\frac{\partial \alpha }{\partial x_{i}}%
+\;\sin \frac{\alpha }{2}\frac{\partial \widehat{n}}{\partial x_{i}}\right]
\nonumber \\
&&+\left( \cos \frac{\alpha }{2}-\widehat{n}\sin \frac{\alpha }{2}\right)
\left( f_{1}(r)\frac{\partial \widehat{n}}{\partial x_{i}}+f_{2}(r)\widehat{n%
}\frac{\partial \widehat{n}}{\partial x_{i}}+f_{3}(r)\widehat{n}n_{i}\right)
\left( \cos \frac{\alpha }{2}+\widehat{n}\sin \frac{\alpha }{2}\right)
\nonumber \\
&=&\left( \widehat{n}\cos ^{2}\frac{\alpha }{2}+\widehat{n}\sin ^{2}\frac{%
\alpha }{2}\right) \frac{1}{2}\frac{\partial \alpha }{\partial x_{i}}+\cos
\frac{\alpha }{2}\sin \frac{\alpha }{2}\frac{\partial \widehat{n}}{\partial
x_{i}}-\sin ^{2}\frac{\alpha }{2}\widehat{n}\frac{\partial \widehat{n}}{%
\partial x_{i}}  \nonumber \\
&&+\left( f_{1}\cos \frac{\alpha }{2}\frac{\partial \widehat{n}}{\partial
x_{i}}+f_{2}\cos \frac{\alpha }{2}\widehat{n}\frac{\partial \widehat{n}}{%
\partial x_{i}}+f_{3}\cos \frac{\alpha }{2}\widehat{n}n_{i}-f_{1}\sin \frac{%
\alpha }{2}\widehat{n}\frac{\partial \widehat{n}}{\partial x_{i}}\right.
\nonumber \\
&&\left. +f_{2}\sin \frac{\alpha }{2}\frac{\partial \widehat{n}}{\partial
x_{i}}+f_{3}\sin \frac{\alpha }{2}n_{i}\right) \left( \cos \frac{\alpha }{2}+%
\widehat{n}\sin \frac{\alpha }{2}\right) \;.  \nonumber \\
&&  \label{eq29}
\end{eqnarray}
Since
\begin{equation}
\frac{\partial \alpha }{\partial x_{i}}=\alpha ^{\prime }(r)n_{i}\;,
\label{eq30}
\end{equation}
we obtain
\begin{eqnarray}
\widetilde{A}_{i} &=&\frac{1}{2}\alpha ^{\prime }(r)\widehat{n}n_{i}+\frac{1%
}{2}\sin \alpha \frac{\partial \widehat{n}}{\partial x_{i}}-\sin ^{2}\frac{%
\alpha }{2}\widehat{n}\frac{\partial \widehat{n}}{\partial x_{i}}  \nonumber
\\
&&+f_{1}\left( \cos ^{2}\frac{\alpha }{2}\frac{\partial \widehat{n}}{%
\partial x_{i}}+\cos \frac{\alpha }{2}\sin \frac{\alpha }{2}\left[ \frac{%
\partial \widehat{n}}{\partial x_{i}},\widehat{n}\right] -\sin ^{2}\frac{%
\alpha }{2}\widehat{n}\frac{\partial \widehat{n}}{\partial x_{i}}\widehat{n}%
\right)  \nonumber \\
&&+f_{2}\left( \cos ^{2}\frac{\alpha }{2}\widehat{n}\frac{\partial \widehat{n%
}}{\partial x_{i}}+\cos \frac{\alpha }{2}\sin \frac{\alpha }{2}\widehat{n}%
\frac{\partial \widehat{n}}{\partial x_{i}}\widehat{n}+\cos \frac{\alpha }{2}%
\sin \frac{\alpha }{2}\frac{\partial \widehat{n}}{\partial x_{i}}+\sin ^{2}%
\frac{\alpha }{2}\frac{\partial \widehat{n}}{\partial x_{i}}\widehat{n}%
\right)  \nonumber \\
&&+f_{3}\widehat{n}n_{i}\;.  \label{eq31}
\end{eqnarray}
Observing that
\begin{equation}
\frac{\partial n_{k}}{\partial x_{i}}n_{j}(\sigma _{k}\sigma _{j}+\sigma
_{j}\sigma _{k})=2\frac{\partial n_{k}}{\partial x_{i}}n_{j}\delta _{kj}=2%
\frac{\partial n_{k}}{\partial x_{i}}n_{k}=\frac{\partial \left(
n_{k}n_{k}\right) }{\partial x_{i}}=0\;,  \label{eq32}
\end{equation}
one has
\begin{equation}
\frac{\partial \widehat{n}}{\partial x_{i}}\widehat{n}=-\widehat{n}\frac{%
\partial \widehat{n}}{\partial x_{i}}\;,  \label{eq33}
\end{equation}
and
\begin{equation}
\widehat{n}\frac{\partial \widehat{n}}{\partial x_{i}}\widehat{n}=-\widehat{n%
}\widehat{n}\frac{\partial \widehat{n}}{\partial x_{i}}=\frac{\partial
\widehat{n}}{\partial x_{i}}\;.  \label{eq34}
\end{equation}
Therefore
\begin{eqnarray}
\widetilde{A}_{i} &=&\left( f_{1}\left( \cos ^{2}\frac{\alpha }{2}-\sin ^{2}%
\frac{\alpha }{2}\right) +2f_{2}\cos \frac{\alpha }{2}\sin \frac{\alpha }{2}%
\right) \frac{\partial \widehat{n}}{\partial x_{i}}  \label{eq35} \\
&&+\left( f_{2}\left( \cos ^{2}\frac{\alpha }{2}-\sin ^{2}\frac{\alpha }{2}%
\right) -2f_{1}\cos \frac{\alpha }{2}\sin \frac{\alpha }{2}\right) \widehat{n%
}\frac{\partial \widehat{n}}{\partial x_{i}}  \nonumber \\
&&+\left( f_{3}+\frac{1}{2}\alpha ^{\prime }(r)\right) \widehat{n}n_{i}+%
\frac{1}{2}\sin \alpha \frac{\partial \widehat{n}}{\partial x_{i}}-\frac{1}{2%
}\left( 1-\cos \alpha \right) \widehat{n}\frac{\partial \widehat{n}}{%
\partial x_{i}}\;.  \nonumber
\end{eqnarray}
Finally, recalling that
\begin{eqnarray}
\cos ^{2}\frac{\alpha }{2}-\sin ^{2}\frac{\alpha }{2} &=&\cos \alpha \;,
\label{eq36} \\
2\cos \frac{\alpha }{2}\sin \frac{\alpha }{2} &=&\sin \alpha \;,  \nonumber
\end{eqnarray}
we get
\begin{eqnarray}
\widetilde{A}_{i} &=&\left( f_{1}\cos \alpha +\left( f_{2}+\frac{1}{2}%
\right) \sin \alpha \right) \frac{\partial \widehat{n}}{\partial x_{i}}
\nonumber \\
&&+\left( \left( f_{2}+\frac{1}{2}\cos \alpha \right) -f_{1}\sin \alpha -%
\frac{1}{2}\right) \widehat{n}\frac{\partial \widehat{n}}{\partial
x_{i}} +\left( f_{3}+\frac{1}{2}\alpha ^{\prime }(r)\right)
\widehat{n}n_{i}\;. \label{eq37}
\end{eqnarray}
It remains now to work out the condition $\partial _{i}\widetilde{A}%
_{i}=\partial _{i}A_{i}$. After a little algebra, one obtains the
differential equation
\begin{equation}
\alpha ^{\prime \prime }(r)+\frac{2}{r}\alpha ^{\prime }(r)-\frac{4}{r^{2}}%
\left( \left( f_{2}+\frac{1}{2}\right) \sin \alpha +f_{1}\left( \cos \alpha
-1\right) \right) =0\;.  \label{eq38}
\end{equation}
Setting
\begin{equation}
\tau =\log r \;, {\ }{\ }{\ }{\ }r=e^{\tau } \;, \nonumber
\end{equation}
equation $\left( \ref{eq38}\right) $ becomes
\begin{equation}
\frac{\partial ^{2}\alpha (\tau )}{\partial \tau ^{2}}+\frac{\partial \alpha
(\tau )}{\partial \tau }-4\left( \left( f_{2}+\frac{1}{2}\right) \sin \alpha
+f_{1}\left( \cos \alpha -1\right) \right) =0\;,  \label{eq39}
\end{equation}
which is the equation of a pendulum in the presence of a damping term $%
\alpha ^{\prime }(\tau)$, see Fig.1.
\begin{figure}[ht]
\centering \epsfig{file=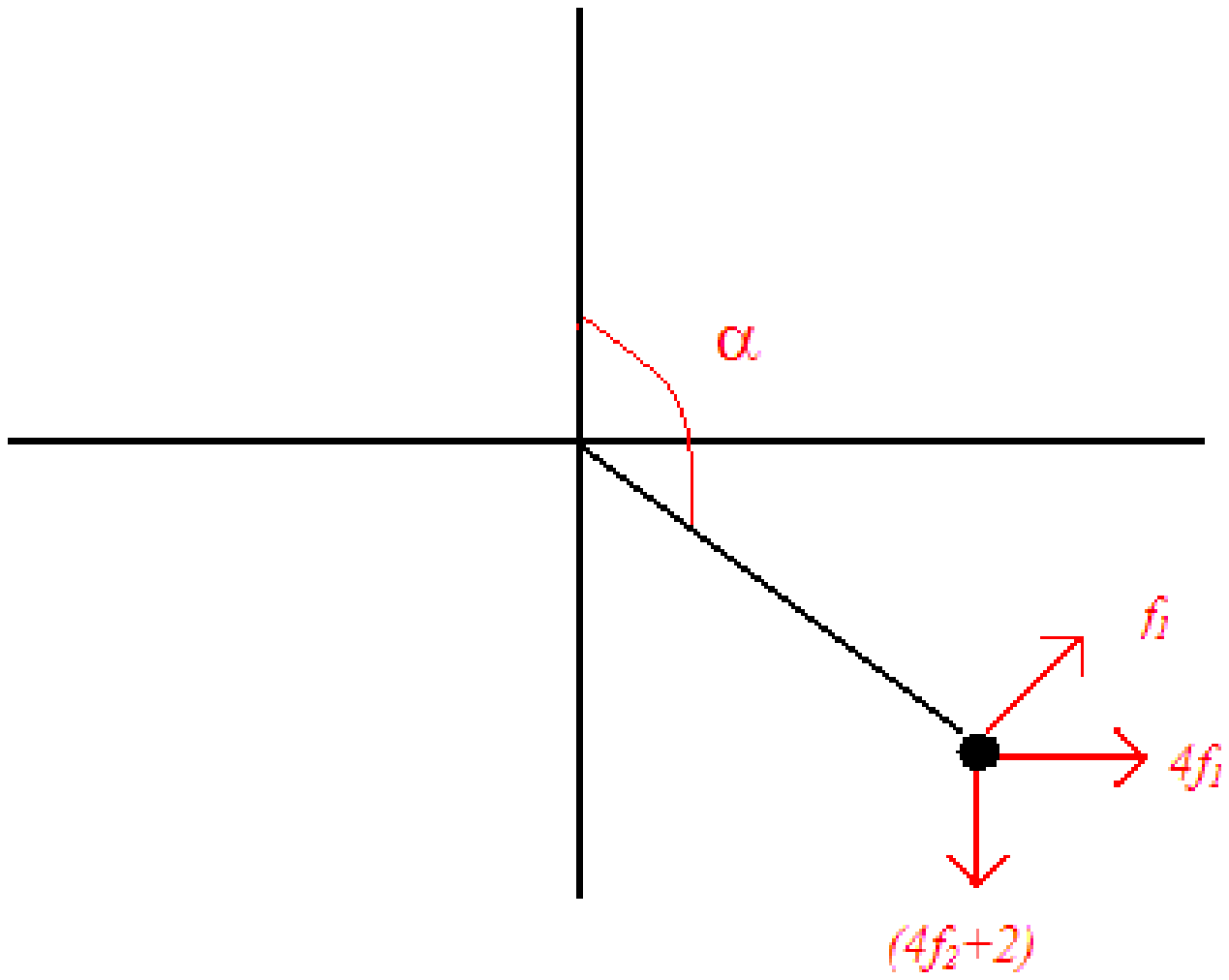,width=10cm} \caption{The Gribov
pendulum}
\end{figure}


\newpage

\begin{itemize}
\item  \textbf{Summary}

\textit{We have started with the most general spherically symmetric $SU(2)$
field
\begin{equation}
A_{i}=f_{1}(r)\frac{\partial \widehat{n}}{\partial x_{i}}+f_{2}(r)\widehat{n}%
\frac{\partial \widehat{n}}{\partial x_{i}}+f_{3}(r)\widehat{n}n_{i}\;.
\label{eq40}
\end{equation}
For the gauge transformed field
\begin{eqnarray}
\widetilde{A}_{i} &=&{S}^{\dagger }\partial _{i}{S}+S^{\dagger }A_{i}{S\;,}
\label{eq41} \\
S &=&e^{\frac{i}{2}\alpha (r)\overrightarrow{n}\cdot \overrightarrow{\sigma }%
}=\cos \frac{\alpha }{2}+i\overrightarrow{n}\cdot \overrightarrow{\sigma }%
\sin \frac{\alpha }{2}\;,  \nonumber
\end{eqnarray}
we have
\begin{eqnarray}
\widetilde{A}_{i} &=&\widetilde{f_{1}}(r)\frac{\partial \widehat{n}}{%
\partial x_{i}}+\widetilde{f_{2}}(r)\widehat{n}\frac{\partial \widehat{n}}{%
\partial x_{i}}+\widetilde{f_{3}}(r)\widehat{n}n_{i}\;,  \label{eq42} \\
\widetilde{f_{1}}(r) &=&f_{1}\cos \alpha +\left( f_{2}+\frac{1}{2}\right)
\sin \alpha \;,  \nonumber \\
\widetilde{f_{2}}(r) &=&f_{2}+\frac{1}{2}\cos \alpha -f_{1}\sin \alpha -%
\frac{1}{2}\;,  \nonumber \\
\widetilde{f_{3}}(r) &=&f_{3}+\frac{1}{2}\alpha ^{\prime }(r)\;.  \nonumber
\end{eqnarray}
Finally, the condition $\partial _{i}\widetilde{A}_{i}=\partial _{i}A_{i}$
yields
\begin{equation}
\frac{\partial ^{2}\alpha (\tau )}{\partial \tau ^{2}}+\frac{\partial \alpha
(\tau )}{\partial \tau }-4\left( \left( f_{2}+\frac{1}{2}\right) \sin \alpha
+f_{1}\left( \cos \alpha -1\right) \right) =0\;.  \label{eq43}
\end{equation}
This is the equation of a pendulum in the presence of a damping
term $\alpha ^{\prime }(\tau)$. The components $f_1, f_2$ of the
gauge field $\left( \ref{eq40}\right) $ correspond to the forces
acting on the pendulum, Fig.1, see also Appendix B. }
\end{itemize}

\subsection{Examples of Gribov's copies}

In this section we shall work out explicit examples of Gribov's copies. We
shall restrict ourselves to the transversality condition
\begin{equation}
\partial _{i}\widetilde{A}_{i}=\partial _{i}A_{i}=0\;.  \label{eq44}
\end{equation}
As initial gauge configuration we shall take
\begin{equation}
A_{i}=\frac{i}{r^{2}}\varepsilon _{ijk}x_{j}\sigma
_{k}f(r)\;,\;\;\;\;\partial _{i}A_{i}=0\;,\;\;\;  \label{eq45}
\end{equation}
corresponding to setting $f_{1}=f_{3}=0$, $f_{2}=f$ in expression $\left(
\ref{eq40}\right) $. The gauge transformed field, eq.$\left( \ref{eq41}%
\right) $, is given by
\begin{equation}
\widetilde{A}_{i}=\left( f+\frac{1}{2}\right) \sin \alpha \frac{\partial
\widehat{n}}{\partial x_{i}}+\left( f+\frac{1}{2}\cos \alpha -\frac{1}{2}%
\right) \widehat{n}\frac{\partial \widehat{n}}{\partial x_{i}}+\frac{1}{2}%
\alpha ^{\prime }(r)\widehat{n}n_{i}\;.  \label{eq46}
\end{equation}
Also, the condition $\partial _{i}\widetilde{A}_{i}=0$, gives
\begin{equation}
\alpha ^{\prime \prime }(r)+\frac{2}{r}\alpha ^{\prime }(r)-\frac{4}{r^{2}}%
\left( f+\frac{1}{2}\right) \sin \alpha =0\;,  \label{eq47}
\end{equation}
or, $\tau =\log r$,
\begin{equation}
\frac{\partial ^{2}\alpha (\tau )}{\partial \tau ^{2}}+\frac{\partial \alpha
(\tau )}{\partial \tau }-4\left( f+\frac{1}{2}\right) \sin \alpha =0\;,
\label{eq48}
\end{equation}
which corresponds to the damped pendulum of Fig.2.

\vspace{1cm}

\begin{figure}[ht]
\centering \epsfig{file=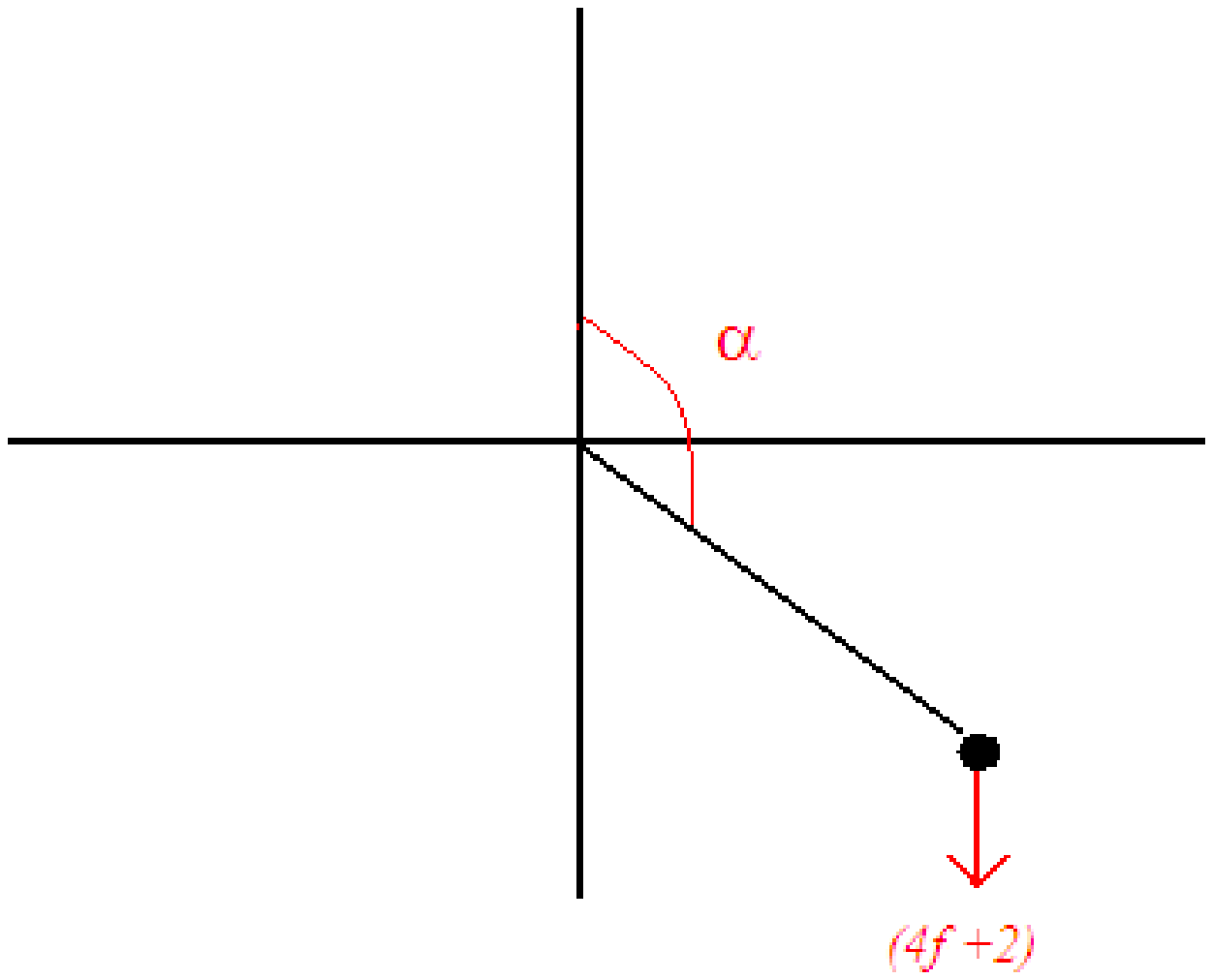,width=10cm} \caption{}
\end{figure}


\newpage

\noindent The presence of the function $f(r)$ in eq.$\left( \ref
{eq45}\right) $ is needed to ensure that expression $\left( \ref{eq45}%
\right) $ is a regular configuration. More precisely, we shall require that $%
A_{i}$ is regular at the origin, $r=0$, and that it goes to zero at
infinity, $r=\infty $. Essentially, we shall consider two types of boundary
conditions, namely the so called weak $\left( WBC\right) $ and strong $%
\left( SBC\right) $ boundary conditions \cite{sciuto,jm}.

\begin{itemize}
\item  For the $WBC$ we have
\begin{eqnarray}
&A_{i}&\;{is\; regular \; at\; the\; origin,} \;\;\;\;\;
f(r)_{r\rightarrow 0} \rightarrow  O(r)\;,  \nonumber \\
&A_{i}&\;{decays\;as\;1/r\;for\; r\rightarrow \infty},\;\;\;\;
f(r)_{r\rightarrow \infty } \rightarrow {constant}\;. \nonumber
\\
\label{eq49}
\end{eqnarray}

\item  For the $SBC$
\begin{eqnarray}
&A_{i}&\;{is\; regular \; at\; the\; origin,} \;\;\;\;\;
f(r)_{r\rightarrow 0} \rightarrow  O(r)\;,  \nonumber \\
&A_{i}&\;decays \; faster \; than\; 1/r \; for \; r \rightarrow
\infty, \;\;\;\;\; f(r)_{r\rightarrow \infty} \rightarrow 0 \;,
\nonumber \\
&(rA_{i})_{r\rightarrow \infty}& \rightarrow 0 \;. \label{eq50}
\end{eqnarray}
\end{itemize}

\noindent Let us begin with the case of $SBC$. Recalling that $\partial
\widehat{n}/\partial x_{i}\sim 1/r$ as $r\rightarrow \infty $, it follows
that the equivalent field $\widetilde{A}_{i}$ in eq.$\left( \ref{eq46}%
\right) $ will obey $SBC$ if
\begin{eqnarray}
\alpha (r) &\rightarrow &2\pi m+\gamma r\;\;\;\;{for\;\;}{%
r\rightarrow 0\;,\;\;\;\;\;}m{\;\;{integer}}  \label{eq51} \\
\alpha (\tau ) &\rightarrow &2\pi m+\gamma e^{\tau }\;\;{for}\;\;{%
\tau \rightarrow -\infty \;,}  \nonumber
\end{eqnarray}
and
\begin{eqnarray}
\alpha (r) &\rightarrow &2\pi n\;\;\;{for}\;\;{r\rightarrow
\infty }\;,\;\;\;\;n\;{{integer}}\;\;  \label{eq52} \\
\alpha (\tau ) &\rightarrow &2\pi n\;\;\;{for}\;\;{%
\tau \rightarrow \infty \;.}  \nonumber
\end{eqnarray}
Two situations$\;$are possible, according to the strength and the
orientation of the force $f$.

\begin{itemize}
\item  $i)\;$The first case corresponds to $\left( 1+2f\;\right) >0$, see
Fig.3.

\vspace{1cm}

\begin{figure}[ht]
\centering \epsfig{file=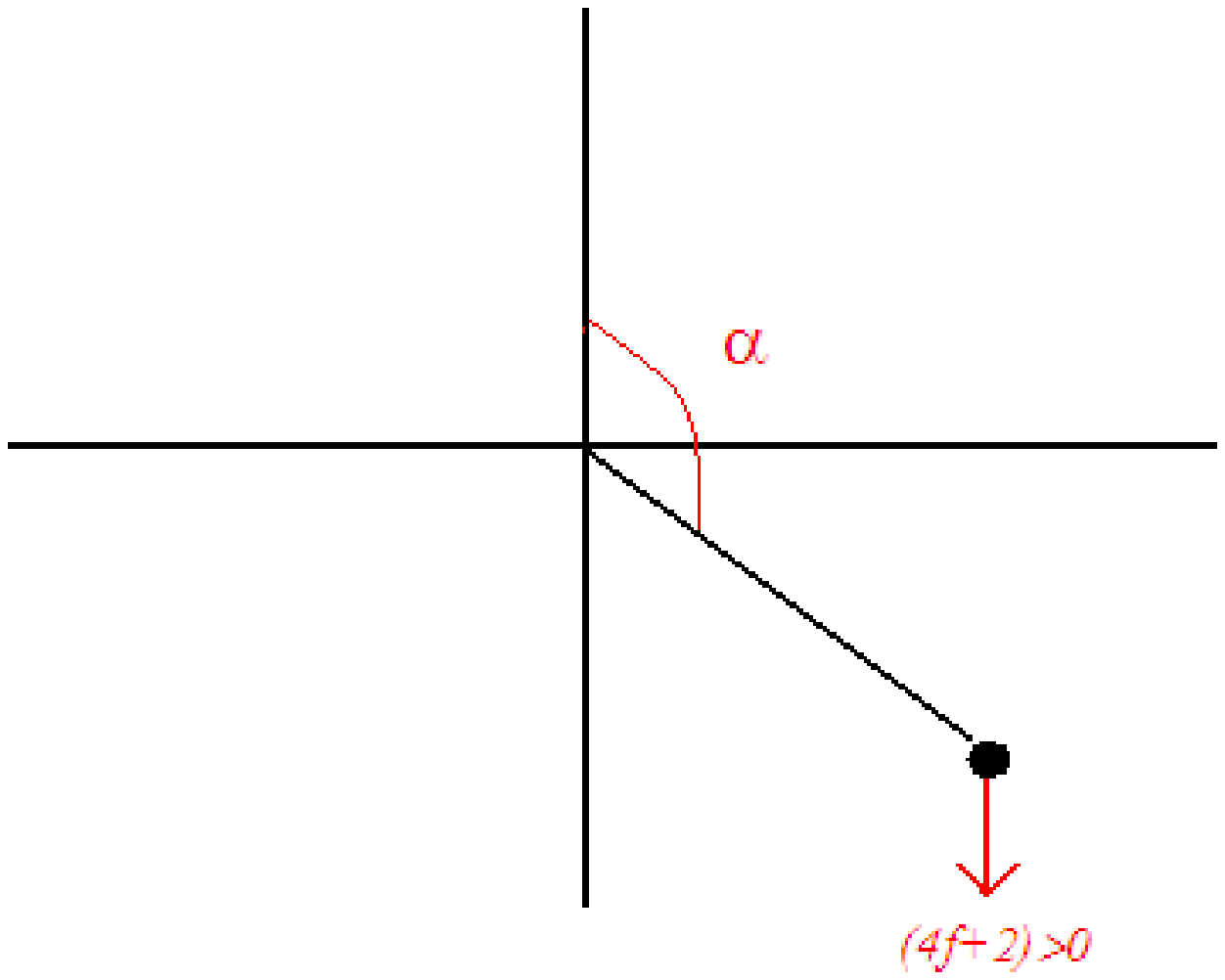,width=7cm} \caption{}
\end{figure}


\newpage

\noindent At $\tau \rightarrow -\infty $, the pendulum starts at a
position of unstable equilibrium, $\alpha =2\pi m$, with velocity
$\alpha ^{\prime }\sim \gamma e^{\tau }$, $\gamma \neq 0$, see
Fig.4.

\vspace{1cm}

\begin{figure}[ht]
\centering \epsfig{file=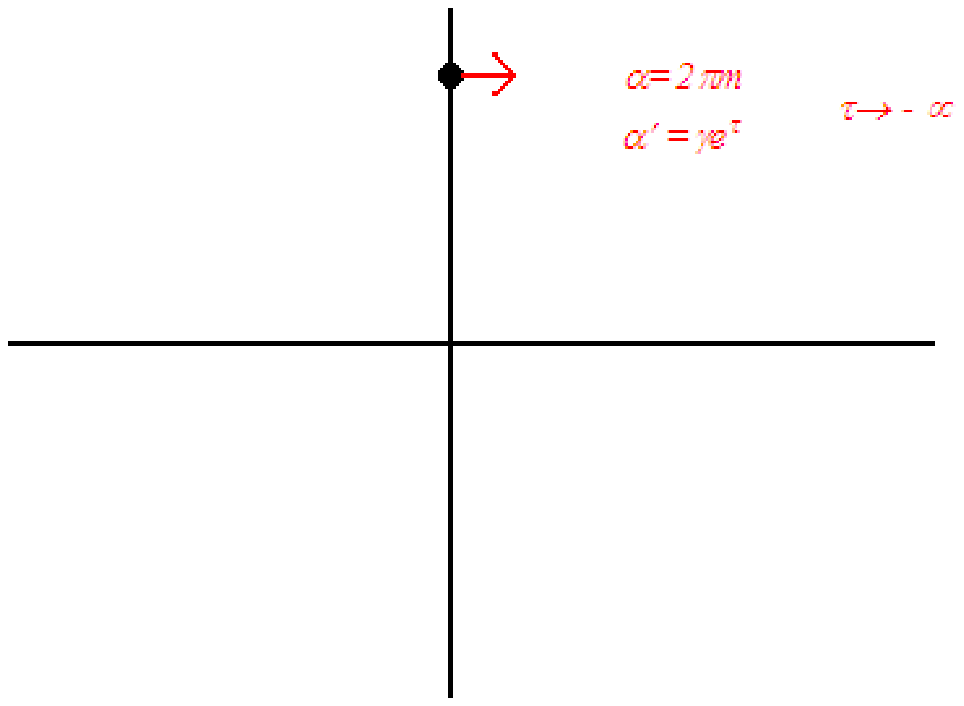,width=9cm} \caption{}
\end{figure}


\vspace{1cm}

\noindent For $\tau \rightarrow \infty$, after a certain number of
oscillations, the pendulum falls down in the
stable equilibrium point, see Fig.5, corresponding to $\alpha =(2p+1)\pi $, $%
p\;$integer, due to the force $\left( 1+2f\;\right) $ and to the
damping term. This situation does not correspond to $SBC$, since
$\alpha =(2p+1)\pi $ for $\tau \rightarrow \infty $. Rather, it
gives a copy $\widetilde{A}_i$ obeying $WBC$. Indeed, from
eq.$\left( \ref{eq46}\right) $ we get
\begin{equation}
\widetilde{A}_{r\rightarrow \infty }\sim \left( const. +
\frac{(\cos (2p+1)\pi) -1}{2}\right) \frac{1}{r}\sim
\frac{1}{r}\;. \label{eq53}
\end{equation}

\begin{figure}[ht]
\centering \epsfig{file=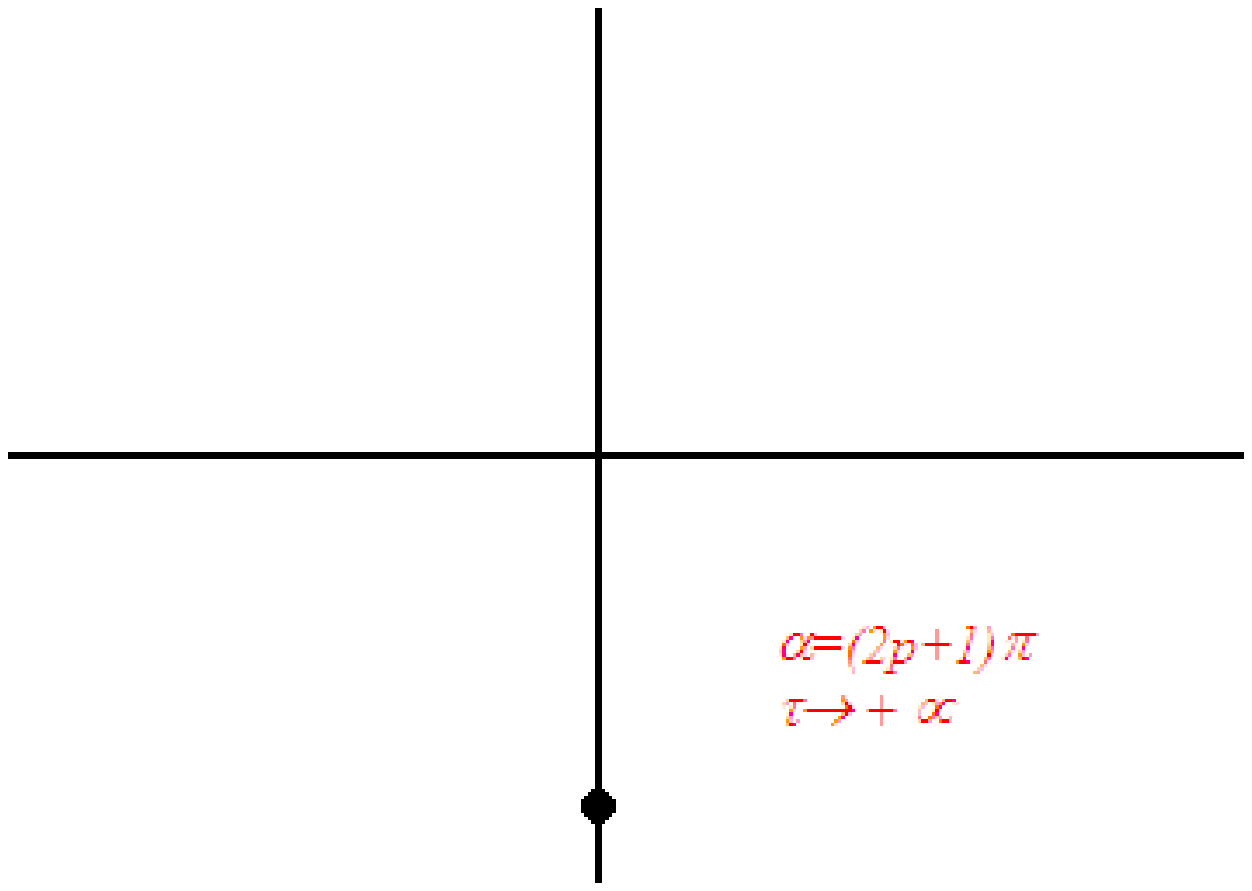,width=7cm} \caption{}
\end{figure}


\newpage

\item  $ii)$ The second case corresponds to a force $f$ which, for a
sufficiently large interval of time $\tau $, is negative enough, \textit{%
i.e., }$f<-1/2$, see Fig.6. In this case we have the following
configuration, see Fig.7. The pendulum starts at $\tau \rightarrow
-\infty $ from an unstable position, $\alpha =2\pi m$, with
velocity $\alpha ^{\prime }\sim \gamma e^{\tau }$. After some
oscillations, it can come back to the unstable position, $2\pi n$,
at $\tau \rightarrow \infty $, under the effect of the force $f$,
and finally it can remain there, see Fig.8. This situation
corresponds to $SBC$, implying that the field $\widetilde{A}_i$
decays faster than $1/r$.

\vspace{1cm}

\begin{figure}[ht]
\centering \epsfig{file=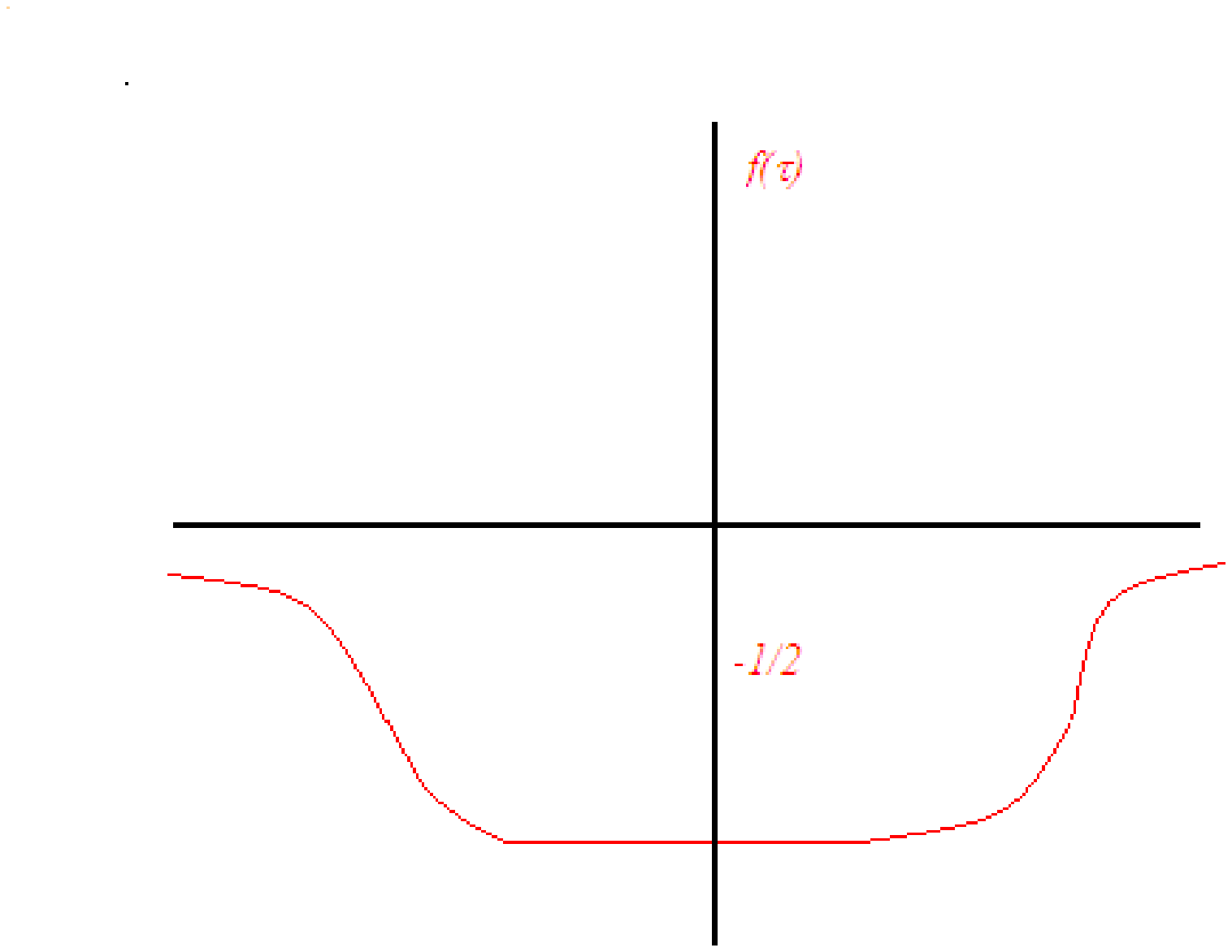,width=9cm} \caption{}
\end{figure}


\begin{figure}[ht]
\centering \epsfig{file=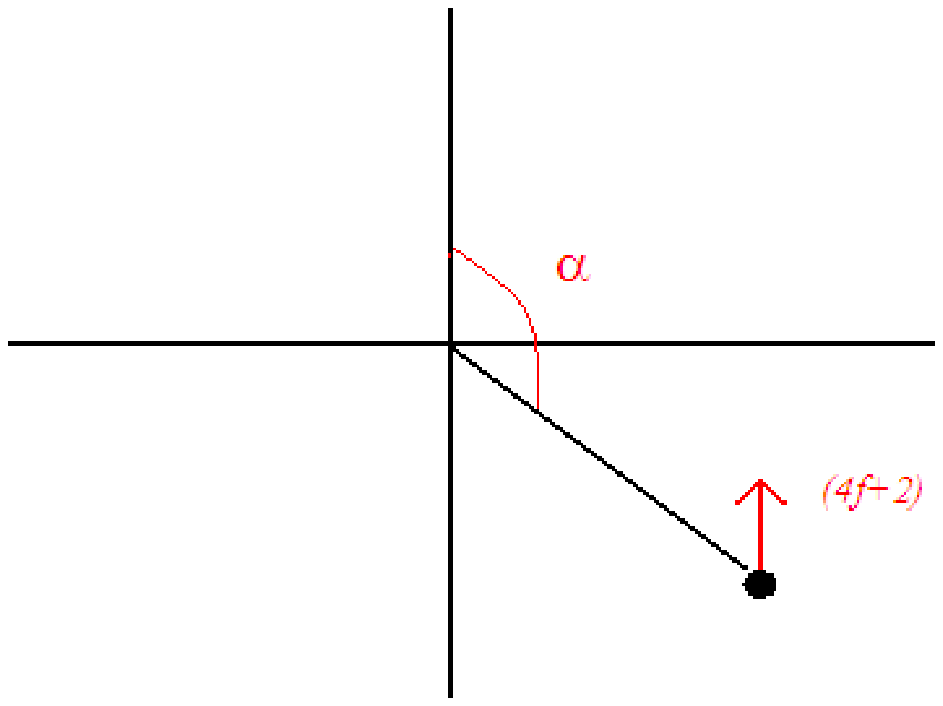,width=7cm} \caption{}
\end{figure}


\newpage

\begin{figure}[ht]
\centering \epsfig{file=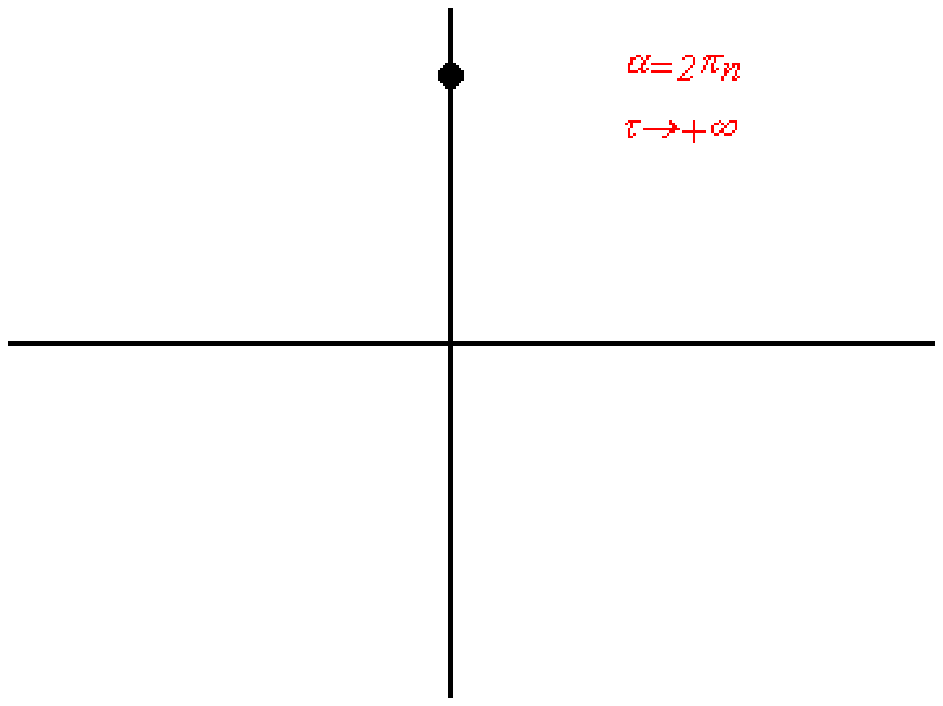,width=8cm} \caption{}
\end{figure}


\vspace{1cm}

\item  \textbf{Summary}

\textit{This example shows that, starting from the configuration
\begin{equation}
A_{i}=\frac{i}{r^{2}}\varepsilon _{ijk}x_{j}\sigma
_{k}f(r)\;,\;\;\;\;\partial _{i}A_{i}=0\;,\;\;  \label{eq54}
\end{equation}
we can obtain an equivalent field
\begin{equation}
\widetilde{A}_{i}={S}^{\dagger }\partial _{i}{S}+S^{\dagger }A_{i}{S\;,\;\;\;%
}\partial _{i}\widetilde{A}_{i}=0\;,  \label{eq55}
\end{equation}
compatible with both $WBC$ and $\;SBC.\;$ }

\textit{In the case of $SBC$%
\begin{eqnarray}
\alpha (r)_{r\rightarrow \infty } &\rightarrow &2\pi
n\;,\;\;\;\;S_{r\rightarrow \infty }\rightarrow \; const.  \label{eq56} \\
\left( r\widetilde{A}\right) _{r\rightarrow \infty } &\rightarrow &0\;.
\nonumber
\end{eqnarray}
}

\textit{For $WBC$
\begin{eqnarray}
&\alpha (r)_{r\rightarrow \infty }&  \rightarrow \left(
2p+1\right) \pi
\;,\;\;\;\;  \nonumber \\
&S_{r\rightarrow \infty }& \rightarrow \;\overrightarrow{\sigma
}\cdot
\overrightarrow{n}\;,\;S\;{depends\;on\;the\;orientation}%
,\;S_{r\rightarrow \infty }\rightarrow S(\theta ,\varphi )  \nonumber \\
&\widetilde{A}_{r\rightarrow \infty }& \sim \frac{1}{r}\;.
\label{eq57}
\end{eqnarray}
}
\end{itemize}

\newpage

\subsubsection{The winding number of the Gribov copies}

Let us evaluate now the winding number, see Appendix C, of the
gauge transformation $S$
\begin{equation}
S=e^{\frac{i}{2}\alpha (r)\overrightarrow{n}\cdot \overrightarrow{\sigma }%
}=\cos \frac{\alpha }{2}+i\overrightarrow{n}\cdot \overrightarrow{\sigma }%
\sin \frac{\alpha }{2}\;.  \label{w1}
\end{equation}
We shall discuss in detail the case of $SBC$, namely
\begin{equation}
\alpha (r)_{r\rightarrow \infty }\rightarrow 2\pi n\;,  \label{w2}
\end{equation}
so that
\begin{equation}
S_{r\rightarrow \infty }\rightarrow \; const.   \label{w3}
\end{equation}
Equation $\left( \ref{w3}\right) $ implies that the space
$\;R^{3}$ turns out to be compactified to the sphere $S^{3}$, by
identifying all points at infinity. We have to evaluate
\begin{equation}
\nu =\frac{1}{24\pi ^{2}}Tr\int dx_{1}dx_{2}dx_{3}\varepsilon _{ijk}\left(
S^{\dagger }\partial _{i}S\right) \left( S^{\dagger }\partial _{j}S\right)
\left( S^{\dagger }\partial _{k}S\right) \;.  \label{w4}
\end{equation}
From
\begin{equation}
S^{\dagger }\partial _{i}S=-\left( \partial _{i}S^{\dagger }\right) S\;,
\label{w5}
\end{equation}
we get
\begin{equation}
\nu =-\frac{1}{24\pi ^{2}}Tr\int d^{3}x\varepsilon _{ijk}\left(
\partial _{i}S\right) \left( S^{\dagger }\partial _{j}S\right)
\left( \partial _{k}S^{\dagger }\right) \;.  \label{w6}
\end{equation}
It is useful to employ the notation
\begin{eqnarray}
S &=&N_{4}+i\overrightarrow{N}\cdot \overrightarrow{\sigma }%
\;,\;\;\;\;\;\;S^{\dagger }=N_{4}-i\overrightarrow{N}\cdot \overrightarrow{%
\sigma }\;,  \nonumber \\
N_{4} &=&\cos \frac{\alpha }{2}\;,\;\;\;\;N_{i}=n_{i}\sin \frac{\alpha }{2}%
\;,\;\;\;\;  \label{w7} \\
N_{a}N_{a} &=&1\;,\;\;\;\;a=1,2,3,4\;.  \nonumber
\end{eqnarray}
In what follows we shall take in due account the surface terms originating
from integration by parts. Thus
\begin{eqnarray}
\nu &=&-\frac{1}{24\pi ^{2}}Tr\int d^{3}x\varepsilon _{ijk}\left( \partial
_{i}N_{4}+i\left( \partial _{i}\overrightarrow{N}\cdot \overrightarrow{%
\sigma }\right) \right) \left( N_{4}-i\overrightarrow{N}\cdot
\overrightarrow{\sigma }\right)  \nonumber \\
&&\times \left( \partial _{j}N_{4}+i\left( \overrightarrow{\partial _{j}N}%
\cdot \overrightarrow{\sigma }\right) \right) \left( \partial
_{k}N_{4}-i\left( \partial _{k}\overrightarrow{N}\cdot \overrightarrow{%
\sigma }\right) \right) \;  \nonumber \\
&=&-\frac{1}{24\pi ^{2}}Tr\int d^{3}x\varepsilon _{ijk}\left[ \left(
\partial _{i}N_{4}\left( N_{4}-i\overrightarrow{N}\cdot \overrightarrow{%
\sigma }\right) +\left( \partial _{i}\overrightarrow{N}\cdot \overrightarrow{%
\sigma }\right) \left( iN_{4}+\overrightarrow{N}\cdot \overrightarrow{\sigma
}\right) \right) \;\right.  \nonumber \\
&&\times \left. \left( \partial _{j}N_{4}\left( \partial
_{k}N_{4}-i\partial _{k}\overrightarrow{N}\cdot
\overrightarrow{\sigma }\right) +\left( \partial
_{j}\overrightarrow{N}\cdot \overrightarrow{\sigma }\right) \left(
i\partial
_{k}N_{4}+\partial _{k}\overrightarrow{N}\cdot \overrightarrow{\sigma }%
\right) \right) \right] \;.  \nonumber \\
&&\;  \label{w88}
\end{eqnarray}
Since
\begin{equation}
\varepsilon _{ijk}\partial _{j}N_{4}\partial _{k}N_{4}=0\;,  \label{w9}
\end{equation}
we have
\begin{eqnarray}
\nu &=&-\frac{Tr}{24\pi ^{2}}\int d^{3}x\varepsilon _{ijk}\left[ \partial
_{i}N_{4}\left( \left( N_{4}-i\overrightarrow{N}\cdot \overrightarrow{\sigma
}\right) \left( \partial _{j}\overrightarrow{N}\cdot \overrightarrow{\sigma }%
\right) \left( \partial _{k}\overrightarrow{N}\cdot \overrightarrow{\sigma }%
\right) \right) \right.  \nonumber \\
&&+N_{4}\left( \partial _{i}\overrightarrow{N}\cdot \overrightarrow{\sigma }%
\right) \left( \left( \partial _{j}N_{4}\right) \left( \partial _{k}%
\overrightarrow{N}\cdot \overrightarrow{\sigma }\right) -\left( \partial
_{k}N_{4}\right) \left( \partial _{j}\overrightarrow{N}\cdot \overrightarrow{%
\sigma }\right) \right)  \nonumber \\
&&+\left( \partial _{i}\overrightarrow{N}\cdot \overrightarrow{\sigma }%
\right) \left( iN_{4}\left( \partial _{j}\overrightarrow{N}\cdot
\overrightarrow{\sigma }\right) \left( \partial _{k}\overrightarrow{N}\cdot
\overrightarrow{\sigma }\right) -i\left( \partial _{j}N_{4}\right) \left(
\overrightarrow{N}\cdot \overrightarrow{\sigma }\right) \left( \partial _{k}%
\overrightarrow{N}\cdot \overrightarrow{\sigma }\right) \right)  \nonumber \\
&&\left. +\left( \partial _{i}\overrightarrow{N}\cdot \overrightarrow{\sigma
}\right) \left( \overrightarrow{N}\cdot \overrightarrow{\sigma }\right)
\left( \partial _{j}\overrightarrow{N}\cdot \overrightarrow{\sigma }\right)
\left( i\partial _{k}N_{4}+\partial _{k}\overrightarrow{N}\cdot
\overrightarrow{\sigma }\right) \right] \;.  \nonumber \\
&&  \label{w10}
\end{eqnarray}
From
\begin{equation}
\sigma _{i}\sigma _{j}=\delta _{ij}+i\varepsilon _{ijk}\sigma _{k}\;,
\label{w10b}
\end{equation}
we find
\begin{eqnarray}
Tr\sigma _{i}\sigma _{j} &=&2\delta _{ij}\;,  \label{w11} \\
Tr\left( \sigma _{i}\sigma _{j}\sigma _{k}\right) &=&2i\varepsilon _{ijk}\;,
\nonumber \\
Tr\left( \sigma _{i}\sigma _{j}\sigma _{k}\sigma _{p}\right) &=&2\left(
\delta _{ij}\delta _{kp}-\delta _{ik}\delta _{jp}+\delta _{ip}\delta
_{jk}\right) \;.  \nonumber
\end{eqnarray}
Furthermore, observing that
\begin{eqnarray}
&&Tr\varepsilon _{ijk}\left( \partial _{i}N_{m}\right) N_{p}\left( \partial
_{j}N_{q}\right) \left( \partial _{k}N_{r}\right) \sigma _{m}\sigma
_{p}\sigma _{q}\sigma _{r}  \nonumber \\
&=&2\varepsilon _{ijk}\left( \partial _{i}N_{m}\right) N_{p}\left( \partial
_{j}N_{q}\right) \left( \partial _{k}N_{r}\right) \left( \delta _{mp}\delta
_{qr}-\delta _{mq}\delta _{pr}+\delta _{mr}\delta _{pq}\right)  \nonumber \\
&=&2\varepsilon _{ijk}\left[ \left( \partial _{i}N_{m}\right) N_{m}\left(
\partial _{j}N_{q}\right) \left( \partial _{k}N_{q}\right) -\left( \partial
_{i}N_{m}\right) N_{p}\left( \partial _{j}N_{m}\right) \left( \partial
_{k}N_{p}\right) \right]  \nonumber \\
&&+2\varepsilon _{ijk}\left( \partial _{i}N_{m}\right) N_{p}\left( \partial
_{j}N_{p}\right) \left( \partial _{k}N_{m}\right)  \nonumber \\
&=&0\;,  \label{w12}
\end{eqnarray}
it follows
\begin{equation}
\nu =\frac{1}{24\pi ^{2}}\int d^{3}x\varepsilon _{ijk}\varepsilon
_{mpq}\left[ 2N_{4}\left( \partial _{i}N_{m}\right) \left( \partial
_{j}N_{p}\right) \left( \partial _{k}N_{q}\right) -6\left( \partial
_{i}N_{4}\right) N_{m}\left( \partial _{j}N_{p}\right) \left( \partial
_{k}N_{q}\right) \right] \;.  \label{w13}
\end{equation}
Integrating by parts the second term, one obtains
\begin{eqnarray}
\nu &=&\frac{1}{3\pi ^{2}}\int d^{3}x\varepsilon _{ijk}\varepsilon
_{mpq}N_{4}\left( \partial _{i}N_{m}\right) \left( \partial _{j}N_{p}\right)
\left( \partial _{k}N_{q}\right)  \nonumber \\
&&-\frac{1}{4\pi ^{2}}\int d^{3}x\varepsilon _{ijk}\partial _{i}\left(
\varepsilon _{mpq}N_{4}N_{m}\left( \partial _{j}N_{p}\right) \left( \partial
_{k}N_{q}\right) \right) \;.  \label{w14}
\end{eqnarray}
Concerning the surface term in eq.$\left( \mathrm{{\ref{w14}}}\right) $, it
turns out that
\begin{eqnarray}
&&\varepsilon _{ijk}\varepsilon _{mpq}N_{4}N_{m}\left( \partial
_{j}N_{p}\right) \left( \partial _{k}N_{q}\right)  \nonumber \\
&=&\cos \frac{\alpha }{2}\sin \frac{\alpha }{2}n_{m}\left( \partial
_{j}\left( n_{p}\sin \frac{\alpha }{2}\right) \right) \left( \partial
_{k}\left( n_{q}\sin \frac{\alpha }{2}\right) \right) \varepsilon
_{ijk}\varepsilon _{mpq}  \nonumber \\
&=&\cos \frac{\alpha }{2}\sin ^{3}\frac{\alpha }{2}n_{m}\left( \partial
_{j}n_{p}\right) \left( \partial _{k}n_{q}\right) \varepsilon
_{ijk}\varepsilon _{mpq}  \nonumber \\
&=&\frac{1}{r^{2}}\cos \frac{\alpha }{2}\sin ^{3}\frac{\alpha }{2}%
n_{m}\left( \delta _{jp}-n_{j}n_{p}\right) \left( \delta
_{kq}-n_{k}n_{q}\right) \varepsilon _{ijk}\varepsilon _{mpq}  \nonumber \\
&=&\frac{1}{r^{2}}\cos \frac{\alpha }{2}\sin ^{3}\frac{\alpha }{2}%
n_{m}\varepsilon _{ijk}\varepsilon _{mjk}=\frac{2}{r^{2}}\cos \frac{\alpha }{%
2}\sin ^{3}\frac{\alpha }{2}n_{i}\;.  \label{w15}
\end{eqnarray}
Thus
\begin{equation}
\int d^{3}x\varepsilon _{ijk}\partial _{i}\left( \varepsilon
_{mpq}N_{4}N_{m}\left( \partial _{j}N_{p}\right) \left( \partial
_{k}N_{q}\right) \right) =\int d^{3}x\partial _{i}\left( \frac{2}{r^{2}}\cos
\frac{\alpha }{2}\sin ^{3}\frac{\alpha }{2}n_{i}\right) \;.  \label{w16}
\end{equation}
Moreover, recalling that $\alpha (r)_{r\rightarrow \infty }\rightarrow 2\pi
n $, it follows that the surface term $\left( \ref{w16}\right) $ vanishes.

\begin{itemize}
\item  \textbf{Remark}

\textit{It is worth noticing that the surface term }$\left( \ref{w16}\right)
$ \textit{also vanishes in the case of $WBC$, due to the presence of }$\cos
\frac{\alpha }{2}$. \textit{Indeed, for $WBC$, }$\alpha (r)_{r\rightarrow
\infty }\rightarrow \left( 2p+1\right) \pi $, \textit{so that }$\cos \frac{%
\alpha (\infty )}{2}=\cos (p\pi +\frac{\pi }{2})=0.\;$\textit{Thus, for both
$SBC$ and $WBC$, we have}
\begin{equation}
\nu =\frac{1}{3\pi ^{2}}\int d^{3}x\varepsilon _{ijk}\varepsilon
_{mpq}N_{4}\left( \partial _{i}N_{m}\right) \left( \partial _{j}N_{p}\right)
\left( \partial _{k}N_{q}\right) \;.  \label{w18}
\end{equation}
\end{itemize}
\noindent Let us proceed then with the evaluation of expression
$\left( \ref {w18}\right) $. We have
\begin{eqnarray}
\nu &=&\frac{1}{3\pi ^{2}}\int d^{3}x\varepsilon _{ijk}\varepsilon
_{mpq}\cos \frac{\alpha }{2}\left( \partial _{i}\left( n_{m}\sin \frac{%
\alpha }{2}\right) \right) \left( \partial _{j}\left( n_{p}\sin \frac{\alpha
}{2}\right) \right) \left( \partial _{k}\left( n_{q}\sin \frac{\alpha }{2}%
\right) \right)  \nonumber \\
&=&\frac{1}{3\pi ^{2}}\int d^{3}x\varepsilon _{ijk}\varepsilon _{mpq}\cos
\frac{\alpha }{2}\left[ \left( \sin \frac{\alpha }{2}\left( \partial
_{i}n_{m}\right) +\frac{n_{m}}{2}\cos \frac{\alpha }{2}\left( \partial
_{i}\alpha \right) \right) \right.  \nonumber \\
&&\left. \times \left( \sin \frac{\alpha }{2}\left( \partial
_{j}n_{p}\right) +\frac{n_{p}}{2}\cos \frac{\alpha }{2}\left( \partial
_{j}\alpha \right) \right) \left( \sin \frac{\alpha }{2}\left( \partial
_{k}n_{q}\right) +\frac{n_{q}}{2}\cos \frac{\alpha }{2}\left( \partial
_{k}\alpha \right) \right) \right] \;  \nonumber \\
&=&\frac{1}{3\pi ^{2}}\int d^{3}x\varepsilon _{ijk}\varepsilon _{mpq}\cos
\frac{\alpha }{2}\left[ \left( \sin ^{2}\frac{\alpha }{2}\left( \partial
_{i}n_{m}\right) \left( \partial _{j}n_{p}\right) +\frac{n_{p}}{2}\sin \frac{%
\alpha }{2}\cos \frac{\alpha }{2}\left( \partial _{i}n_{m}\right) \left(
\partial _{j}\alpha \right) \right. \right.  \nonumber \\
&&\left. \left. +\frac{n_{m}}{2}\sin \frac{\alpha }{2}\cos \frac{\alpha }{2}%
\left( \partial _{i}\alpha \right) \left( \partial _{j}n_{p}\right) \right)
\times \left( \sin \frac{\alpha }{2}\left( \partial _{k}n_{q}\right) +\frac{%
n_{q}}{2}\cos \frac{\alpha }{2}\left( \partial _{k}\alpha \right) \right)
\right] \;  \nonumber \\
&=&\frac{1}{3\pi ^{2}}\int d^{3}x\varepsilon _{ijk}\varepsilon _{mpq}\cos
\frac{\alpha }{2}\left( \sin ^{3}\frac{\alpha }{2}\left( \partial
_{i}n_{m}\right) \left( \partial _{j}n_{p}\right) \left( \partial
_{k}n_{q}\right) \right.  \nonumber \\
&&\;\;\;\;\;\;\;\;\;\;+\sin ^{2}\frac{\alpha }{2}\cos \frac{\alpha }{2}\frac{%
n_{q}}{2}\left( \partial _{i}n_{m}\right) \left( \partial _{j}n_{p}\right)
\left( \partial _{k}\alpha \right)  \nonumber \\
&&\;\;\;\;\;\;\;\;\;\;+\frac{n_{p}}{2}\sin ^{2}\frac{\alpha }{2}\cos \frac{%
\alpha }{2}\left( \partial _{i}n_{m}\right) \left( \partial _{j}\alpha
\right) \left( \partial _{k}n_{q}\right)  \nonumber \\
&&\;\;\;\;\;\;\;\;\;\; \left. +\frac{n_{m}}{2}\sin ^{2}\frac{\alpha }{2}\cos
\frac{\alpha }{2}\left( \partial _{i}\alpha \right) \left( \partial
_{j}n_{p}\right) \left( \partial _{k}n_{q}\right) \right)  \nonumber \\
&=&\frac{1}{3\pi ^{2}}\int d^{3}x\varepsilon _{ijk}\varepsilon _{mpq}\cos
\frac{\alpha }{2}\left( \sin ^{3}\frac{\alpha }{2}\left( \partial
_{i}n_{m}\right) \left( \partial _{j}n_{p}\right) \left( \partial
_{k}n_{q}\right) \;\right.  \nonumber \\
&&\left. +\frac{3}{2}\sin ^{2}\frac{\alpha }{2}\cos \frac{\alpha }{2}%
n_{q}\left( \partial _{i}n_{m}\right) \left( \partial _{j}n_{p}\right)
\left( \partial _{k}\alpha \right) \right)  \nonumber \\
&&  \label{w19}
\end{eqnarray}
Furthermore
\begin{eqnarray}
\nu &=&\frac{1}{3\pi ^{2}}\int d^{3}x\frac{1}{r^{3}}\cos \frac{\alpha }{2}%
\sin ^{3}\frac{\alpha }{2}\varepsilon _{ijk}\varepsilon _{mpq}\left( \delta
_{im}-n_{i}n_{m}\right) \left( \delta _{jp}-n_{j}n_{p}\right) \left( \delta
_{kq}-n_{k}n_{q}\right)  \nonumber \\
&&+\frac{1}{2\pi ^{2}}\int d^{3}x\sin ^{2}\frac{\alpha }{2}\cos ^{2}\frac{%
\alpha }{2}\alpha ^{\prime }\frac{1}{r^{2}}\varepsilon _{ijk}\varepsilon
_{mpq}n_{q}n_{k}\left( \delta _{im}-n_{i}n_{m}\right) \left( \delta
_{jp}-n_{j}n_{p}\right) \;  \nonumber \\
&=&\frac{1}{3\pi ^{2}}\int d^{3}x\frac{1}{r^{3}}\cos \frac{\alpha
}{2}\sin ^{3}\frac{\alpha }{2}\varepsilon _{ijk}\varepsilon
_{mpq}\left( \left( \delta _{im}\delta _{jp}-\delta
_{im}n_{j}n_{p}-\delta _{jp}n_{i}n_{m}\right) \left( \delta
_{kq}-n_{k}n_{q}\right) \right)
\nonumber \\
&&+\frac{1}{2\pi ^{2}}\int d^{3}x\sin ^{2}\frac{\alpha }{2}\cos ^{2}\frac{%
\alpha }{2}\alpha ^{\prime }\frac{1}{r^{2}}\varepsilon _{ijk}\varepsilon
_{mpq}n_{q}n_{k}\left( \delta _{im}\delta _{jp}-\delta
_{im}n_{j}n_{p}-\delta _{jp}n_{i}n_{m}\right)  \nonumber \\
&=&\frac{1}{3\pi ^{2}}\int d^{3}x\frac{1}{r^{3}}\cos \frac{\alpha
}{2}\sin ^{3}\frac{\alpha }{2}\varepsilon _{ijk}\varepsilon
_{mpq}\left( \delta _{im}\left( \delta _{jp}\delta _{kq}-\delta
_{jp}n_{k}n_{q}-\delta _{kq}n_{j}n_{p}\right) -\delta _{jp}\delta
_{kq}n_{i}n_{m}\right)  \nonumber \\
&&+\frac{1}{2\pi ^{2}}\int d^{3}x\sin ^{2}\frac{\alpha }{2}\cos ^{2}\frac{%
\alpha }{2}\alpha ^{\prime }\frac{1}{r^{2}}\varepsilon
_{ijk}\varepsilon
_{mpq}n_{q}n_{k}\delta _{im}\delta _{jp}  \nonumber \\
&=&\frac{1}{3\pi ^{2}}\int d^{3}x\frac{1}{r^{3}}\cos \frac{\alpha
}{2}\sin ^{3}\frac{\alpha }{2}\varepsilon _{ijk}\varepsilon
_{mpq}\left( \delta _{im}\delta _{jp}\delta _{kq}-3\delta
_{im}\delta _{jp}n_{k}nq\right)
\nonumber \\
&&+\frac{1}{2\pi ^{2}}\int d^{3}x\sin ^{2}\frac{\alpha }{2}\cos ^{2}\frac{%
\alpha }{2}\alpha ^{\prime }\frac{1}{r^{2}}\varepsilon
_{ijk}\varepsilon
_{ijq}n_{q}n_{k}  \nonumber \\
&=&\frac{1}{3\pi ^{2}}\int d^{3}x\frac{1}{r^{3}}\cos \frac{\alpha
}{2}\sin ^{3}\frac{\alpha }{2}\left( \varepsilon _{ijk}\varepsilon
_{ijk}-3\varepsilon _{ijk}\varepsilon _{ijq}n_{k}nq\right)  \nonumber \\
&&+\frac{1}{\pi ^{2}}\int d^{3}x\sin ^{2}\frac{\alpha }{2}\cos ^{2}\frac{%
\alpha }{2}\alpha ^{\prime }\frac{1}{r^{2}}  \nonumber \\
%
%
&=&\frac{1}{\pi ^{2}}\int d^{3}x\sin ^{2}\frac{\alpha }{2}\cos ^{2}\frac{%
\alpha }{2}\alpha ^{\prime }\frac{1}{r^{2}}  \label{w20}
\end{eqnarray}
In order to evaluate this last integral we shall make use of polar
coordinates
\begin{eqnarray}
\nu &=&\frac{4}{\pi }\int_{0}^{\infty }dr\sin ^{2}\frac{\alpha }{2}\cos ^{2}%
\frac{\alpha }{2}\alpha ^{\prime }=-\frac{1}{4\pi }\int_{0}^{\infty
}dr\alpha ^{\prime }\left( e^{i\frac{\alpha }{2}}+e^{-i\frac{\alpha }{2}%
}\right) ^{2}\left( e^{i\frac{\alpha }{2}}-e^{-i\frac{\alpha }{2}}\right)
^{2}  \nonumber \\
&=&-\frac{1}{4\pi }\int_{0}^{\infty }dr\alpha ^{\prime }\left( e^{i\alpha
}+e^{-i\alpha }+2\right) \left( e^{i\alpha }+e^{-i\alpha }-2\right)
\nonumber \\
&=&\frac{1}{2\pi }\int_{0}^{\infty }dr\alpha ^{\prime }\left( 1-\cos 2\alpha
\right) \;.  \label{w21}
\end{eqnarray}
\noindent Finally, for the winding number $\nu $ associated to the
gauge transformation $S$, we find
\begin{equation}
\nu =\frac{1}{2\pi }\left( \alpha (\infty )-\alpha (0)\right) -\frac{1}{4\pi
}\left( \sin 2\alpha (\infty )-\sin 2\alpha (0)\right) \;.  \label{w22}
\end{equation}

\vspace{1cm}

\begin{itemize}
\item  \textbf{Summary}

\textit{Let us consider first the case of }$SBC.$\textit{\ From }
\begin{eqnarray}
\alpha (r)_{r\rightarrow \infty } &\rightarrow &2\pi n\;,\;\;\;n\;\;\;\;{%
integer}\;  \nonumber \\
\alpha (r)_{r\rightarrow 0} &\rightarrow &2\pi m\;,\;\;\;m\;\;{integer%
}\;,  \label{w23}
\end{eqnarray}
\textit{it follows}
\begin{equation}
\nu _{SBC}=\;\;{{integer}}  \label{w24}
\end{equation}

\vspace{1cm}

\textit{\vspace{1cm} For $WBC$
\begin{eqnarray}
\alpha (r)_{r\rightarrow \infty } &\rightarrow &\left( 2p+1\right) \pi
\;,\;\;\;\;p\;\;{integer}\;  \nonumber \\
\alpha (r)_{r\rightarrow 0} &\rightarrow &2\pi m\;,\;\;\;\;\;\;\;\;\;\;m\;\;{\;%
}{integer}{\;},  \label{w25}
\end{eqnarray}
we have
\begin{equation}
\nu _{WBC}=\;\left( {integer}+\frac{1}{2}\right) \;. \label{w26}
\end{equation}}

\item  \textbf{Remark}

{\it It should be noted that, in the case of WBC, the expression $\left( \ref
{w4}\right)$ does not have the meaning of a winding number. In fact, due to the behavior of the gauge
transformation $S$
at infinity, eq.$\left( \ref
{eq57}\right)$, the space $\;R^{3}$ cannot be compactified to the sphere $S^{3}$. As a consequence, expression $\left( \ref
{w4}\right)$ may now take non integer values.
}
\end{itemize}

\newpage

\subsubsection{The Gribov copies of the vacuum}

Let us discuss now the existence of the Gribov copies of the vacuum, $%
A_{i}=0 $, which obviously satisfies $\partial _{i}A_{i}=0$. We look at the
equivalent field
\begin{eqnarray}
\widetilde{A}_{i} &=&{S}^{\dagger }\partial _{i}{S\;=}\frac{1}{2}\sin \alpha
\frac{\partial \widehat{n}}{\partial x_{i}}+\left( \frac{1}{2}\cos \alpha -%
\frac{1}{2}\right) \widehat{n}\frac{\partial \widehat{n}}{\partial x_{i}}+%
\frac{1}{2}\alpha ^{\prime }(r)\widehat{n}n_{i}\;,  \nonumber \\
S &=&e^{\frac{i}{2}\alpha (r)\overrightarrow{n}\cdot \overrightarrow{\sigma }%
}=\cos \frac{\alpha }{2}+i\overrightarrow{n}\cdot \overrightarrow{\sigma }%
\sin \frac{\alpha }{2}\;,  \label{eq58}
\end{eqnarray}
which corresponds to a pure gauge
\begin{equation}
\widetilde{A}_i=S^{\dagger}\partial_i S \;, \;\;\;\;
F_{ij}(\widetilde{A})=0 \;. \nonumber
\end{equation}
The gauge condition $\partial _{i}\widetilde{A}_{i}=0$ reads now
\begin{equation}
\alpha ^{\prime \prime }(r)+\frac{2}{r}\alpha ^{\prime }(r)-\frac{2}{r^{2}}%
\sin \alpha =0\;,  \label{eq59}
\end{equation}
or, $\tau =\log r$,
\begin{equation}
\frac{\partial ^{2}\alpha (\tau )}{\partial \tau ^{2}}+\frac{\partial \alpha
(\tau )}{\partial \tau }-2\sin \alpha =0\;,  \label{eq60}
\end{equation}
corresponding to the damped pendulum of Fig.9.

\vspace{1cm}

\begin{figure}[ht]
\centering \epsfig{file=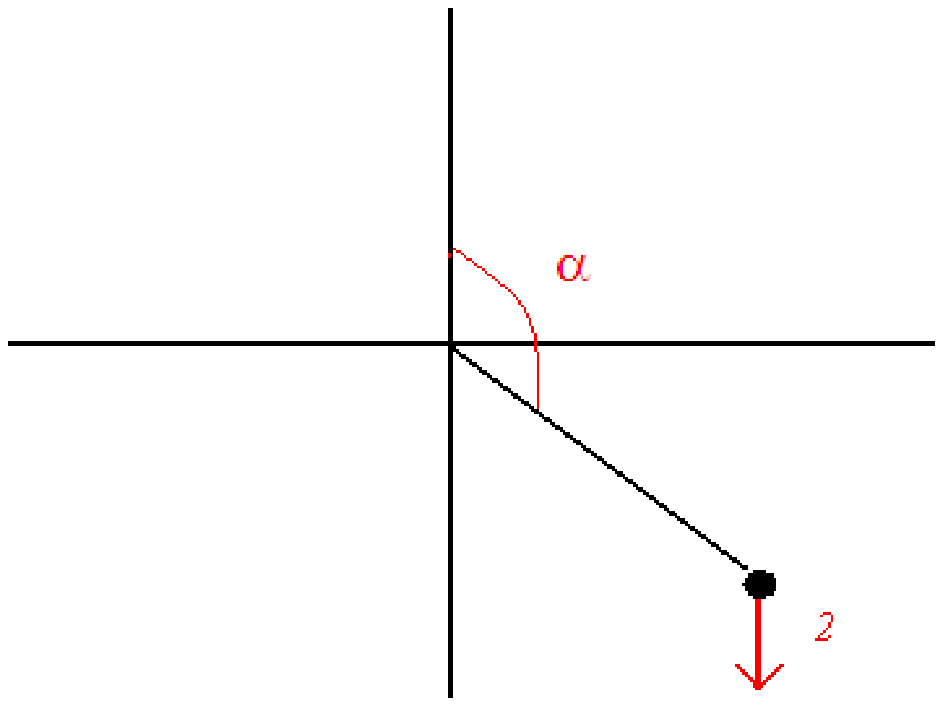,width=8cm} \caption{}
\end{figure}


\vspace{1cm}

\noindent The regularity condition of the equivalent field
$\widetilde{A}_{i}$ at the origin, $r=0$, $\tau =-\infty $, gives
\begin{eqnarray}
\alpha (r) &\rightarrow &2\pi m+\gamma r\;\;{for\;} {\
r\rightarrow 0} \;,\;\;\;\;\;m\;\;{{integer}}  \label{eq61} \\
\alpha (\tau ) &\rightarrow &2\pi m+\gamma e^{\tau }\;{for}\; {%
\tau \rightarrow -\infty \;,}  \nonumber
\end{eqnarray}
which holds for both $WBC$ and $SBC$. Since the force $f$ is now absent, the
pendulum starts at $\tau \rightarrow -\infty $ from the unstable position, $%
\alpha =2\pi m$, with velocity $\alpha ^{\prime }=\gamma e^{\tau }$, see
Fig.10.

\vspace{1cm}

\begin{figure}[ht]
\centering \epsfig{file=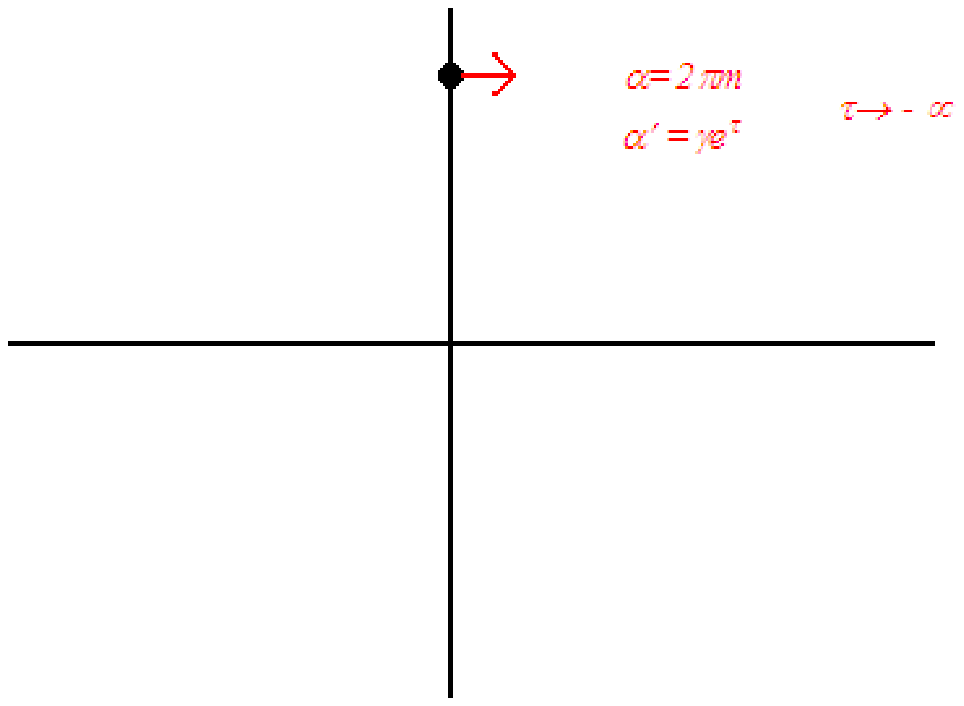,width=7cm} \caption{}
\end{figure}


\vspace{1cm}


\noindent After a certain number of oscillations it comes to the stable position $%
\alpha _{\tau \rightarrow \infty }=\left( 2p+1\right) \pi $, due
to the constant force and to the damping, see Fig.11.
\vspace{1cm}
\begin{figure}[ht]
\centering \epsfig{file=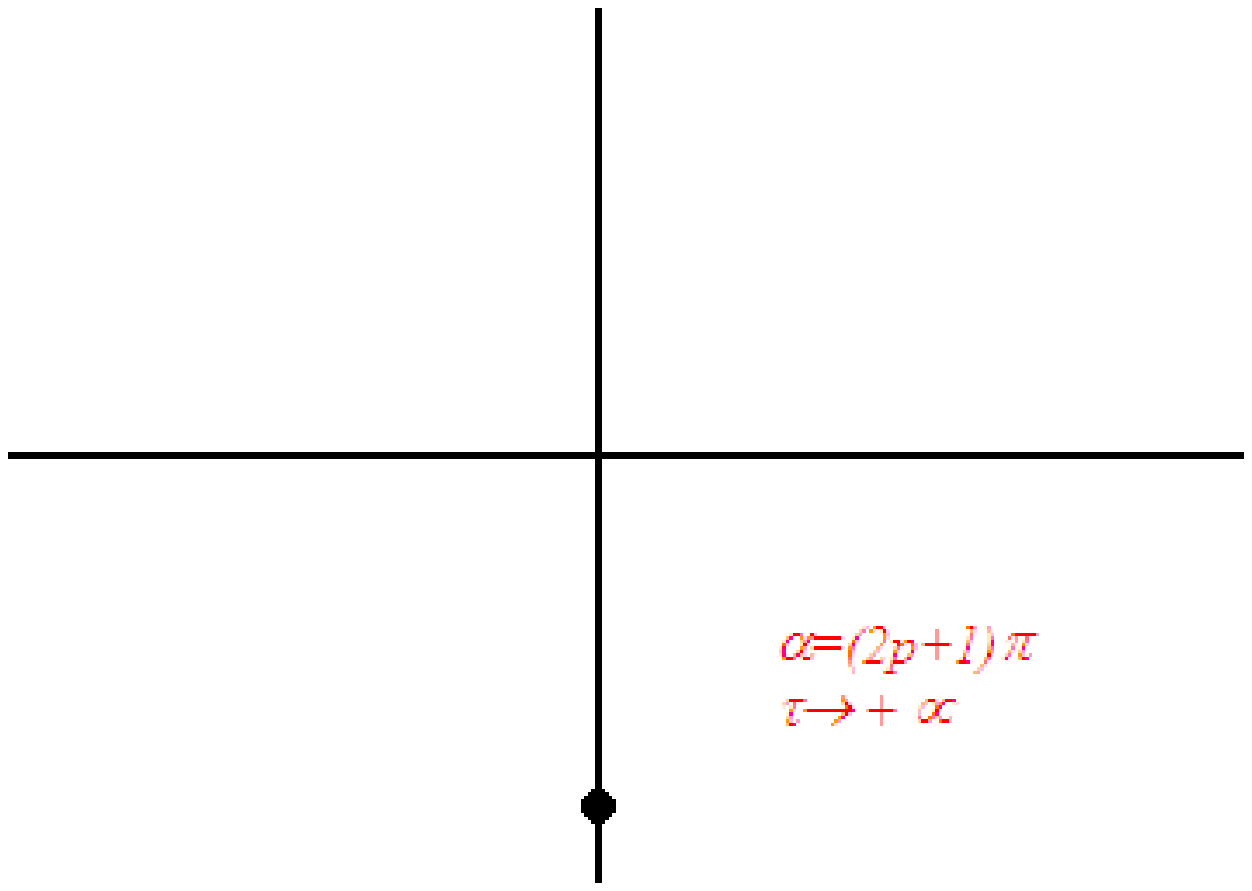,width=7cm} \caption{}
\end{figure}


\vspace{2cm}

\noindent Thus, in the absence of the force $f$, there are no
Gribov copies of the vacuum if $SBC$ are imposed. However, if
$WBC$ are adopted, even the vacuum has Gribov copies. As we have
already seen, this corresponds to an equivalent field
$\widetilde{A}_{i}$ which, for $\alpha _{\tau \rightarrow \infty
}=\left( 2p+1\right) \pi $, behaves as
\begin{equation}
\widetilde{A}_{r\rightarrow \infty }\sim \left( \cos (2p+1)\pi
-1\right) \frac{1}{r}\sim \frac{1}{r}\;.  \label{eq62}
\end{equation}

\pagebreak

\subsection{The Henyey example}

In this section we shall discuss a class of Gribov's copies proposed by
Henyey \cite{hen}. The relevance of Henyey's work is due to the fact that
Gribov's copies with vanishing winding number and which fall off faster than
$1/r$, for $r\rightarrow \infty $, are explicitly obtained.

\noindent The starting configuration is
\begin{equation}
\overrightarrow{A}=i\overrightarrow{a}\sigma _{3}\;,\;\;\;\partial
_{i}A_{i}=0\;.  \label{h1}
\end{equation}
Notice that the gauge configuration, eq.$\left( \ref{h1}\right) $,
lies in the abelian diagonal subgroup of $SU(2)$. We shall
consider the set of gauge transformations parametrized by
\begin{equation}
S=e^{i\alpha \overrightarrow{f}\cdot \overrightarrow{\sigma }}=\cos \alpha +i%
\overrightarrow{f}\cdot \overrightarrow{\sigma }\sin \alpha \;,  \label{h2}
\end{equation}
where $\overrightarrow{f}$ is the unit vector

\begin{eqnarray}
\overrightarrow{f}\cdot \overrightarrow{f} &=&1\;,  \label{h3} \\
\overrightarrow{f} &=&\overrightarrow{e}_{x}\cos \beta +\overrightarrow{e}%
_{y}\sin \beta \;,  \nonumber
\end{eqnarray}
and $\left( \overrightarrow{e}_{x},\overrightarrow{e}_{y},\overrightarrow{e}%
_{z}\right) $ are the unit orthogonal cartesian vectors.
Introducing the vector $\overrightarrow{g}$

\begin{equation}
\overrightarrow{g}=-\overrightarrow{e}_{x}\sin \beta +\overrightarrow{e}%
_{y}\cos \beta \;,  \label{h4}
\end{equation}
it turns out that the set $\left( \overrightarrow{f},\overrightarrow{g},%
\overrightarrow{e}_{z}\right) $ yields a right-handed orthonormal
triad which rotates about the $z$-axis. Also,

\begin{equation}
\partial _{i}\overrightarrow{f}=\overrightarrow{g}\partial _{i}\beta
\;,\;\;\;\;\;\partial _{i}\overrightarrow{g}=-\overrightarrow{f}\partial
_{i}\beta \;.  \label{h5}
\end{equation}
Let us evaluate the gauge transformed field $\widetilde{A}_{i}$

\begin{eqnarray}
\widetilde{A}_{i} &=&{S}^{\dagger }\partial _{i}{S}+S^{\dagger }A_{i}{S}
\label{h6} \\
&=&i\left( \cos \alpha -i\overrightarrow{f}\cdot \overrightarrow{\sigma }%
\sin \alpha \right) a_{i}\sigma _{3}\left( \cos \alpha +i\overrightarrow{f}%
\cdot \overrightarrow{\sigma }\sin \alpha \right) \;  \nonumber \\
&&+\left( \cos \alpha -i\overrightarrow{f}\cdot \overrightarrow{\sigma }\sin
\alpha \right) \left( -\sin \alpha +i\overrightarrow{f}\cdot \overrightarrow{%
\sigma }\cos \alpha \right) \partial _{i}\alpha  \nonumber \\
&&+i\left( \cos \alpha -i\overrightarrow{f}\cdot \overrightarrow{\sigma }%
\sin \alpha \right) \overrightarrow{g}\cdot \overrightarrow{\sigma }\sin
\alpha \partial _{i}\beta \;.  \nonumber
\end{eqnarray}
Recalling that $\left( \overrightarrow{f}\cdot \overrightarrow{\sigma }%
\right) ^{2}=1$, it follows that

\begin{eqnarray}
\widetilde{A}_{i} &=&ia_{i}\left( \cos ^{2}\alpha \sigma _{3}+i\sin \alpha
\cos \alpha f_{k}\left[ \sigma _{3},\sigma _{k}\right] +\sin ^{2}\alpha
f_{k}f_{m}\sigma _{k}\sigma _{3}\sigma _{m}\right)  \label{h7} \\
&&+i\partial _{i}\alpha \overrightarrow{f}\cdot \overrightarrow{\sigma }%
+\partial _{i}\beta \left( i\sin \alpha \cos \alpha \overrightarrow{g}\cdot
\overrightarrow{\sigma }+\sin ^{2}\alpha \left( \overrightarrow{f}\cdot
\overrightarrow{\sigma }\right) \left( \overrightarrow{g}\cdot
\overrightarrow{\sigma }\right) \right) \;  \nonumber \\
&=&ia_{i}\left( \cos ^{2}\alpha \sigma _{3}-2\sin \alpha \cos \alpha
\varepsilon _{3km}\sigma _{m}f_{k}+\sin ^{2}\alpha f_{k}f_{m}\sigma
_{k}\left( \delta _{3m}+i\varepsilon _{3mp}\sigma _{p}\right) \right)
\nonumber \\
&&+i\partial _{i}\alpha \overrightarrow{f}\cdot \overrightarrow{\sigma }%
+\partial _{i}\beta \left( i\sin \alpha \cos \alpha \overrightarrow{g}\cdot
\overrightarrow{\sigma }+\sin ^{2}\alpha f_{i}g_{k}\left( \delta
_{ik}+i\varepsilon _{ikm}\sigma _{m}\right) \right) \;.  \nonumber
\end{eqnarray}
From
\begin{eqnarray}
\varepsilon _{3km}\sigma _{m}f_{k} &=&\left( \overrightarrow{e}_{z}\times
\overrightarrow{f}\right) \cdot \overrightarrow{\sigma }=\left(
\overrightarrow{e}_{z}\times \left( \overrightarrow{e}_{x}\cos \beta +%
\overrightarrow{e}_{y}\sin \beta \right) \right) \cdot \overrightarrow{%
\sigma }  \nonumber \\
&=&\left( \overrightarrow{e}_{y}\cos \beta -\overrightarrow{e}_{x}\sin \beta
\right) \cdot \overrightarrow{\sigma }=\overrightarrow{g}\cdot
\overrightarrow{\sigma }\;,  \nonumber \\
f_{i}g_{k}\varepsilon _{ikm}\sigma _{m} &=&\left( \overrightarrow{f}\times
\overrightarrow{g}\right) \cdot \overrightarrow{\sigma }=\left( \left(
\overrightarrow{e}_{x}\cos \beta +\overrightarrow{e}_{y}\sin \beta \right)
\times \left( -\overrightarrow{e}_{x}\sin \beta +\overrightarrow{e}_{y}\cos
\beta \right) \right) \cdot \overrightarrow{\sigma }\;  \nonumber \\
&=&\overrightarrow{e}_{z}\cdot \overrightarrow{\sigma }\;=\sigma _{3}\;,
\nonumber \\
f_{k}f_{m}\varepsilon _{3mp}\sigma _{k}\sigma _{p} &=&f_{k}f_{m}\varepsilon
_{3mp}\left( \delta _{kp}+i\varepsilon _{kpq}\sigma _{q}\right)
=f_{k}f_{m}\varepsilon _{3mk}-if_{k}f_{m}\varepsilon _{3mp}\varepsilon
_{pkq}\sigma _{q}  \nonumber \\
&=&-if_{k}f_{m}\left( \delta _{3k}\delta _{mq}-\delta _{3q}\delta
_{mk}\right) \sigma _{q}=i\sigma _{3\;,}  \nonumber \\
f_{3} &=&0\;,\;\;\;\overrightarrow{f}\cdot \overrightarrow{g}=0\mathrm{{\ }%
,\;}  \label{h8}
\end{eqnarray}
it follows
\begin{eqnarray}
\widetilde{A}_{i} &=&\;ia_{i}\left( \cos ^{2}\alpha \sigma _{3}-2\sin \alpha
\cos \alpha \overrightarrow{g}\cdot \overrightarrow{\sigma }-\sin ^{2}\alpha
\sigma _{3}\right) +i\partial _{i}\alpha \overrightarrow{f}\cdot
\overrightarrow{\sigma }  \nonumber \\
&&+\partial _{i}\beta \left( i\sin \alpha \cos \alpha \overrightarrow{g}%
\cdot \overrightarrow{\sigma }+i\sin ^{2}\alpha \sigma _{3}\right) \;.
\label{h9}
\end{eqnarray}
Finally, from $\left( \cos ^{2}\alpha -\sin ^{2}\alpha \right) =\cos 2\alpha
$,\ we get
\begin{eqnarray}
\widetilde{A}_{i} &=&i\sigma _{3}\left( a_{i}\cos 2\alpha +\partial
_{i}\beta \sin ^{2}\alpha \right) +i\partial _{i}\alpha \overrightarrow{f}%
\cdot \overrightarrow{\sigma }  \nonumber \\
&&+i\overrightarrow{g}\cdot \overrightarrow{\sigma }\left( -a_{i}\sin
2\alpha +\frac{1}{2}\partial _{i}\beta \sin 2\alpha \right) \;.  \label{h10}
\end{eqnarray}
Let us turn now to the condition $\partial
_{i}\widetilde{A}_{i}=0$. Computing the divergence of
$\widetilde{A}_{i}$, yields
\begin{eqnarray}
\partial _{i}\widetilde{A}_{i} &=&i\sigma _{3}\left( -2a_{i}\partial
_{i}\alpha \sin 2\alpha +\partial ^{2}\beta \sin ^{2}\alpha +2\sin \alpha
\cos \alpha \partial _{i}\beta \partial _{i}\alpha \right)  \nonumber \\
&&+i\partial ^{2}\alpha \overrightarrow{f}\cdot \overrightarrow{\sigma }%
+i\partial _{i}\alpha \partial _{i}\beta \overrightarrow{g}\cdot
\overrightarrow{\sigma }\;  \nonumber \\
&&-i\partial _{i}\beta \overrightarrow{f}\cdot \overrightarrow{\sigma }%
\left( -a_{i}\sin 2\alpha +\frac{1}{2}\partial _{i}\beta \sin 2\alpha \right)
\nonumber \\
&&+i\overrightarrow{g}\cdot \overrightarrow{\sigma }\left(
-2a_{i}\partial _{i}\alpha \cos 2\alpha +\frac{1}{2}\partial
^{2}\beta \sin 2\alpha +\partial _{i}\beta \partial _{i}\alpha
\cos 2\alpha \right)=0 \;.  \nonumber
\\
&&  \label{h11}
\end{eqnarray}
Following Henyey \cite{hen}, we shall impose\footnote{%
These conditions are more restrictive than necessary.}

\begin{equation}
\partial ^{2}\beta =\partial _{i}\partial _{i}\beta
=0\;,\;\;\;\;a_{i}\partial _{i}\alpha =0\;,\;\;\;\;\partial _{i}\beta
\partial _{i}\alpha =0\;.  \label{h12}
\end{equation}
Thus, the condition $\partial _{i}\widetilde{A}_{i}=0$ reduces to

\begin{equation}
\partial ^{2}\alpha +a_{i}\partial _{i}\beta \sin 2\alpha -\frac{1}{2}%
\partial _{i}\beta \partial _{i}\beta \sin 2\alpha =0\;.  \label{h13}
\end{equation}

\newpage
\begin{itemize}
\item  \textbf{Summary}
\textit{\begin{equation}
\overrightarrow{A}=i\overrightarrow{a}\sigma _{3}\;,\;\;\;\partial
_{i}A_{i}=\partial _{i}a_{i}=0\;.  \label{sh1}
\end{equation}
\begin{eqnarray}
S &=&e^{i\alpha \overrightarrow{f}\cdot \overrightarrow{\sigma }}=\cos
\alpha +i\overrightarrow{f}\cdot \overrightarrow{\sigma }\sin \alpha \;,
\nonumber \\
\overrightarrow{f} &=&\overrightarrow{e}_{x}\cos \beta +\overrightarrow{e}%
_{y}\sin \beta \;,\;\;\;\;\overrightarrow{f}\cdot \overrightarrow{f}=1\;.
\label{sh2}
\end{eqnarray}
For the gauge transformed field we get
\begin{eqnarray}
\widetilde{A}_{i} &=&{S}^{\dagger }\partial _{i}{S}+S^{\dagger }A_{i}{S}
\nonumber \\
&=&i\sigma _{3}\left( a_{i}\cos 2\alpha +\partial _{i}\beta \sin ^{2}\alpha
\right) +i\partial _{i}\alpha \overrightarrow{f}\cdot \overrightarrow{\sigma
}  \nonumber \\
&&+i\overrightarrow{g}\cdot \overrightarrow{\sigma }\left( -a_{i}\sin
2\alpha +\frac{1}{2}\partial _{i}\beta \sin 2\alpha \right) \;.  \label{sh3}
\end{eqnarray}
The condition for the existence of Gribov's copies, $\partial _{i}\widetilde{%
A}_{i}=0$ gives
\begin{equation}
\partial ^{2}\beta =\partial _{i}\partial _{i}\beta
=0\;,\;\;\;\;a_{i}\partial _{i}\alpha =0\;,\;\;\;\;\partial _{i}\beta
\partial _{i}\alpha =0\;,\;  \label{sh4}
\end{equation}
and
\begin{equation}
\partial ^{2}\alpha +a_{i}\partial _{i}\beta \sin 2\alpha -\frac{1}{2}%
\partial _{i}\beta \partial _{i}\beta \sin 2\alpha =0\;.  \label{sh5}
\end{equation}}
\end{itemize}

\subsubsection{Henyey's solution}

In order to solve the eqs.$\left( \ref{sh4}\right) $, $\left( \ref{sh5}%
\right) $ it is useful to adopt polar coordinates \cite{hen}, see Appendix
D. We set

\begin{equation}
\overrightarrow{a}=a(r,\theta )\overrightarrow{e}_{\varphi }\;,\;\;\;\;\beta
=\beta (\varphi )\;,\;\;\;\;\alpha =\alpha (r,\theta )\;.  \label{h14}
\end{equation}
The equations

\begin{eqnarray}
\overrightarrow{\nabla }\cdot \overrightarrow{a} &=&\frac{1}{r\sin \theta }%
\frac{\partial }{\partial \varphi }a(r,\theta )=0\;,  \label{h15} \\
\overrightarrow{a}\cdot \overrightarrow{\nabla }\alpha &=&a(r,\theta )\frac{1%
}{r\sin \theta }\frac{\partial }{\partial \varphi }\alpha (r,\theta )=0\;,
\nonumber \\
\overrightarrow{\nabla }\alpha \cdot \overrightarrow{\nabla }\beta &=&0\;,
\nonumber
\end{eqnarray}
are fulfilled. From
\begin{equation}
\overrightarrow{\nabla }^{2}\beta =\frac{1}{r^{2}\sin ^{2}\theta }\frac{%
\partial ^{2}}{\partial \varphi ^{2}}\beta =0\;,  \label{h16}
\end{equation}
it follows that we may take
\begin{equation}
\beta =\varphi \;.  \label{h17}
\end{equation}
It remains to solve the condition $\left( \ref{sh5}\right) $, which now
reads
\begin{equation}
\frac{1}{r^{2}}\frac{\partial }{\partial r}\left( r^{2}\frac{\partial \alpha
}{\partial r}\right) +\frac{1}{r^{2}\sin \theta }\frac{\partial }{\partial
\theta }\left( \sin \theta \frac{\partial \alpha }{\partial \theta }\right)
+\sin 2\alpha \left( \frac{a}{r\sin \theta }-\frac{1}{2r^{2}\sin ^{2}\theta }%
\right) =0\;.  \label{h18}
\end{equation}
Setting
\begin{equation}
\alpha (r,\theta )=rb(r)\sin \theta \;,  \label{h19}
\end{equation}
equation $\left( \ref{h18}\right) $ becomes
\begin{equation}
b+\left( r^{2}b^{\prime \prime }+4rb^{\prime }\right) \sin ^{2}\theta
+\left( a-\frac{1}{2r\sin \theta }\right) \sin \left( 2rb\sin \theta \right)
=0\;.  \label{h20}
\end{equation}
The strategy adopted by Henyey \cite{hen} in order to solve this
equation is that of expressing $a$ in terms of $b$, and then
searching for a suitable $b$ which gives the desired behavior for
$a$. Accordingly, we write
\begin{equation}
a=\frac{1}{2r\sin \theta }-\frac{1}{\sin \left( 2rb\sin \theta \right) }%
\left( b+r^{2}\sin ^{2}\theta \left( b^{\prime \prime }+4\frac{b^{\prime }}{r%
}\right) \right) \;.  \label{h21}
\end{equation}
We look then for a function $b$ such that:
\begin{eqnarray}
a\;{\ }{\ }{\ }\mathrm{is\;regular\;at\;the\;origin},\;\;r &\rightarrow &0\;,
\label{h22} \\
a\;{\ }{\ }{\ }\mathrm{decays\;faster\;than}\;1/r\;\mathrm{for}\;r
&\rightarrow &\infty \;.  \nonumber
\end{eqnarray}
First of all we notice that expression $\left( \ref{h21}\right) $
is not singular at $\theta =0$. Although each term of eq.$\left(
\ref{h21}\right) $ is singular, their sum has no singularity at
$\theta=0$, \textit{i.e.}
\begin{equation}
a_{\theta \rightarrow 0}\approx \frac{1}{2r\theta }-\frac{1}{2rb\theta }%
\left( b+r^{2}\theta ^{2}\left( b^{\prime \prime }+4\frac{b^{\prime }}{r}%
\right) \right) =0\;.  \label{h23}
\end{equation}
Also, in order to avoid possible singularities in the term
$\left(1/\sin \left( 2rb\sin \theta \right) \right)$ we require
\begin{equation}
2rb(r)<\pi \;.  \label{h24}
\end{equation}
Since $\sin \theta \leq 1$, the condition $\left( \ref{h24}\right) $ implies
that the argument of $\sin \left( 2rb\sin \theta \right) $ cannot be equal
to $\pi $, \textit{i.e. }$2rb\sin \theta <\pi $.

\noindent Let us look now at $r=0$. We have
\begin{equation}
a_{r\rightarrow 0}\approx \frac{1}{2r\sin \theta }-\frac{1}{2rb\sin \theta }%
\left( b+r^{2}\sin ^{2}\theta \left( b^{\prime \prime }+4\frac{b^{\prime }}{r%
}\right) \right) \;.  \label{h25}
\end{equation}
Requiring that
\begin{equation}
\left. b(r)\right. _{r\rightarrow 0}\approx
b_{0}+b_{2}r^{2}+O(r^{3})\;,\;\;\;\;\;\;\mathrm{with}\;\;b_{0}\neq 0\;,
\label{h26}
\end{equation}
it follows
\begin{equation}
a_{r\rightarrow 0}\approx \frac{1}{2r\sin \theta }-\frac{1}{2rb\sin \theta }%
+O(r)\approx O(r)\;,  \label{h27}
\end{equation}
so that $a$ is regular at the origin. It remains now to discuss the limit $%
r\rightarrow \infty $. In this case we search for a $b(r)$ such that $%
r^{2}a\rightarrow 0$ for $r\rightarrow \infty $. Let us set
\begin{equation}
\left. b(r)\right. _{r\rightarrow \infty }\approx \frac{1}{r^{n}}\;,
\label{h28}
\end{equation}
and let us determine $n$. From
\begin{eqnarray}
\left. r^{2}a\right. _{r\rightarrow \infty } &\approx &r^{2}\left( \frac{1}{%
2r\sin \theta }-\frac{1}{r^{n}}\frac{1}{\sin \left( \frac{2\sin \theta }{%
r^{n-1}}\right) }\right) -\frac{r^{4}\sin ^{2}\theta }{\sin \left( \frac{%
2\sin \theta }{r^{n-1}}\right) }\left( n(n+1)-4n)\right) \frac{1}{r^{n+2}}
\nonumber \\
&\approx &r^{2}\left( \frac{1}{2r\sin \theta }-\frac{1}{r^{n}}\frac{1}{%
\left( \frac{2\sin \theta }{r^{n-1}}\right) +O\left( \frac{1}{r^{3n-3}}%
\right) }\right) -n(n-3)\frac{r^{4}}{r^{n+2}}\frac{\sin ^{2}\theta }{\sin
\left( \frac{2\sin \theta }{r^{n-1}}\right) }  \nonumber \\
&\approx &O\left( \frac{1}{r^{2n-5}}\right) -n(n-3)\frac{r\sin \theta }{2}\;.
\label{h29}
\end{eqnarray}
Thus, if $n=3$,
\begin{eqnarray}
\left. b(r)\right. _{r\rightarrow \infty } &\approx &\frac{1}{r^{3}}\;,
\label{h30} \\
\left. r^{2}a(r)\right. _{r\rightarrow \infty } &\approx &O\left( \frac{1}{r}%
\right) \;.  \nonumber
\end{eqnarray}

\begin{itemize}
\item  \textbf{Summary}

\textit{In summary, any function $b(r)$ such that }

\textit{
\begin{eqnarray}
\left. b(r)\right. _{r\rightarrow 0} &\approx
&b_{0}+b_{2}r^{2}+O(r^{3})\;,\;\;\;\;\;\;\mathrm{with}\;\;b_{0}\neq 0\;,
\nonumber \\
2rb(r) &<&\pi \;,  \nonumber \\
\left. b(r)\right. _{r\rightarrow \infty } &\approx &\frac{1}{r^{3}}\;,
\label{h31}
\end{eqnarray}
will give a gauge field  $\overrightarrow{A}$%
\begin{eqnarray}
\overrightarrow{A} &=&i\overrightarrow{a}\sigma
_{3}\;,\;\;\;\partial
_{i}A_{i}=\partial _{i}a_{i}=0\;.  \label{h32} \\
\overrightarrow{a} &=&a(r,\theta )\overrightarrow{e}_{\varphi }  \nonumber \\
a(r,\theta ) &=&\frac{1}{2r\sin \theta }-\frac{1}{\sin \left( 2rb\sin \theta
\right) }\left( b+r^{2}\sin ^{2}\theta \left( b^{\prime \prime }+4\frac{%
b^{\prime }}{r}\right) \right)  \nonumber
\end{eqnarray}
which is regular at the origin, $r=0$, and decays faster than $1/r^{2}$ at
infinity, $r\rightarrow \infty $%
\begin{eqnarray}
a_{r\rightarrow 0} &\approx &O(r)\;,  \nonumber \\
\left. r^{2}a(r)\right. _{r\rightarrow \infty } &\approx &O\left( \frac{1}{r}%
\right) \;.  \label{h33}
\end{eqnarray}
An example of such a function $b(r)$ is given by
\begin{eqnarray}
b(r) &=&\frac{k}{\left( r^{2}+r_{0}^{2}\right) ^{3/2}}\;,  \nonumber \\
k &<&\frac{3^{3/2}\pi }{4}r_{0}^{2}\;,  \label{h34}
\end{eqnarray}
where the condition $k<\left( 3^{3/2}\pi r_{0}^{2}\right) /4\;$stems from
the requirement $2rb(r)<\pi .$ }
\end{itemize}

\subsubsection{The winding number of Henyey's solution}

It remains now to discuss the winding number of Henyey's solution. Let us
begin with the evaluation of the winding number corresponding to the gauge
transformation $\left( \ref{h2}\right) $%
\begin{equation}
S=e^{i\alpha \overrightarrow{f}\cdot \overrightarrow{\sigma }}=\cos \alpha +i%
\overrightarrow{f}\cdot \overrightarrow{\sigma }\sin \alpha \;.  \label{h35}
\end{equation}
From expression $\left( \ref{w14}\right) $, we have
\begin{eqnarray}
\nu &=&\frac{1}{3\pi ^{2}}\int d^{3}x\varepsilon _{ijk}\varepsilon
_{mpq}N_{4}\left( \partial _{i}N_{m}\right) \left( \partial _{j}N_{p}\right)
\left( \partial _{k}N_{q}\right)  \nonumber \\
&&-\frac{1}{4\pi ^{2}}\int d^{3}x\varepsilon _{ijk}\partial _{i}\left(
\varepsilon _{mpq}N_{4}N_{m}\left( \partial _{j}N_{p}\right) \left( \partial
_{k}N_{q}\right) \right) \;,  \label{h36}
\end{eqnarray}
with
\begin{equation}
N_{4}=\cos \alpha \;,\;\;\;\;N_{i}=f_{i}\sin \alpha \;.  \label{h37}
\end{equation}
Recalling that
\begin{equation}
\partial _{i}\overrightarrow{f}=\overrightarrow{g}\partial _{i}\beta
\;,\;\;\;\;\overrightarrow{f}\cdot \overrightarrow{g}=0\;,  \label{h38}
\end{equation}
it follows
\begin{eqnarray}
\nu &=&\frac{1}{3\pi ^{2}}\int d^{3}x\varepsilon _{ijk}\varepsilon
_{mpq}\cos \alpha \left[ \left( \sin \alpha g_{m}\partial _{i}\beta
+f_{m}\cos \alpha \partial _{i}\alpha \right) \left( \sin \alpha
g_{p}\partial _{j}\beta +f_{p}\cos \alpha \partial _{j}\alpha \right) \right.
\nonumber \\
&&\;\;\;\;\;\;\;\;\;\;\;\;\;\;\;\;\left. \;\times \left( \sin \alpha
g_{q}\partial _{k}\beta +f_{q}\cos \alpha \partial _{k}\alpha \right) \right]
\nonumber \\
&&-\frac{1}{4\pi ^{2}}\int d^{3}x\varepsilon _{ijk}\partial _{i}\left[
\varepsilon _{mpq}\sin \alpha \cos \alpha f_{m}\left( \sin \alpha
g_{p}\partial _{j}\beta +f_{p}\cos \alpha \partial _{j}\alpha \right) \right.
\nonumber \\
&&\;\;\;\;\;\;\;\;\;\;\;\;\;\;\;\;\left. \times \left( \sin \alpha
g_{q}\partial _{k}\beta +f_{q}\cos \alpha \partial _{k}\alpha \right) \right]
\\
&=&\frac{1}{3\pi ^{2}}\int d^{3}x\varepsilon _{ijk}\varepsilon _{mpq}\cos
\alpha \left[ \left( \sin \alpha \cos \alpha g_{m}f_{p}\partial _{i}\beta
\partial _{j}\alpha +\sin \alpha \cos \alpha f_{m}g_{p}\partial _{i}\alpha
\partial _{j}\beta \right) \right.  \nonumber \\
&&\;\;\;\;\;\;\;\;\;\;\;\;\;\;\;\;\left. \times \left( \sin \alpha
g_{q}\partial _{k}\beta +f_{q}\cos \alpha \partial _{k}\alpha \right) \right]
\nonumber \\
&&-\frac{1}{4\pi ^{2}}\int d^{3}x\varepsilon _{ijk}\partial _{i}\left(
\varepsilon _{mpq}\sin \alpha \cos \alpha f_{m}\left( \sin \alpha
g_{p}\partial _{j}\beta \right) \left( \sin \alpha g_{q}\partial _{k}\beta
\right) \right)  \nonumber \\
&=&0\;.  \label{h39}
\end{eqnarray}
We proceed now with the computation of the winding number of the starting
gauge configuration $\left( \ref{h2}\right) $ and of the gauge copy $\left(
\ref{h6}\right) $. To obtain the winding number of the expression $\left(
\ref{h2}\right) $, we start from the Pontryagin index, eqs.$\left( \ref{c50}%
\right) -\left( \ref{c52}\right) \;$of Appendix C,
\begin{equation}
\nu \;=-\frac{1}{8\pi ^{2}}Tr\int_{S_{\infty }^{3}}dS_{\mu }\varepsilon
_{\mu \nu \rho \sigma }\left( A_{\nu }\partial _{\rho }A_{\sigma }+\frac{2}{3%
}A_{\nu }A_{\rho }A_{\sigma }\right) \;.  \label{h40}
\end{equation}
From
\begin{equation}
\overrightarrow{A}=i\overrightarrow{a}\sigma _{3}\;,\;\;\;\partial
_{i}A_{i}=0\;,\;\;  \label{h41}
\end{equation}
and
\begin{equation}
A_{4}=0\;,  \label{h42}
\end{equation}
expression $\left( \ref{h40}\right) $ becomes
\begin{equation}
\nu \;=-\frac{1}{8\pi ^{2}}Tr\int_{S_{\infty }^{3}}dS_{4}\varepsilon
_{4ijk}\left( A_{i}\partial _{j}A_{k}+\frac{2}{3}A_{i}A_{j}A_{k}\right) \;.
\label{h43}
\end{equation}
Since the starting field is in the abelian subgroup of $SU(2)$, it follows
that
\begin{equation}
TrA_{i}A_{j}A_{k}=-ia_{i}a_{j}a_{k}Tr\left( \sigma _{3}\sigma _{3}\sigma
_{3}\right) =0\;.  \label{h44}
\end{equation}
Also
\begin{equation}
\varepsilon _{4ijk}Tr\left( A_{i}\partial _{j}A_{k}\right) =-2\varepsilon
_{4ijk}a_{i}\partial _{j}a_{k}=-2\overrightarrow{a}\cdot \left(
\overrightarrow{\nabla }\times \overrightarrow{a}\right) \;.  \label{h45}
\end{equation}
Furthermore, since $\overrightarrow{a}=a(r,\theta )\overrightarrow{e}%
_{\varphi }$, eq.$\left( \ref{h32}\right) $,$\;$has component only along the
direction $\overrightarrow{e}_{\varphi }$ and $\overrightarrow{\nabla }%
\times \overrightarrow{a}$ has components along the directions $%
\overrightarrow{e}_{r}$, $\overrightarrow{e}_{\theta }$, it follows that
\begin{equation}
\overrightarrow{a}\cdot \left( \overrightarrow{\nabla }\times
\overrightarrow{a}\right) =0\;.  \label{h46}
\end{equation}
We see therefore that the Pontryagin density $\varepsilon _{4ijk}\left(
A_{i}\partial _{j}A_{k}+\frac{2}{3}A_{i}A_{j}A_{k}\right) $ associated to
the starting gauge configuration vanishes. Finally, for the winding number
of the copy $\left( \ref{h6}\right) $ one has
\begin{equation}
\nu _{\mathrm{copy}}=\nu _{\mathrm{original}}+\nu _{\mathrm{transformation}%
}=0\;.  \label{h47}
\end{equation}
Therefore, we have shown that it is possible to construct Gribov
copies with vanishing winding number and which fall off faster than $1/r$ for $%
r\rightarrow \infty $. This concludes the discussion of Henyey's
example.

\newpage

\section{Part II: The Gribov Horizons}

\subsection{Generalities}

In order to introduce the notion of Gribov horizon let us look at
the eigenvalues of the Faddeev-Popov operator, {\it i.e.}
\begin{equation}
-\partial _{\mu }\left( \partial _{\mu }\psi +\left[ A_{\mu },\psi
\right] \right) =\epsilon (A)\psi \;.  \label{ho1}
\end{equation}
As already underlined in Part I, equation $\left( \ref{ho1}\right)
$ can be seen as a kind of Schr\"{o}dinger equation, with $A_{\mu
}$ playing the role
of the potential. For small values of $A_{\mu }$, eq.$\left( \ref{ho1}%
\right) $ is solvable for positive $\epsilon $ only. More
precisely, denoting by $\epsilon_1(A), \epsilon_2(A),
\epsilon_3(A), ....,$ the eigenvalues corresponding to a given
field configuration $A_{\mu}$, one has that, for small $A_{\mu }$,
all $\epsilon_i(A)$ are positive, {\it i.e.} $\epsilon_i(A)>0$.
However, for a sufficiently large value of the field $A_{\mu}$,
one of the eigenvalues, say $\epsilon_1(A)$, turns out to vanish,
becoming negative as the field increases further. As in the case
of the Schr\"{o}dinger equation, this means that the field
$A_{\mu}$ is large enough to ensure the existence of negative
energy solutions, {\it i.e.} bound states. For a greater magnitude
of the field $A_{\mu}$, a second eigenvalue, say $\epsilon_2(A)$,
will vanish, becoming negative as the field increases again.
Following Gribov \cite{g}, we may thus divide the functional space
of the fields into regions $C_{0}, C_{1}, C_{2}, ..., C_{n}$ over
which the Faddeev-Popov operator, $-\partial _{\mu }\left(
\partial _{\mu }\cdot +\left[ A_{\mu },\cdot \right] \right) $,
has $0, 1, 2,...., n$ negative eigenvalues, see Fig.12.

\vspace{1cm}

\begin{figure}[ht]
\centering \epsfig{file=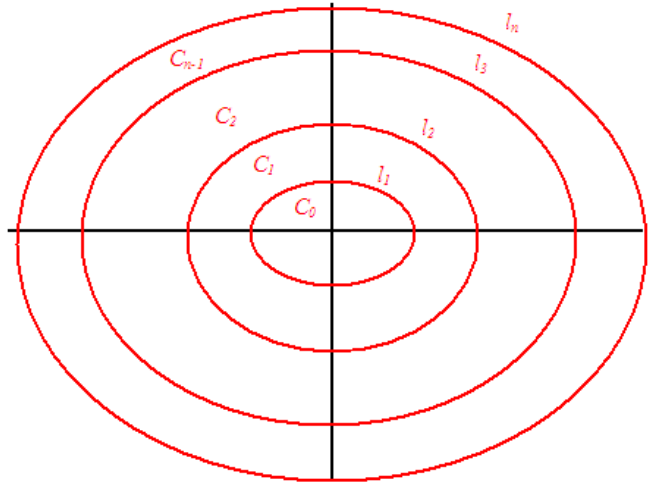,width=10cm} \caption{The
Gribov horizons}
\end{figure}


\vspace{1cm}

\noindent These regions are separated by lines $l_{1}, l_{2},
l_{3}, ..., l_{n}$ on which the Faddeev-Popov operator has zero
energy solutions. The meaning of Fig.12 is as follows. In the
region $C_{0}$ all eigenvalues of the Faddeev-Popov operator are
positive, {\it i.e.} $-\partial _{\mu }\left( \partial _{\mu
}\cdot +\left[ A_{\mu },\cdot \right] \right) >0 $. At the
boundary $l_{1}$ of the region $C_{0}$ the first vanishing
eigenvalue appears, namely on $l_{1}$ the Faddeev-Popov operator
possesses a normalizable zero mode $\chi $
\begin{equation}
\partial _{\mu }\left( \partial _{\mu }\chi +\left[ A_{\mu },\chi \right]
\right) =0\;.  \label{ho2}
\end{equation}
In the region $C_{1}$ the Faddeev-Popov operator has one bound
state, {\it i.e.} one negative energy solution. At the boundary
$l_{2}$, a zero eigenvalue reappears. In the region $C_{2}\;$the
Faddeev-Popov operator has two bound states, {\it i.e.} two
negative energy solutions. On $l_{3}$ a zero eigenvalue shows up
again, and so on. The boundaries $l_{1},\;l_{2},\;l_{3},%
\;....,\;l_{n}$,$\;$ on which the Faddeev-Popov operator has zero
eigenvalues are called Gribov horizons. In particular, the
boundary $l_{1}$ where the first vanishing eigenvalue appears is
called the first horizon.

\begin{itemize}
\item \textbf{Remark}

{\it It is useful to emphasize that in the region $C_0$ the
Faddeev-Popov operator has only positive eigenvalues. Therefore,
this region can be defined as the set of all transverse fields for
which the Faddeev-Popov operator is positive definite, namely
\begin{equation}
C_0=\{ A_{\mu},\; \partial A=0, \; -\partial _{\mu }\left(
\partial _{\mu }\cdot +\left[ A_{\mu },\cdot \right] \right) >0 \}
\nonumber \;.
\end{equation}
}
\end{itemize}

\begin{itemize}
\item  \textbf{Remark}

\textit{In order to obtain a better understanding of the notion of
the Gribov horizons, let us remark that there is a close
relationship between the horizons and the existence of Gribov
copies. In part I, we have discussed the existence of equivalent
fields by considering finite gauge transformations}
\begin{equation}
\widetilde{A}_{\mu }={S}^{\dagger }\partial _{\mu }{S}+S^{\dagger }A_{\mu }{S%
}\;.  \label{ho3}
\end{equation}
\textit{The requirement that the field }$\widetilde{A}_{\mu
}$\textit{\ obeys the same transversality condition as }$A_{\mu }$
\begin{equation}
\partial \widetilde{A}=\partial A=0\;,  \label{ho4}
\end{equation}
\textit{yields the equation}
\begin{equation}
\partial _{\mu }S^{\dagger }\partial _{\mu }{S+S}^{\dagger }\partial _{\mu
}\partial _{\mu }{S+\partial }_{\mu }S^{\dagger }A_{\mu
}{S+}S^{\dagger }A_{\mu }{\partial }_{\mu }{S=0\;.}  \label{ho5}
\end{equation}
\textit{For }$S$\textit{\ close to unit, }$S=1+\alpha ,$\textit{\
}$\alpha \ll 1$\textit{, expression }$\left( \ref{ho5}\right)
$\textit{\ reduces to }
\begin{equation}
\partial _{\mu }\left( \partial _{\mu }\alpha +\left[ A_{\mu },\alpha
\right] \right) =0\;.  \label{ho6}
\end{equation}
\textit{We see therefore that the condition for the existence of
an
equivalent field }$\widetilde{A}_{\mu }$\textit{\ close to }$A_{\mu }$%
\textit{, i.e. }
\begin{equation}
\widetilde{A}_{\mu }=A_{\mu }+\left( \partial _{\mu }\alpha
+\left[ A_{\mu },\alpha \right] \right) \;,  \label{ho77}
\end{equation}
\textit{relies on the existence of a zero mode for the
Faddeev-Popov operator. }

\item  \textbf{Remark}

{\it The transversality condition of the Landau gauge, $\partial
_{\mu }A_{\mu }=0$, implies that the space-time
derivative $\partial _{\mu }$ and the covariant derivative $%
D_{\mu }(A)=\left( \partial _{\mu }\cdot +\left[ A_{\mu },\cdot
\right] \right) $ obey the following commutation relation
\begin{equation}
\partial _{\mu }D_{\mu }(A)=D_{\mu }(A)\partial _{\mu }\;.  \label{l1}
\end{equation}
As a consequence, the Faddeev-Popov operator
$\mathcal{M}=-\partial _{\mu
}D_{\mu }(A)$ turns out to be Hermitian, $\mathcal{M}=\mathcal{M}^{\dagger }$%
. Its eigenvalues are thus real. In fact, integrating by parts,
one has
\begin{eqnarray}
\int d^{4}x\left( \left( \partial _{\mu }D_{\mu }\psi \right)
^{a}\right) ^{\dagger }\varphi ^{a} &=&\int d^{4}x\left( \partial
_{\mu }\partial _{\mu }\psi ^{a\dagger }+f^{abc}A_{\mu
}^{b}\partial _{\mu }\psi ^{c\dagger
}\right) \varphi ^{a}  \nonumber \\
&=&\int d^{4}x\left( \psi ^{a\dagger }\partial _{\mu }\partial
_{\mu }\varphi ^{a}-f^{abc}\psi ^{c\dagger }A_{\mu }^{b}\partial
_{\mu }\varphi
^{a}\right)   \nonumber \\
&=&\int d^{4}x\left( \psi ^{a\dagger }\partial _{\mu }\partial
_{\mu }\varphi ^{a}+f^{abc}\psi ^{a\dagger }A_{\mu }^{b}\partial
_{\mu }\varphi
^{c}\right)   \nonumber \\
&=&\int d^{4}x\psi ^{a\dagger }\left( \partial _{\mu }D_{\mu
}\varphi ^{a}\right) \;.  \label{l2}
\end{eqnarray}
}
\end{itemize}

\subsection{Example of a zero mode of the Faddeev-Popov operator}

It is useful to provide here an example of a normalizable zero
mode of the Faddeev-Popov operator. We shall work in three
dimensions, the gauge group being $SU(2)$. The aim is to obtain a
normalizable zero mode $\chi $, solution of
\begin{equation}
\overrightarrow{\nabla }^{2}\chi +\left[ A_{i},\partial _{i}\chi
\right] =0\;.  \label{ho8}
\end{equation}
We shall follow Henyey's strategy \cite{hen}, see also \cite{vb},
and write
\begin{equation}
\overrightarrow{A}=i\overrightarrow{a}\sigma _{3}\;,\;\;\;\partial
_{i}A_{i}=0\;.  \label{ho9}
\end{equation}
Adopting polar coordinates, see App.D, we shall set
\begin{equation}
\overrightarrow{a}=a(r,\theta )\overrightarrow{e}_{\varphi
}\;.\;\;\; \label{ho10}
\end{equation}
We look now at a zero mode $\chi $ of the form
\begin{equation}
\chi =\alpha (r,\theta )\left( \sigma _{1}\cos \varphi +\sigma
_{2}\sin \varphi \right) \;.  \label{ho11}
\end{equation}
From
\begin{equation}
\left[ \sigma _{i},\sigma _{j}\right] =2i\varepsilon _{ijk}\sigma
_{k}\;, \label{ho12}
\end{equation}
it follows

\begin{equation}
\left[ \sigma _{1},\sigma _{2}\right] =2i\sigma
_{3}\;,\;\;\;\;\;\;\;\left[ \sigma _{3},\sigma _{1}\right]
=2i\sigma _{2}\;,\;\;\;\;\;\;\;\;\left[ \sigma _{3},\sigma
_{2}\right] =-2i\sigma _{1}\;.  \label{ho13}
\end{equation}
Therefore
\begin{eqnarray}
\left[ A_{i},\partial _{i}\chi \right]  &=&i\;\left[ \sigma
_{3},\sigma _{1}\right] a_{i}\partial _{i}\left( \alpha \cos
\varphi \right) +i\;\left[ \sigma _{3},\sigma _{2}\right]
a_{i}\partial _{i}\left( \alpha \sin \varphi
\right)   \nonumber \\
&=&-2\sigma _{2}a_{i}\partial _{i}\left( \alpha \cos \varphi
\right) +2\sigma _{1}a_{i}\partial _{i}\left( \alpha \sin \varphi
\right)   \nonumber
\\
&=&-2\sigma _{2}\frac{a}{r\sin \theta }\frac{\partial \left(
\alpha \cos
\varphi \right) }{\partial \varphi }+2\sigma _{1}\frac{a}{r\sin \theta }%
\frac{\partial \left( \alpha \sin \varphi \right) }{\partial
\varphi }\;.
\nonumber \\
&&  \label{ho14}
\end{eqnarray}
Setting
\begin{equation}
\alpha (r,\theta )=rb(r)\sin \theta \;,  \label{ho15}
\end{equation}
we get

\begin{equation}
\left[ A_{i},\partial _{i}\chi \right] =2ab\sin \varphi \sigma
_{2}+2ab\cos \varphi \sigma _{1}\;.  \label{ho16}
\end{equation}
Also,
\begin{eqnarray}
\overrightarrow{\nabla }^{2}\chi &=&\frac{1}{r^{2}}\frac{\partial
}{\partial r}\left( r^{2}\frac{\partial \chi }{\partial r}\right)
+\frac{1}{r^{2}\sin \theta }\frac{\partial }{\partial \theta
}\left( \sin \theta \frac{\partial
\chi }{\partial \theta }\right) +\frac{1}{r^{2}\sin ^{2}\theta }\frac{%
\partial ^{2}\chi }{\partial \varphi ^{2}}\;  \nonumber \\
&=&\sin \theta \left( \sigma _{1}\cos \varphi +\sigma _{2}\sin
\varphi
\right) \frac{1}{r^{2}}\frac{\partial }{\partial r}\left( r^{2}\frac{%
\partial \left( rb(r)\right) }{\partial r}\right)  \nonumber \\
&&+\left( \sigma _{1}\cos \varphi +\sigma _{2}\sin \varphi \right) \frac{b(r)%
}{r\sin \theta }\frac{\partial }{\partial \theta }\left( \sin \theta \frac{%
\partial \sin \theta }{\partial \theta }\right)  \nonumber \\
&&+\frac{b(r)}{r\sin \theta }\frac{\partial ^{2}\left( \sigma
_{1}\cos \varphi +\sigma _{2}\sin \varphi \right) \;}{\partial
\varphi ^{2}}\;
\nonumber \\
&=&\sin \theta \left( \sigma _{1}\cos \varphi +\sigma _{2}\sin
\varphi \right) \frac{1}{r^{2}}\frac{\partial }{\partial r}\left(
r^{3}b^{\prime
}+r^{2}b\right)  \nonumber \\
&&+\left( \sigma _{1}\cos \varphi +\sigma _{2}\sin \varphi \right) \frac{b}{%
r\sin \theta }\frac{\partial }{\partial \theta }\left( \sin \theta
\cos
\theta \right)  \nonumber \\
&&-\frac{b(r)}{r\sin \theta }\left( \sigma _{1}\cos \varphi
+\sigma _{2}\sin
\varphi \right)  \nonumber \\
&=&\sin \theta \left( \sigma _{1}\cos \varphi +\sigma _{2}\sin
\varphi \right) \frac{1}{r^{2}}(4r^{2}b^{\prime }+r^{3}b^{\prime
\prime }+2rb)
\nonumber \\
&&+\left( \sigma _{1}\cos \varphi +\sigma _{2}\sin \varphi \right) \frac{b}{%
r\sin \theta }\left( \cos ^{2}\theta -\sin ^{2}\theta \right)  \nonumber \\
&&-\frac{b(r)}{r\sin \theta }\left( \sigma _{1}\cos \varphi
+\sigma _{2}\sin
\varphi \right)  \nonumber \\
&=&(4b^{\prime }+rb^{\prime \prime })\sin \theta \left( \sigma
_{1}\cos
\varphi +\sigma _{2}\sin \varphi \right)  \nonumber \\
&&+\frac{b}{r}\left( 2\sin \theta +\frac{\left( \cos ^{2}\theta
-\sin ^{2}\theta \right) }{\sin \theta }\right) \left( \sigma
_{1}\cos \varphi
+\sigma _{2}\sin \varphi \right)  \nonumber \\
&&-\frac{b(r)}{r\sin \theta }\left( \sigma _{1}\cos \varphi
+\sigma _{2}\sin
\varphi \right)  \nonumber \\
&=&(4b^{\prime }+rb^{\prime \prime })\sin \theta \left( \sigma
_{1}\cos
\varphi +\sigma _{2}\sin \varphi \right)  \nonumber \\
&&+\frac{b}{r\sin \theta }\left( \sigma _{1}\cos \varphi +\sigma
_{2}\sin \varphi \right) -\frac{b(r)}{r\sin \theta }\left( \sigma
_{1}\cos \varphi
+\sigma _{2}\sin \varphi \right)  \nonumber \\
&=&(4b^{\prime }+rb^{\prime \prime })\sin \theta \left( \sigma
_{1}\cos \varphi +\sigma _{2}\sin \varphi \right)  \label{ho17}
\end{eqnarray}
Finally,
\begin{equation}
\overrightarrow{\nabla }^{2}\chi =(4b^{\prime }+rb^{\prime \prime
})\sin \theta \left( \sigma _{1}\cos \varphi +\sigma _{2}\sin
\varphi \right) \;. \label{hoo18}
\end{equation}
The equation $\left( \ref{ho8}\right) $ is thus equivalent to
\begin{equation}
(4b^{\prime }+rb^{\prime \prime })\sin \theta \left( \sigma
_{1}\cos \varphi +\sigma _{2}\sin \varphi \right) +2ab\sin \varphi
\sigma _{2}+2ab\cos \varphi \sigma _{1}=0\;,  \label{ho18}
\end{equation}
which yields

\begin{equation}
(4b^{\prime }+rb^{\prime \prime })\sin \theta +2ab=0\;.
\label{ho19}
\end{equation}
Following Henyey \cite{hen}, for $a$ we obtain
\begin{equation}
a=-\frac{r}{2b}\sin \theta \left( b^{\prime \prime }+\frac{4b^{\prime }}{r}%
\right) \;.  \label{ho20}
\end{equation}
We now look for a function $b$ which yields a field configuration $%
a(r,\theta )$ which is regular at the origin, $r=0$, and which
decays faster than $1/r$ for $r\rightarrow \infty $. As done in
Part I, for $b$ we choose
\begin{equation}
b(r)=\frac{k}{\left( r^{2}+r_{0}^{2}\right) ^{3/2}}\;,\;\;\;\;\;k\;\;\;%
\mathrm{const.\;\;and\;}r_{0}\neq 0\;.  \label{hoo21}
\end{equation}
Therefore, for $a(r,\theta )$ we find \cite{vb}

\begin{eqnarray}
a(r,\theta ) &=&-\frac{r}{2}\left( r^{2}+r_{0}^{2}\right)
^{3/2}\sin \theta \left( -3\frac{\partial }{\partial
r}\frac{r}{\left( r^{2}+r_{0}^{2}\right)
^{5/2}}-\frac{12}{\left( r^{2}+r_{0}^{2}\right) ^{5/2}}\right)  \nonumber \\
&=&-\frac{r}{2}\left( r^{2}+r_{0}^{2}\right) ^{3/2}\sin \theta \left( 15%
\frac{r^{2}}{\left( r^{2}+r_{0}^{2}\right)
^{7/2}}-\frac{15}{\left(
r^{2}+r_{0}^{2}\right) ^{5/2}}\right)  \nonumber \\
&=&\frac{15}{2}\frac{rr_{0}^{2}}{\left( r^{2}+r_{0}^{2}\right)
^{2}}\sin \theta \;.  \label{ho22}
\end{eqnarray}
Observe that
\begin{eqnarray}
a(r,\theta ) &\rightarrow &0\;\;\mathrm{for\;\;}r\rightarrow 0\;,
\label{ho23} \\
a(r,\theta ) &\sim &\frac{1}{r^{3}}\;\;\mathrm{for\;}r\rightarrow
\infty \;. \nonumber
\end{eqnarray}
It remains to check that $\chi $ is a normalizable zero mode with
respect to the scalar product
\begin{equation}
\left\langle X\mid Y\right\rangle =Tr\int d^{3}xX^{\dagger }Y\;.
\label{hoo24}
\end{equation}
From
\begin{equation}
\chi =rb\sin \theta \left( \sigma _{1}\cos \varphi +\sigma
_{2}\sin \varphi \right) \;,  \label{ho25}
\end{equation}
we have

\begin{equation}
\chi ^{\dagger }=rb\sin \theta \left( \sigma _{1}\cos \varphi
+\sigma _{2}\sin \varphi \right) \;,  \label{ho26}
\end{equation}

\noindent and
\begin{equation}
Tr\left( \chi ^{\dagger }\chi \right) =2r^{2}b^{2}\sin ^{2}\theta
\;. \label{ho27}
\end{equation}
Thus
\begin{equation}
\left\langle \chi \mid \chi \right\rangle =2\int
d^{3}xr^{2}b^{2}\sin ^{2}\theta =4\pi k^{2}\int_{0}^{\pi }\sin
^{3}\theta d\theta \int_{0}^{\infty }dr\frac{r^{4}}{\left(
r^{2}+r_{0}^{2}\right) ^{3}}<\infty \;,  \label{h028}
\end{equation}
showing that we have found a normalizable zero mode of the
Faddeev-Popov operator.

\newpage

\begin{itemize}
\item  \textbf{Summary}

\textit{For the zero mode we have
\begin{equation}
\overrightarrow{\nabla }^{2}\chi +\left[ A_{i},\partial _{i}\chi
\right] =0\;,  \label{ho29}
\end{equation}
with
\begin{eqnarray}
\overrightarrow{A}=i\overrightarrow{a}\sigma _{3}\;,\;\;\;\partial
_{i}A_{i}=0\;,  \nonumber \\
\overrightarrow{a}=a(r,\theta )\overrightarrow{e}_{\varphi }\;,
\label{ho30}
\end{eqnarray}
and
\begin{equation}
\chi =rb\sin \theta \left( \sigma _{1}\cos \varphi +\sigma
_{2}\sin \varphi \right) \;,  \label{ho31}
\end{equation}
\begin{equation}
b(r)=\frac{k}{\left( r^{2}+r_{0}^{2}\right) ^{3/2}}\;,\;\;\;\;\;k\;\mathrm{{%
\mathrm{const.}}\mathrm{\;\;and\;}r_{0}\neq 0\;.}  \label{ho32}
\end{equation}
\begin{equation}
\left\langle \chi \mid \chi \right\rangle =\int d^{3}xTr\left(
\chi ^{\dagger }\chi \right) <\infty \;\;.  \label{ho33}
\end{equation}
For }$a(r,\theta )$ \textit{\ one finds
\begin{equation}
a=-\frac{r}{2b}\sin \theta \left( b^{\prime \prime }+\frac{4b^{\prime }}{r}%
\right) \;=\frac{15}{2}\frac{rr_{0}^{2}}{\left( r^{2}+r_{0}^{2}\right) ^{2}}%
\sin \theta \;,  \label{ho34}
\end{equation}
\begin{eqnarray}
a(r,\theta )\rightarrow 0\;\;\mathrm{for\;\;}r\rightarrow 0\;,
\label{ho35}
\\
a(r,\theta )\sim \frac{1}{r^{3}}\;\;\mathrm{for\;}r\rightarrow
\infty \;. \nonumber
\end{eqnarray}
Also, }$\overrightarrow{a}$ \textit{is transverse}
\begin{equation}
\partial _{i}a_{i}=\frac{1}{r\sin \theta }\frac{\partial a(r,\theta )}{%
\partial \varphi }=0\;.  \label{ho36}
\end{equation}
\end{itemize}

\newpage

\subsection{An important statement}

Let us prove the following statement \cite{g}:

\begin{itemize}
\item  \textbf{Statement}

\textit{For any field located within the region }$C_{1}$,
\textit{close to the boundary } $l_{1}$, \textit{there is an
equivalent field within the region }$C_{0}$, \textit{close to the
same boundary } $l_{1}$, \textit{see Fig.13}
\end{itemize}

\begin{figure}[ht]
\centering \epsfig{file=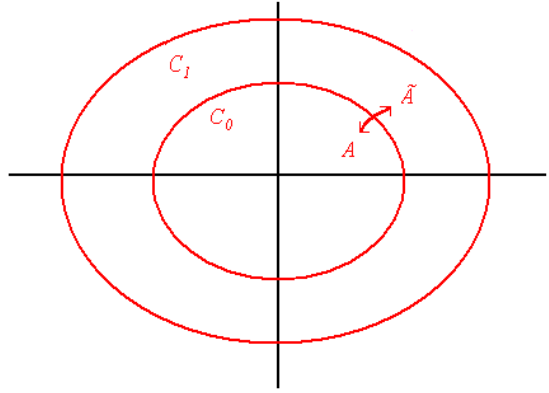,width=10cm} \caption{The
Equivalent Fields}
\end{figure}



\vspace{0.5cm}

\noindent Let $A_{\mu }$ be the field located in $C_{0}$, close to the first horizon $%
l_{1}$. We write
\begin{equation}
A_{\mu }=C_{\mu }+a_{\mu }\;,\;\;\;\;\;a_{\mu }\;\mathrm{small\;with%
\;respect\;to\;}C_{\mu }\;,  \label{s1}
\end{equation}
where the transverse field $C_{\mu }$, $\partial C=0$, lies on the
Gribov horizon $l_{1}$, \textit{i.e.} it exists a normalizable
zero mode $\varphi _{0}$ such that
\begin{equation}
\partial _{\mu }\left( \partial _{\mu }\varphi _{0}+\left[ C_{\mu },\varphi
_{0}\right] \right) =0\;.  \label{s2}
\end{equation}

\begin{itemize}
\item  \textbf{Remark}

{\it Notice also that, by definition, the field $A_{\mu }$ is
located in the region $C_{0}$, so that
\begin{equation}
-\partial _{\mu }\left( \partial _{\mu }\cdot +\left[ A_{\mu
},\cdot \right] \right) > 0\;.  \label{s4}
\end{equation}
}
\end{itemize}
The eigenvalues problem for the Faddeev-Popov operator
corresponding to the field $A_{\mu }$ can be easily handled by
means of the perturbation theory of quantum mechanics. Let us look
indeed at the eigenvalues equation
\begin{equation}
-\partial _{\mu }\left( \partial _{\mu }\psi +\left[ C_{\mu
}+a_{\mu },\psi \right] \right) =\epsilon (a)\psi \;,  \label{s5}
\end{equation}
{\it i.e.}
\begin{equation}
-\left( \partial ^{2}\psi +\left[ C_{\mu },\partial _{\mu }\psi
\right] +\partial _{\mu }\left[ a_{\mu },\psi \right] \right)
=\epsilon (a)\psi \;. \label{s6}
\end{equation}
One might think of $a_{\mu }$ as a small perturbation.
Accordingly, the shift $\epsilon (a)$ of the eigenvalue of the
Faddeev-Popov operator from zero can be obtained in perturbation
theory by evaluating the expectation value of the term
$\partial_{\mu}\left[a_{\mu},\cdot\right]$, namely
\begin{equation}
\epsilon (a)=-\frac{\left\langle \varphi _{0}\mid \partial _{\mu
}\left[a_{\mu },\varphi _{0}\right] \right\rangle }{\left\langle
\varphi _{0}\mid \varphi _{0}\right\rangle }=-\frac{Tr\int
d^{4}x\left( \varphi _{0}\partial _{\mu }\left[ a_{\mu },\varphi
_{0}\right] \right) }{Tr\int d^{4}x\left( \varphi _{0}\varphi
_{0}\right) }\;.  \label{s7}
\end{equation}
Since $\partial _{\mu }\left( \partial _{\mu }\varphi _{0}+\left[
C_{\mu },\varphi _{0}\right] \right) =0$, it is easily verified
that the field
\begin{equation}
\widetilde{A}_{\mu }=A_{\mu }+D_{\mu }(C)\varphi _{0}\;,
\label{s8}
\end{equation}
has the same divergence of $A_{\mu }$. In fact
\begin{equation}
\partial _{\mu }\widetilde{A}_{\mu }=\partial _{\mu }A_{\mu }+\partial _{\mu
}D_{\mu }(C)\varphi _{0}=\partial _{\mu }A_{\mu }\;.\;  \label{s9}
\end{equation}
The fields $\widetilde{A}_{\mu }$ and $A_{\mu }$ can thus define
equivalent fields. For that, we have to find a gauge
transformation $S$ such that
\begin{equation}
\widetilde{A}_{\mu }={S}^{\dagger }\partial _{\mu }{S}+S^{\dagger }A_{\mu }{S%
}\;.  \label{s10}
\end{equation}
We shall look at $S$ close to unit, in the form
\begin{equation}
S=1+\alpha +\frac{1}{2}\alpha ^{2}+....  \label{s11}
\end{equation}
Also
\begin{equation}
S^{\dagger }=1-\alpha +\frac{1}{2}\alpha ^{2}+....  \label{s12}
\end{equation}
Indeed
\begin{eqnarray}
SS^{\dagger } &=&\left( 1+\alpha +\frac{1}{2}\alpha ^{2}\right)
\left(
1-\alpha +\frac{1}{2}\alpha ^{2}\right) +O(\alpha ^{3})  \nonumber \\
&=&1-\alpha +\frac{1}{2}\alpha ^{2}+\alpha -\alpha
^{2}+\frac{1}{2}\alpha
^{2}+O(\alpha ^{3})  \nonumber \\
&=&1+O(\alpha ^{3})\;.  \label{s13}
\end{eqnarray}
For $\widetilde{A}_{\mu }$ we get
\begin{eqnarray}
\widetilde{A}_{\mu } &=&\left( 1-\alpha +\frac{1}{2}\alpha
^{2}\right)
A_{\mu }\left( 1+\alpha +\frac{1}{2}\alpha ^{2}\right) +\left( 1-\alpha +%
\frac{1}{2}\alpha ^{2}\right) \partial _{\mu }\left( 1+\alpha +\frac{1}{2}%
\alpha ^{2}\right)  \nonumber \\
&=&\left( 1-\alpha +\frac{1}{2}\alpha ^{2}\right) \left( A_{\mu
}+A_{\mu
}\alpha +\frac{1}{2}A_{\mu }\alpha ^{2}\right)  \nonumber \\
&&+\left( 1-\alpha +\frac{1}{2}\alpha ^{2}\right) \left( \partial
_{\mu
}\alpha +\frac{1}{2}\left( \partial _{\mu }\alpha \right) \alpha +\frac{1}{2}%
\alpha \left( \partial _{\mu }\alpha \right) \right)  \nonumber \\
&=&A_{\mu }+A_{\mu }\alpha +\frac{1}{2}A_{\mu }\alpha ^{2}-\alpha
A_{\mu }-\alpha A_{\mu }\alpha +\frac{1}{2}\alpha ^{2}A_{\mu
}+\partial _{\mu
}\alpha  \nonumber \\
&&+\frac{1}{2}\left( \partial _{\mu }\alpha \right) \alpha +\frac{1}{2}%
\alpha \left( \partial _{\mu }\alpha \right) -\alpha \partial
_{\mu }\alpha
+O(\alpha ^{3})  \nonumber \\
&=&A_{\mu }+D_{\mu }(A)\alpha +\frac{1}{2}\left( \partial _{\mu
}\alpha \right) \alpha -\frac{1}{2}\alpha \left( \partial _{\mu
}\alpha \right)
\nonumber \\
&&+\frac{1}{2}A_{\mu }\alpha ^{2}-\alpha A_{\mu }\alpha
+\frac{1}{2}\alpha ^{2}A_{\mu }+O(\alpha ^{3})\;.  \label{s14}
\end{eqnarray}
From
\begin{equation}
\left[ \alpha ,\partial _{\mu }\alpha +\left[ A_{\mu },\alpha
\right] \right] =\alpha \left( \partial _{\mu }\alpha \right)
-\left( \partial _{\mu }\alpha \right) \alpha +\alpha A_{\mu
}\alpha -\alpha ^{2}A_{\mu }-A_{\mu }\alpha ^{2}+\alpha A_{\mu
}\alpha \;,  \label{s15}
\end{equation}
it follows
\begin{equation}
\widetilde{A}_{\mu }=A_{\mu }+D_{\mu }(A)\alpha -\frac{1}{2}\left[
\alpha ,\partial _{\mu }\alpha +\left[ A_{\mu },\alpha \right]
\right] +O(\alpha ^{3})\;.  \label{s16}
\end{equation}

\newpage

\begin{itemize}
\item  \textbf{Remark}

\textit{We must retain the second order term in the equation
}$\left( \ref {s11}\right) $ \textit{\ otherwise, from }
\begin{equation}
S=1+\alpha \;,  \label{s17}
\end{equation}
\textit{we would obtain}
\begin{equation}
\widetilde{A}_{\mu }=A_{\mu }+D_{\mu }(A)\alpha \;.  \label{s18}
\end{equation}
\textit{Condition }$\left( \ref{s9}\right) $\textit{, }$\partial \widetilde{A%
}=\partial A$\textit{, would thus imply }
\begin{equation}
\partial _{\mu }D_{\mu }(A)\alpha =0\;,  \label{s19}
\end{equation}
\textit{which has no solution for }$\alpha $\textit{ since }$A_{\mu }$%
\textit{\ is not located on a horizon, see eq.}$\left( \ref{s4}\right) $%
\textit{. }
\end{itemize}

\noindent From equation $\left( \ref{s16}\right) $ one obviously
has
\begin{equation}
D_{\mu }(C)\varphi _{0}=D_{\mu }(A)\alpha -\frac{1}{2}\left[
\alpha ,\partial _{\mu }\alpha +\left[ A_{\mu },\alpha \right]
\right] +O(\alpha ^{3})\;.  \label{s20}
\end{equation}
Tacking the divergence of both sides of eq.$\left(
\ref{s20}\right) $, we
obtain the condition to be fulfilled in order that $A_{\mu }\;$and $%
\widetilde{A}_{\mu }$ have the same divergence, $\partial \widetilde{A}%
=\partial A$, \textit{i.e. }
\begin{equation}
\partial _{\mu }D_{\mu }(A)\alpha -\frac{1}{2}\partial _{\mu }\left[ \alpha
,\partial _{\mu }\alpha +\left[ A_{\mu },\alpha \right] \right]
=0\;. \label{s21}
\end{equation}
This condition can be analyzed iteratively by setting
\begin{eqnarray}
\alpha  &=&\varphi _{0}+\widetilde{\alpha }\;,\,\,\,\,\,\,\,\,\,\,\,\,\,\,\,%
\;\widetilde{\alpha }\;\mathrm{small\;with\;respect\;to\;}\varphi
_{0}\;,
\nonumber \\
A_{\mu } &=&C_{\mu }+a_{\mu }\;,\;\;\;\;a_{\mu }\;\mathrm{%
small\;with\;respect\;to\;}A_{\mu }\;,  \label{s22}
\end{eqnarray}
It is useful to introduce an expansion parameter $\lambda $ which
we shall
set to one at the end\footnote{%
The introduction of $\lambda $ turns out to be useful in order to
analyze the condition $\left( \ref{s21} \right) $. Observe indeed
that, if $\lambda <1$, then $\lambda ^{2}<\lambda $. This means
that $a_{\mu }$ is smaller than $C_{\mu }$, and that
$\widetilde{\alpha }\;$is smaller than $\varphi _{0}$.}
\begin{eqnarray}
\alpha  &=&\lambda \varphi _{0}+\lambda ^{2}\widetilde{\alpha }\;,
\nonumber
\\
A_{\mu } &=&C_{\mu }+\lambda a_{\mu }\;.\,  \label{s23}
\end{eqnarray}
\noindent Condition $\left( \ref{s21}\right) $ gives
\begin{eqnarray}
\partial ^{2}\left( \lambda \varphi _{0}+\lambda ^{2}\widetilde{\alpha }%
\right) + &\partial _{\mu }\left[ \;C_{\mu }+\lambda a_{\mu
},\lambda \varphi _{0}+\lambda ^{2}\widetilde{\alpha }\right]
=\frac{1}{2}\partial
_{\mu }\left[ \lambda \varphi _{0}+\lambda ^{2}\widetilde{\alpha }%
\;,\partial _{\mu }\left( \lambda \varphi _{0}+\lambda
^{2}\widetilde{\alpha
}\right) \right]   \nonumber &\\[1mm]
& +\frac{1}{2}\partial _{\mu }\left[ \lambda \varphi _{0}+\lambda ^{2}%
\widetilde{\alpha }\;,\left[ C_{\mu }+\lambda a_{\mu },\lambda
\varphi
_{0}+\lambda ^{2}\widetilde{\alpha }\right] \right]   \nonumber \\
&&  \label{s24}
\end{eqnarray}
so that, up to terms of the order $\lambda ^{4}$
\begin{eqnarray}
&\lambda ^{2}\partial _{\mu }\left( \partial _{\mu }\widetilde{\alpha }%
+\left[ \;C_{\mu },\widetilde{\alpha }\right] \right) +\lambda
^{2}\partial _{\mu }\left[ \;a_{\mu },\varphi _{0}\right] +\lambda
^{3}\partial _{\mu
}\left[ \;a_{\mu },\widetilde{\alpha }\right] =\frac{\lambda ^{2}}{2}%
\partial _{\mu }\left[ \varphi _{0}\;,\partial _{\mu }\varphi _{0}+\left[
C_{\mu },\varphi _{0}\right] \right]   \nonumber   &\\[1mm]
&+\frac{\lambda ^{3}}{2}\partial _{\mu }\left[ \varphi _{0},\partial _{\mu }%
\widetilde{\alpha }+\left[ C_{\mu },\widetilde{\alpha }\right] \right] +%
\frac{\lambda ^{3}}{2}\partial _{\mu }\left[ \widetilde{\alpha
},\partial
_{\mu }\varphi _{0}+\left[ C_{\mu },\varphi _{0}\right] \right] +\frac{%
\lambda ^{3}}{2}\partial _{\mu }\left[ \varphi _{0},\left[ a_{\mu
},\varphi
_{0}\right] \right] +O(\lambda ^{4})\;.  \nonumber \\
&& \label{s25}
\end{eqnarray}
In particular, to the first order $\lambda ^{2}$, we find
\begin{equation}
\partial _{\mu }D_{\mu }(C)\widetilde{\alpha }+\partial _{\mu }\left[
\;a_{\mu },\varphi _{0}\right] =\frac{1}{2}\partial _{\mu }\left[
\varphi _{0},D_{\mu }(C)\varphi _{0}\right] \;,  \label{s226}
\end{equation}
from which it follows that
\begin{equation}
Tr\int d^{4}x\left( \varphi _{0}\partial _{\mu }D_{\mu
}(C)\widetilde{\alpha
}+\varphi _{0}\partial _{\mu }\left[ \;a_{\mu },\varphi _{0}\right] \right) =%
\frac{1}{2}Tr\int d^{4}x\left( \varphi _{0}\partial _{\mu }\left[
\varphi _{0},D_{\mu }(C)\varphi _{0}\right] \right) \;.
\label{s2226}
\end{equation}
Moreover, due to property $\partial_{\mu}D_{\mu }(C)=D_{\mu
}(C)\partial_{\mu}$, we have
\begin{equation}
Tr\int d^{4}x\varphi _{0}\partial _{\mu }D_{\mu }(C)\widetilde{\alpha }%
=Tr\int d^{4}x\left( \partial _{\mu }D_{\mu }(C)\varphi
_{0}\right) \widetilde{\alpha }=0\;.  \label{s126}
\end{equation}
As a consequence, condition $\left( \ref{s2226}\right) $ reads
\begin{equation}
Tr\int d^{4}x\left( \varphi _{0}\partial _{\mu }\left[ \;a_{\mu
},\varphi _{0}\right] \right) =\frac{1}{2}Tr\int d^{4}x\left(
\varphi _{0}\partial _{\mu }\left[ \varphi _{0},D_{\mu }(C)\varphi
_{0}\right] \right) \;. \label{sf26}
\end{equation}

\begin{itemize}
\item  \textbf{Summary}

\textit{The fields }$A_{\mu }\;$\textit{and }$\widetilde{A}_{\mu
}$\textit{\ }
\begin{equation}
\widetilde{A}_{\mu }=A_{\mu }+D_{\mu }(C)\varphi _{0}\;,
\label{s27}
\end{equation}
\textit{are equivalent fields }
\begin{equation}
\partial _{\mu }\widetilde{A}_{\mu }=\partial _{\mu }A_{\mu }+\partial _{\mu
}D_{\mu }(C)\varphi _{0}=\partial _{\mu }A_{\mu }\;.\;
\label{s28}
\end{equation}
\textit{For the gauge transformation }$S$%
\begin{equation}
\widetilde{A}_{\mu }={S}^{\dagger }\partial _{\mu }{S}+S^{\dagger }A_{\mu }{S%
}\;,  \label{s29}
\end{equation}
\textit{we have }
\begin{equation}
S=1+\alpha +\frac{1}{2}\alpha ^{2}+O(\alpha ^{3})\;,  \label{s30}
\end{equation}
\textit{so that }
\begin{equation}
\widetilde{A}_{\mu }=A_{\mu }+D_{\mu }(A)\alpha -\frac{1}{2}\left[
\alpha ,\partial _{\mu }\alpha +\left[ A_{\mu },\alpha \right]
\right] +O(\alpha ^{3})\;.  \label{s31}
\end{equation}
\textit{The condition \thinspace }$\partial \widetilde{A}=\partial A$\textit{%
\ gives }
\begin{equation}
\partial _{\mu }D_{\mu }(A)\alpha =\frac{1}{2}\partial _{\mu }\left[ \alpha
,\partial _{\mu }\alpha +\left[ A_{\mu },\alpha \right] \right]
+O(\alpha ^{3})\;.  \label{s32}
\end{equation}
\textit{Setting }
\begin{eqnarray}
\alpha  &=&\varphi _{0}+\widetilde{\alpha }\;,\,\,\,\,\,\,\,\,\,\,\,\,\,\,\,%
\;\widetilde{\alpha }\;\mathrm{small\;with\;respect\;to\;}\varphi
_{0}\;,
\nonumber \\
A_{\mu } &=&C_{\mu }+a_{\mu }\;,\;\;\;\;a_{\mu }\;\mathrm{%
small\;with\;respect\;to\;}A_{\mu }\;,  \label{s33}
\end{eqnarray}
\textit{we obtain, to the first order, }
\end{itemize}

\begin{equation}
\partial _{\mu }D_{\mu }(C)\widetilde{\alpha }+\partial _{\mu }\left[
\;a_{\mu },\varphi _{0}\right] =\frac{1}{2}\partial _{\mu }\left[
\varphi _{0},D_{\mu }(C)\varphi _{0}\right] \;.  \label{s34}
\end{equation}

\vspace{.5cm} \noindent It remains now to check on which side of
the horizon
$l_{1}$ the equivalent field $\widetilde{A}_{\mu }$ lies. Let us rewrite $%
\widetilde{A}_{\mu }$ as
\begin{equation}
\widetilde{A}_{\mu }=A_{\mu }+D_{\mu }(C)\varphi _{0}\;=C_{\mu
}+a_{\mu }+D_{\mu }(C)\varphi _{0}=C_{\mu }+a_{\mu }^{\prime }\;.
\label{s35}
\end{equation}
As done before, we evaluate the shift $\epsilon (a^{\prime })$ of
the
eigenvalue of the Faddeev-Popov operator from zero. Treating the field $%
a_{\mu }^{\prime }=a_{\mu }+D_{\mu }(C)\varphi _{0}\;$as a
perturbation, one obtains

\begin{equation}
\epsilon (a^{\prime })=-\frac{\left\langle \varphi _{0}\mid
\partial _{\mu }\left[ a_{\mu }^{\prime },\varphi _{0}\right]
\right\rangle }{\left\langle \varphi _{0}\mid \varphi
_{0}\right\rangle }=-\frac{Tr\int d^{4}x\left( \varphi
_{0}\partial _{\mu }\left[ a_{\mu },\varphi _{0}\right] +\varphi
_{0}\partial _{\mu }\left[ D_{\mu }(C)\varphi _{0},\varphi
_{0}\right] \right) }{Tr\int d^{4}x\left( \varphi _{0}\varphi
_{0}\right) }\;. \label{s36}
\end{equation}
Furthermore, from eq.$\left( \ref{sf26}\right) $, it follows

\begin{equation}
\epsilon (a^{\prime })=\frac{Tr\int d^{4}x\left( \varphi
_{0}\partial _{\mu }\left[ a_{\mu },\varphi _{0}\right] \right)
}{Tr\int d^{4}x\left( \varphi _{0}\varphi _{0}\right)
}\;=-\epsilon (a)\;.  \label{s37}
\end{equation}
Thus, if $A_{\mu }$, close to $l_{1}$, is located in $C_{0}$,
$\epsilon (a)>0 $, there is an equivalent field,
$\widetilde{A}_{\mu }=A_{\mu }+D_{\mu }(C)\varphi _{0}$, close to
$l_{1}$, which is located in $C_{1}$, $\epsilon (a^{\prime
})=-\epsilon (a)<0$. Also, it is worth mentioning that this
derivation can be generalized to fields close to any horizon
$l_{n}$. This concludes the proof of the statement.


\subsection{Restriction of the domain of integration to the first
horizon}

The previous statement suggests that the domain of integration in
the path integral should be restricted to the first horizon,
\textit{i.e.} to the region $C_{0}$ where the Faddeev-Popov
operator is positive definite.

\begin{itemize}
\item  \textbf{Remark}

{\it More precisely, two additional requirements should be
fulfilled to justify the hypothesis that the domain of integration
should be restricted to the region $C_{0}$. The first one is that
one should be able to prove that not only for small neighborhoods
close to the horizons $l_{n}$, but also for any field in the
region $C_{n}$ there is an equivalent field in the region
$C_{n-1}$. This would ensure that it is always possible to find a
chain of gauge transformations which brings a field in the region
$C_{n}$ to the corresponding equivalent field in the region
$C_{0}$. The second requirement is that one has to be sure that
the region $C_{0}$ is free from Gribov copies. We shall come back
on these issues later on in the Conclusion. For the time being, we
shall assume that the significant range of integration in the
path-integral is determined by the region $C_{0}$. }
\end{itemize}

\noindent Accordingly, for the partition function $\mathcal{Z}$ we
write

\begin{eqnarray}
\mathcal{Z} &=&\mathcal{N}\int DA_{\mu }DcD\overline{c}\delta
(\partial A)e^{-\left( S_{YM}+\int d^{4}x\overline{c}^{a}\partial
_{\mu }D_{\mu
}^{ab}c^{b}\right) }\mathcal{V}(C_{0})  \nonumber \\
&=&\mathcal{N}\int DA_{\mu }\delta (\partial A)e^{-S_{YM}}\det
\left( -\partial _{\mu }(\partial _{\mu }\delta
^{ab}-f^{abc}A_{\mu }^{c})\right) \mathcal{V}(C_{0})\;,
\label{r1}
\end{eqnarray}
where the presence of the factor $\mathcal{V}(C_{0})$ means that
the integration is performed only over the region $C_{0}$. In
order to characterize the quantity $\mathcal{V}(C_{0})$, we look
at the connected
two-point ghost function $\left\langle \overline{c}^{a}(x)c^{b}(y)\right%
\rangle _{\mathrm{c}}$. We have
\begin{eqnarray}
\left\langle \overline{c}^{a}(x)c^{b}(y)\right\rangle _{\mathrm{c}} &=&%
\mathcal{N}\int DA_{\mu }DcD\overline{c}\delta (\partial A)\overline{c}%
^{a}(x)c^{b}(y)e^{-\left( S_{YM}+\int
d^{4}x\overline{c}^{a}\partial _{\mu
}D_{\mu }^{ab}c^{b}\right) }\mathcal{V}(C_{0})  \nonumber \\
&=&\mathcal{N}\int DA_{\mu }\delta (\partial A)e^{-S_{YM}}\det
\left( -\partial _{\mu }D_{\mu }\right) \left[ \left( \partial
_{\mu }D_{\mu
}\right) ^{-1}\right] _{xy}^{ab}\mathcal{V}(C_{0})\;.  \nonumber \\
&&  \label{r2}
\end{eqnarray}
The presence of the factor $\mathcal{V}(C_{0})$ in eq.$\left( \ref
{r2}\right)$ implies that $\left\langle
\overline{c}^{a}(x)c^{b}(y)\right\rangle _{\mathrm{c}}$ can become
large only when approaching the horizon $l_{1}$. On the other
hand, the behavior
of the two-point ghost function obtained from perturbation theory, \textit{%
i.e.} with $\mathcal{V}(C_{0})=1$, is given by
\begin{eqnarray}
\left\langle \overline{c}^{a}(x)c^{b}(y)\right\rangle
_{\mathrm{c}}
&=&\delta ^{ab}\int \frac{d^{4}k}{\left( 2\pi \right) ^{4}}\mathcal{G}%
(k)e^{ik(x-y)}\;,  \nonumber \\
\mathcal{G}(k) &=&\frac{1}{k^{2}}\frac{1}{\left( 1-\frac{11g^{2}N}{48\pi ^{2}%
}\log \frac{\Lambda ^{2}}{k^{2}}\right) ^{\frac{9}{44}}}\;,
\label{r3}
\end{eqnarray}
where $\Lambda $ is the ultraviolet cutoff and $N$ is the Casimir
of the
adjoint representation of the gauge group $SU(N)$%
\begin{equation}
f^{acd}f^{bcd}=N\delta ^{ab}\;.  \label{r4}
\end{equation}
The expression of $\mathcal{G}(k)$ in eq.$\left( \ref{r3}\right) $
displays
two singularities, at $k^{2}=0$ and at $k_{c}^{2}=\Lambda ^{2}\exp \left( -%
\frac{1}{g^{2}}\frac{48\pi ^{2}}{11N}\right) $. However, the
presence of the
factor $\mathcal{V}(C_{0})$ makes it impossible for a singularity of $%
\mathcal{G}(k)$ to exist at nonvanishing $k$. Indeed, below the
singularity
position, \textit{i.e. }for $k^{2}<k_{c}^{2}$, the quantity $\left( 1-\frac{%
11g^{2}N}{48\pi ^{2}}\log \frac{\Lambda ^{2}}{k^{2}}\right) $ is
negative. Thus, for $k^{2}<k_{c}^{2}$, $\mathcal{G}(k)$ becomes
complex, indicating that the Faddeev-Popov operator $\left(
-\partial _{\mu }D_{\mu }\right) $ has ceased to be a positive
defined quantity, namely one has left the region $C_{0}$. It
remains the singularity at $k^{2}=0$. This singularity
has a simple interpretation. It means that we are approaching the boundary $%
l_{1}$ of $C_{0}$. In other words, at $k^{2}=0$ we feel the fields
on the horizon $l_{1}$.

\noindent Therefore, following \cite{g}, a possible characterization of the factor $%
\mathcal{V}(C_{0})$ can be obtained by computing the connected
two-point ghost function and by requiring that it has no poles at
nonvanishing $k$.

\begin{itemize}
\item  \textbf{Remark}

{\it For a better understanding of the previous statement about
the characterization of the factor $\mathcal{V}(C_{0})$, it is
useful to remind here that the region $C_{0}$ is defined as the
set of all transverse connections for which the Faddeev-Popov
operator is positive definite, namely
\begin{equation}
C_0=\{ A_{\mu},\; \partial A=0, \; -\partial _{\mu }\left(
\partial _{\mu }\delta^{ab} -f^{abc}A_{\mu }^c \right) >0 \}
\nonumber \;.
\end{equation}
In the region $C_{0}$, the Faddeev-Popov operator is invertible,
its inverse $\left[-\partial _{\mu }\left(
\partial _{\mu }\delta^{ab} -f^{abc}A_{\mu }^c
\right)\right]^{-1}$ becoming large only when approaching the
horizon $l_{1}$, due to the existence of a zero mode. \newline
\noindent Therefore, denoting by $\mathcal{G}(k;A)$ the color
singlet Fourier transform of $\left[-\partial _{\mu }\left(
\partial _{\mu }\delta^{ab} -f^{abc}A_{\mu }^c
\right)\right]^{-1}$,
\begin{equation}
\mathcal{G}(k;A)=\sum_{ab}\frac{\delta ^{ab}}{N^2-1}
<k\mid\left[-\partial _{\mu }\left(
\partial _{\mu }\delta^{ab} -f^{abc}A_{\mu }^c
\right)\right]^{-1}\mid k > \;, \label{cosi}
\end{equation}
we shall require that $\mathcal{G}(k;A)$ has no poles for a given
nonvanishing value of the momentum $k$, except for the singularity
at $k=0$, corresponding to the Gribov horizon $l_{1}$. Finally, we
remark that expression $\left( \ref{cosi}\right) $ can be
evaluated order by order in perturbation theory. It can be
obtained by computing the connected two-point ghost function in
the background of the gauge field $A_{\mu}^a$, which plays the
role of an external field. This will be task of the next section.}
\end{itemize}

\subsection{Characterization of $\mathcal{V}(C_{0})$}

In order to characterize $\mathcal{V}(C_{0})$, we start with the
expression of the connected, color singlet, two-point ghost
function
\begin{eqnarray}
\sum_{ab}\frac{\delta ^{ab}\left\langle \overline{c}^{a}(x)c^{b}(y)\right%
\rangle _{\mathrm{c}}}{N^{2}-1} &=&\mathcal{N}\int DA_{\mu }DcD\overline{c}%
\delta (\partial A)\frac{\overline{c}^{a}(x)c^{a}(y)}{\left( N^{2}-1\right) }%
e^{-\left( S_{YM}+\int d^{4}x\overline{c}^{a}\partial _{\mu
}D_{\mu
}^{ab}c^{b}\right) }  \nonumber \\
&=&\mathcal{N}\int DA_{\mu }\delta (\partial A)e^{-S_{YM}}\mathcal{G}%
(x,y;A)\;,  \nonumber \\
&&  \label{r5}
\end{eqnarray}
where $\mathcal{G}(x,y;A)$ stands for the connected ghost
two-point function with the gauge field $A_{\mu }^{a}$ considered
as an external classical field, namely

\begin{eqnarray}
\mathcal{G}(x,y;A) &=&\frac{1}{N^{2}-1}\int DcD\overline{c}\;\overline{c}%
^{a}(x)c^{a}(y)e^{-\int d^{4}x\overline{c}^{a}\partial _{\mu
}D_{\mu
}^{ab}c^{b}}\;.  \nonumber \\
&&  \label{r6}
\end{eqnarray}
We shall evaluate $\mathcal{G}(x,y;A)$ up to the second order in
perturbation theory. Making use of the Wick theorem, we obtain
\begin{eqnarray}
& \left\langle \overline{c}^{a}(x)c^{b}(y)\right\rangle
_{\mathrm{c}} =\left\langle
\overline{c}_{0}^{a}(x)c_{0}^{b}(y)\left( 1+\int d^{4}x_{1}\left(
\partial _{x_{1}}^{\mu }\overline{c}_{0}^{m}(x_{1})\right)
f^{mnp}A_{\mu }^{n}(x_{1})c_{0}^{p}(x_{1})\right. \right. &\cr &
\left.
\left. +\frac{1}{2}\int d^{4}x_{1}d^{4}x_{2}\left( \partial _{x_{1}}^{\mu }%
\overline{c}_{0}^{m}(x_{1})\right) f^{mnp}A_{\mu
}^{n}(x_{1})c_{0}^{p}(x_{1})\left( \partial _{x_{2}}^{\nu }\overline{c}%
_{0}^{q}(x_{2})\right) f^{qrt}A_{\mu
}^{r}(x_{2})c_{0}^{t}(x_{2})\right)
\right\rangle _{\mathrm{c}}  \nonumber \\
&&  \label{r7}
\end{eqnarray}
where $\overline{c}_{0}^{a}(x)$, $c_{0}^{b}(y)$ stand for free
fields. Performing the Wick contraction, yields

\begin{eqnarray}
\left\langle \overline{c}^{a}(x)c^{b}(y)\right\rangle
_{\mathrm{c}} =\left\langle
\overline{c}_{0}^{a}(x)c_{0}^{b}(y)\right\rangle -\int
d^{4}x_{1}f^{mnp}A_{\mu }^{n}(x_{1})\left\langle \overline{c}%
_{0}^{a}(x)c_{0}^{p}(x_{1})\right\rangle \partial _{x_{1}}^{\mu
}\left\langle \overline{c}_{0}^{m}(x_{1})c_{0}^{b}(y)\right\rangle & &\cr %
+\int d^{4}x_{1}d^{4}x_{2}f^{mnp}A_{\mu }^{n}(x_{1})f^{qrt}A_{\nu
}^{r}(x_{2})\left\langle \overline{c}_{0}^{a}(x)c_{0}^{p}(x_{1})\right%
\rangle \partial _{x_{1}}^{\mu }\left\langle \overline{c}%
_{0}^{m}(x_{1})c_{0}^{t}(x_{2})\right\rangle \partial
_{x_{2}}^{\nu }\left\langle
\overline{c}_{0}^{q}(x_{2})c_{0}^{b}(y)\right\rangle  \nonumber
\\
&&  \label{r8}
\end{eqnarray}
From

\begin{equation}
\left\langle \overline{c}_{0}^{a}(x)c_{0}^{b}(y)\right\rangle =\delta ^{ab}%
\mathcal{G}_{0}(x-y)\;,  \label{r9}
\end{equation}
with $\mathcal{G}_{0}(x-y)$ being the free ghost propagator, we
get

\begin{eqnarray}
\left\langle \overline{c}^{a}(x)c^{b}(y)\right\rangle
_{\mathrm{c}} =\delta ^{ab}\mathcal{G}_{0}(x-y)-\int
d^{4}x_{1}\mathcal{G}_{0}(x-x_{1})\partial
_{x_{1}}^{\mu }\mathcal{G}_{0}(x_{1}-y)f^{bna}A_{\mu }^{n}(x_{1}) & &\cr %
+\int d^{4}x_{1}d^{4}x_{2}\mathcal{G}_{0}(x-x_{1})\partial _{x_{1}}^{\mu }%
\mathcal{G}_{0}(x_{1}-x_{2})\partial _{x_{2}}^{\nu }\mathcal{G}%
_{0}(x_{2}-y)f^{tna}A_{\mu }^{n}(x_{1})f^{brt}A_{\nu
}^{r}(x_{2})\;.
\nonumber \\
&&  \label{r10}
\end{eqnarray}
Finally, for $\mathcal{G}(x,y;A)$ we obtain
\begin{eqnarray}
\mathcal{G}(x,y;A) =\mathcal{G}_{0}(x-y)-\frac{1}{N^{2}-1}\int d^{4}x_{1}%
\mathcal{G}_{0}(x-x_{1})\partial _{x_{1}}^{\mu }\mathcal{G}%
_{0}(x_{1}-y)f^{ana}A_{\mu }^{n}(x_{1}) & &\cr
+\frac{1}{N^{2}-1}\int
d^{4}x_{1}d^{4}x_{2}\mathcal{G}_{0}(x-x_{1})\partial _{x_{1}}^{\mu }\mathcal{%
G}_{0}(x_{1}-x_{2})\partial _{x_{2}}^{\nu }\mathcal{G}%
_{0}(x_{2}-y)f^{tna}A_{\mu }^{n}(x_{1})f^{art}A_{\nu
}^{r}(x_{2})\;,
\nonumber \\
&&\;  \label{r11}
\end{eqnarray}
which, due to
\begin{equation}
f^{aab}=0\;,\;\;\;\;\;f^{acd}f^{bcd}=N\delta ^{ab}\;,  \label{r12}
\end{equation}
reads
\begin{eqnarray}
& \mathcal{G}(x,y;A) =\mathcal{G}_{0}(x-y) &  \nonumber \\[1mm]
&-\frac{N}{N^{2}-1}\int
d^{4}x_{1}d^{4}x_{2}\mathcal{G}_{0}(x-x_{1})\partial
_{x_{1}}^{\mu }\mathcal{G}_{0}(x_{1}-x_{2})\partial _{x_{2}}^{\nu }\mathcal{G%
}_{0}(x_{2}-y)A_{\mu }^{a}(x_{1})A_{\nu }^{a}(x_{2})\;.  \nonumber \\
&&  \label{r13}
\end{eqnarray}
It remains now to take the Fourier transformation of expression
$\left( \ref {r13}\right) $. It turns out to be useful to work in
a finite volume $V$, taken here to be a four-dimensional hypercube
$V=L^{4}$, and to perform the thermodynamic limit, $V\rightarrow
\infty $, at the end.

\begin{itemize}
\item  \textbf{Remark}

\textit{The following conventions will be adopted in a finite
volume. For
the Fourier transformation of the fields $\varphi ^{a}=\left( A_{\mu }^{a},%
\overline{c}^{a},c^{b}\right) $ we have
\begin{equation}
\varphi ^{a}(x)=\frac{1}{\sqrt{V}}\sum_{q}\varphi
^{a}(q)e^{iqx}\;, \label{r14}
\end{equation}
with
\begin{equation}
q=\frac{2\pi }{L}\left( n_{1},n_{2},n_{3},n_{4}\right) \;,\;\;\;\;\;n_{i}\;%
\; integers.  \label{rr15}
\end{equation}
The inverse of the Fourier transformation is easily obtained by
making use of
\begin{equation}
\int_{V}d^{4}xe^{i(q-q^{\prime })x}=V\delta _{qq^{\prime }}\;.
\label{r16}
\end{equation}
Thus
\begin{equation}
\varphi ^{a}(q)=\frac{1}{\sqrt{V}}\int_{V}d^{4}x\varphi
^{a}(x)e^{-iqx}\;. \label{r17}
\end{equation}
Also,
\begin{equation}
\sum_{q}\equiv \sum_{n_{1},n_{2},n_{3},n_{4}}\;,  \label{r18}
\end{equation}
and, in the thermodynamic limit, $V\rightarrow \infty $,
\begin{equation}
\sum_{q}=\sum_{n_{i}}\rightarrow \int d^{4}n=V\int
\frac{d^{4}q}{\left( 2\pi \right) ^{4}}\;.\;  \label{r19}
\end{equation}
}
\end{itemize}

\noindent Let us evaluate first the Fourier transformation of the
free ghost propagator
\begin{eqnarray}
\left\langle \overline{c}_{0}^{a}(x)c_{0}^{b}(y)\right\rangle &=&\mathcal{N}%
\int DcD\overline{c}\;\overline{c}_{0}^{a}(x)c_{0}^{b}(y)e^{-\int d^{4}x%
\overline{c}^{a}\partial ^{2}c^{a}}  \nonumber \\
&=&\frac{1}{V}\sum_{q,p}e^{i(qx+py)}\mathcal{N}\int DcD\overline{c}\;%
\overline{c}_{0}^{a}(q)c_{0}^{b}(p)e^{-\sum_{k}\overline{c}%
^{a}(k)k^{2}c^{a}(-k)}\;  \nonumber \\
&=&\frac{1}{V}\sum_{q,p}e^{i(qx+py)}\frac{1}{q^{2}}\delta
^{ab}\delta
_{q(-p)}  \nonumber \\
&=&\frac{1}{V}\sum_{q}e^{iq(x-y)}\frac{1}{q^{2}}\delta ^{ab}\;.
\label{r20}
\end{eqnarray}
Therefore, for $\mathcal{G}_{0}(x-y)$ we have
\begin{equation}
\mathcal{G}_{0}(x-y)=\frac{1}{V}\sum_{q}e^{iq(x-y)}\frac{1}{q^{2}}\;.
\label{r21}
\end{equation}
We are now ready to evaluate the Fourier transformation of $\mathcal{G}%
(x,y;A)$. Setting

\begin{equation}
\mathcal{G}(k;A)=\frac{1}{V}\int
d^{4}xd^{4}ye^{ik(x-y)}\mathcal{G}(x,y;A)\;, \label{r22}
\end{equation}
we get
\begin{eqnarray}
\mathcal{G}(k;A) &=&\frac{1}{V}\int_{V}d^{4}xd^{4}ye^{ik(x-y)}\mathcal{G}%
_{0}(x-y)  \nonumber \\
&&-\frac{N}{N^{2}-1}\frac{1}{V}%
\int_{V}d^{4}xd^{4}yd^{4}x_{1}d^{4}x_{2}e^{ik(x-y)}\left( \mathcal{G}%
_{0}(x-x_{1})\partial _{x_{1}}^{\mu
}\mathcal{G}_{0}(x_{1}-x_{2})\right.
\nonumber \\
&&\times \left. \partial _{x_{2}}^{\nu
}\mathcal{G}_{0}(x_{2}-y)A_{\mu
}^{a}(x_{1})A_{\nu }^{a}(x_{2})\;\right) \;.  \nonumber \\
&&  \label{rr23}
\end{eqnarray}
Thus
\begin{eqnarray}
\mathcal{G}(k;A) &=&\frac{1}{V^{2}}\sum_{q}%
\int_{V}d^{4}xd^{4}ye^{ix(k+q)}e^{-iy(k+q)}\frac{1}{q^{2}}\;  \nonumber \\
&&+\frac{N}{N^{2}-1}\frac{1}{V^{5}}\sum_{qplur}%
\int_{V}d^{4}xd^{4}yd^{4}x_{1}d^{4}x_{2}\left(
e^{ix(k+q)}e^{-i(k+l)}e^{ix_{1}(p-q+u)}\right.  \nonumber \\
&&\times \left. e^{ix_{2}(l-p+r)}\frac{1}{q^{2}}\frac{p_{\mu }}{p^{2}}\frac{%
l_{\nu }}{l^{2}}A_{\mu }^{a}(u)A_{\nu }^{a}(r)\right) \;  \nonumber \\
&=&\frac{1}{V}\sum_{q}\int_{V}d^{4}ye^{-iy(k+q)}\frac{1}{q^{2}}\delta
_{(q+k)0}
\nonumber \\
&&+\frac{N}{N^{2}-1}\frac{V^{4}}{V^{5}}\sum_{qplur}\delta
_{q-k}\delta
_{l-k}\delta _{(p-q+u)0}\delta _{(l-p+r)0}\frac{1}{q^{2}}\frac{p_{\mu }}{%
p^{2}}\frac{l_{\nu }}{l^{2}}A_{\mu }^{a}(u)A_{\nu }^{a}(r)  \nonumber \\
&=&\frac{1}{k^{2}}+\frac{N}{N^{2}-1}\frac{1}{V}\frac{1}{k^{2}}\frac{-k_{\nu }%
}{k^{2}}\sum_{p}\frac{p_{\mu }}{p^{2}}A_{\mu }^{a}(-p-k)A_{\nu
}^{a}(p+k)
\nonumber \\
&=&\frac{1}{k^{2}}+\frac{N}{N^{2}-1}\frac{1}{V}\frac{1}{k^{2}}\frac{-k_{\nu }%
}{k^{2}}\sum_{p}\frac{\left( p-k\right) _{\mu }}{\left( p-k\right) ^{2}}%
A_{\mu }^{a}(-p)A_{\nu }^{a}(p)  \nonumber \\
&=&\frac{1}{k^{2}}\left( 1+\frac{N}{N^{2}-1}\frac{1}{V}\frac{1}{k^{2}}%
\sum_{p}\frac{\left( k-p\right) _{\mu }k_{\nu }}{\left( k-p\right) ^{2}}%
A_{\mu }^{a}(-p)A_{\nu }^{a}(p)\right)  \nonumber \\
&=&\frac{1}{k^{2}}\left( 1+\sigma (k,A)\right) \approx \frac{1}{k^{2}}\frac{1%
}{\left( 1-\sigma (k,A)\right) }\;.  \label{r24}
\end{eqnarray}
Finally

\begin{equation}
\mathcal{G}(k;A)\approx \frac{1}{k^{2}}\frac{1}{\left( 1-\sigma
(k,A)\right) }\;,  \label{r25}
\end{equation}
with
\begin{equation}
\sigma
(k,A)=\frac{N}{N^{2}-1}\frac{1}{V}\frac{1}{k^{2}}\sum_{q}\frac{\left(
k-q\right) _{\mu }k_{\nu }}{\left( k-q\right) ^{2}}A_{\mu
}^{a}(-q)A_{\nu }^{a}(q)\;.  \label{r26}
\end{equation}

\begin{itemize}
\item  \textbf{Summary}

\textit{For the Fourier transformation of $\mathcal{G}(x,y;A)$ we
have }

\textit{
\begin{equation}
\mathcal{G}(k;A)=\frac{1}{V}\int
d^{4}xd^{4}ye^{ik(x-y)}\mathcal{G}(x,y;A)\;, \label{r27}
\end{equation}
\begin{equation}
\mathcal{G}(k;A)\approx \frac{1}{k^{2}}\frac{1}{\left( 1-\sigma
(k,A)\right) }\;,  \label{r28}
\end{equation}
with
\begin{equation}
\sigma
(k,A)=\frac{N}{N^{2}-1}\frac{1}{V}\frac{1}{k^{2}}\sum_{q}\frac{\left(
k-q\right) _{\mu }k_{\nu }}{\left( k-q\right) ^{2}}A_{\mu
}^{a}(-q)A_{\nu }^{a}(q)\;.  \label{r29}
\end{equation}
In the thermodynamic limit, $V\rightarrow \infty $,
\begin{equation}
\sigma (k,A)=\frac{N}{N^{2}-1}\frac{1}{k^{2}}\int
\frac{d^{4}q}{\left( 2\pi
\right) ^{4}}\frac{\left( k-q\right) _{\mu }k_{\nu }}{\left( k-q\right) ^{2}}%
A_{\mu }^{a}(-q)A_{\nu }^{a}(q)\;.  \label{r30}
\end{equation}
}
\end{itemize}

\begin{itemize}
\item  \textbf{Remark}

{\it Notice also that, according to equation $\left(
\ref{cosi}\right) $, the expression for the Fourier transformation
$\left( \ref{r27}\right) $ implies that the propagation of the
ghosts in the field $A^{a}_{\mu}$ occurs with conservation of the
ghost momentum $k$. }
\end{itemize}

\subsection{The no-pole condition}

We can now establish the no-pole condition for the two-point ghost
function. From expression $\left( \ref{r28}\right) $ it follows
that the no-pole condition at finite nonvanishing $k$ can be
stated as
\begin{equation}
\sigma (k,A)<1\;.  \label{r31}
\end{equation}
Moreover, following \cite{g}, condition $\left( \ref{r31}\right) $
can be simplified by recalling that in the Landau gauge the field
$A_{\mu }^{a}(q)$ is transverse, namely
\begin{equation}
q_{\mu }A_{\mu }^{a}(q)=0\;.  \label{r33}
\end{equation}
From
\begin{equation}
q_{\mu }A_{\mu }^{a}(-q)A_{\nu }^{a}(q)=q_{\nu }A_{\mu
}^{a}(-q)A_{\nu }^{a}(q)=0\;,  \label{r34}
\end{equation}
we can set
\begin{equation}
A_{\mu }^{a}(-q)A_{\nu }^{a}(q)=\omega (A)\left( \delta _{\mu \nu }-\frac{%
q_{\mu }q_{\nu }}{q^{2}}\right) \;.  \label{r35}
\end{equation}
Contracting both sides with $\delta _{\mu \nu }$, it follows
\begin{equation}
\omega (A)=\frac{1}{3}A_{\lambda }^{a}(q)A_{\lambda }^{a}(-q)\;.
\label{r36}
\end{equation}
Therefore, for $\sigma (k,A)$ we obtain
\begin{eqnarray}
\sigma (k,A) &=&\frac{1}{3}\frac{N}{N^{2}-1}\frac{1}{k^{2}}\frac{1}{V}%
\sum_{q}\frac{\left( k-q\right) _{\mu }k_{\nu }}{\left( k-q\right) ^{2}}%
\left( A_{\lambda }^{a}(q)A_{\lambda }^{a}(-q)\right) \left(
\delta _{\mu
\nu }-\frac{q_{\mu }q_{\nu }}{q^{2}}\right)   \nonumber \\
&=&\frac{1}{3}\frac{N}{N^{2}-1}\frac{k_{\mu }k_{\nu }}{k^{2}}\frac{1}{V}%
\sum_{q}\frac{1}{\left( k-q\right) ^{2}}\left( A_{\lambda
}^{a}(q)A_{\lambda
}^{a}(-q)\right) \left( \delta _{\mu \nu }-\frac{q_{\mu }q_{\nu }}{q^{2}}%
\right) \;\;.  \nonumber \\
&&  \label{r37}
\end{eqnarray}
In the thermodynamic limit
\begin{eqnarray}
\sigma (k,A) &=&\frac{1}{3}\frac{N}{N^{2}-1}\frac{k_{\mu }k_{\nu }}{k^{2}}%
\mathcal{I}_{\mu \nu }(k)\;,  \nonumber \\
\mathcal{I}_{\mu \nu }(k) &=&\int \frac{d^{4}q}{\left( 2\pi \right) ^{4}}%
\frac{1}{\left( k-q\right) ^{2}}\left( A_{\lambda
}^{a}(q)A_{\lambda
}^{a}(-q)\right) \left( \delta _{\mu \nu }-\frac{q_{\mu }q_{\nu }}{q^{2}}%
\right) \;.  \label{r38}
\end{eqnarray}
As it will be checked later on, the quantity $\left( A_{\lambda
}^{a}(q)A_{\lambda }^{a}(-q)\right) $ turns out to decrease with
$q^2$, so that $\sigma (k,A)$ decreases as $k^2$ increases. Hence,
as no-pole condition one can take
\begin{equation}
\sigma (0,A)<1\;,  \label{r45}
\end{equation}
where
\begin{equation}
\sigma (0,A)=\frac{1}{4}\frac{N}{N^{2}-1}\int \frac{d^{4}q}{\left(
2\pi \right) ^{4}}\frac{1}{q^{2}}\left( A_{\lambda
}^{a}(q)A_{\lambda }^{a}(-q)\right) \;.  \label{r44}
\end{equation}
This expression follows by observing that
\begin{eqnarray}
\mathcal{I}_{\mu \nu }(0) &=&\int \frac{d^{4}q}{\left( 2\pi \right) ^{4}}%
\frac{1}{q^{2}}\left( A_{\lambda }^{a}(q)A_{\lambda
}^{a}(-q)\right) \left(
\delta _{\mu \nu }-\frac{q_{\mu }q_{\nu }}{q^{2}}\right)   \nonumber \\
&=&\frac{3}{4}\delta _{\mu \nu }\int \frac{d^{4}q}{\left( 2\pi \right) ^{4}}%
\frac{1}{q^{2}}\left( A_{\lambda }^{a}(q)A_{\lambda
}^{a}(-q)\right) \;, \label{r40}
\end{eqnarray}
where the last equality follows from Lorentz covariance. In fact,
setting
\begin{equation}
\int \frac{d^{4}q}{\left( 2\pi \right) ^{4}}\frac{1}{q^{2}}\left(
A_{\lambda }^{a}(q)A_{\lambda }^{a}(-q)\right) \left( \delta _{\mu
\nu }-\frac{q_{\mu }q_{\nu }}{q^{2}}\right) =\mathcal{J}\delta
_{\mu \nu }\;,  \label{r41}
\end{equation}
and contracting both sides with $\delta _{\mu \nu }$, we get
\begin{equation}
\mathcal{J=}\frac{3}{4}\int \frac{d^{4}q}{\left( 2\pi \right) ^{4}}\frac{1}{%
q^{2}}\left( A_{\lambda }^{a}(q)A_{\lambda }^{a}(-q)\right) \;.
\label{r42}
\end{equation}

\begin{itemize}
\item  \textbf{Summary}

\textit{In the Landau gauge, the factor $\sigma (k,A)$ is given by
the expression
\begin{equation}
\sigma (k,A)=\frac{1}{3}\frac{N}{N^{2}-1}\frac{k_{\mu }k_{\nu }}{k^{2}}\frac{%
1}{V}\sum_{q}\frac{1}{\left( k-q\right) ^{2}}\left( A_{\lambda
}^{a}(q)A_{\lambda }^{a}(-q)\right) \left( \delta _{\mu \nu
}-\frac{q_{\mu }q_{\nu }}{q^{2}}\right) \;.  \label{r46}
\end{equation}
For the no-pole condition for the two-point ghost function we have
\begin{equation}
\sigma (0,A)<1\;,  \label{r47}
\end{equation}
with
\begin{equation}
\sigma (0,A)=\frac{1}{4}\frac{N}{N^{2}-1}\frac{1}{V}\sum_{q}\frac{1}{q^{2}}%
\left( A_{\lambda }^{a}(q)A_{\lambda }^{a}(-q)\right) \;,
\label{r48}
\end{equation}
which, in the thermodynamic limit, reads
\begin{equation}
\sigma (0,A)=\frac{1}{4}\frac{N}{N^{2}-1}\int \frac{d^{4}q}{\left(
2\pi \right) ^{4}}\frac{1}{q^{2}}\left( A_{\lambda
}^{a}(q)A_{\lambda }^{a}(-q)\right) \;.  \label{r49}
\end{equation}
}
\end{itemize}

\subsection{An expression for $\mathcal{V}(C_{0})$}

According to \cite{g}, the expression for the factor
$\mathcal{V}(C_{0})$ which implements the no-pole condition
$\left( \ref{r45}\right) $ in the path integral can be taken as
\begin{equation}
\mathcal{V}(C_{0})=\theta (1-\sigma (0,A))\;,  \label{r50}
\end{equation}
where $\theta (x)$ stands for the step function\footnote{$\theta (x)=1$ for $%
x>0$, $\theta (x)=0$ for $x<0$.}. Therefore, for the partition function $%
\mathcal{Z}$ we have

\begin{equation}
\mathcal{Z}=\mathcal{N}\int DA_{\mu }\delta (\partial
A)e^{-S_{YM}}\det \left( -\partial _{\mu }D_{\mu }\right) \theta
(1-\sigma (0,A))\mathcal{\;}. \label{r51}
\end{equation}
Using the integral representation for the step function
\begin{equation}
\theta (x)=\int_{-i\infty +\varepsilon }^{i\infty +\varepsilon
}\frac{d\beta }{2\pi i\beta }e^{\beta x}\;,  \label{r52}
\end{equation}
we arrive at the expression
\begin{equation}
\mathcal{Z}=\mathcal{N}\int \frac{d\beta }{2\pi i\beta }DA_{\mu
}\delta (\partial A)e^{\beta (1-\sigma (0,A))}e^{-S_{YM}}\det
\left( -\partial _{\mu }D_{\mu }\right) \mathcal{\;},  \label{r53}
\end{equation}
which is suitable for analyzing the structure of the gauge
propagator. This will be the task of the next section.


\subsection{The gluon propagator\ in the Landau gauge}

In order to work out the gluon propagator, it is sufficient to
retain only the quadratic terms in the expression for the
partition function $\left( \ref {r53}\right) $, \textit{i.e.} we
start from
\begin{eqnarray}
\mathcal{Z}_{\mathrm{quadr}} &=&\mathcal{N}\int \frac{d\beta }{2\pi i\beta }%
DA_{\mu }e^{\beta (1-\sigma (0,A))}e^{-\frac{1}{4g^{2}}\int
d^{4}x\left(
\partial _{\mu }A_{\nu }^{a}-\partial _{\nu }A_{\mu }^{a}\right) ^{2}-\frac{1%
}{2g^{2}\alpha }\int d^{4}x\left( \partial _{\mu }A_{\mu }^{a}\right) ^{2}}%
\mathcal{\;},  \nonumber \\
&&  \label{r54}
\end{eqnarray}
where the limit $\alpha \rightarrow 0$ has to be taken at the end
in order to recover the Landau gauge. Passing to momentum space,
we get
\begin{eqnarray}
\mathcal{Z}_{\mathrm{quadr}} &=&\mathcal{N}\int \frac{d\beta }{2\pi i\beta }%
DA_{\mu }e^{-\frac{1}{2g^{2}}\sum_{q}A_{\mu }^{a}(q)\left(
q^{2}\delta _{\mu \nu }+\left( \frac{1}{\alpha }-1\right) q_{\mu
}q_{\nu }\right) A_{\nu
}^{a}(-q)\;}\mathcal{\;}  \nonumber \\
&&\times e^{\beta }e^{-\beta \frac{N}{N^{2}-1}\frac{1}{4}\frac{1}{V}\sum_{q}%
\frac{1}{q^{2}}\left( A_{\lambda }^{a}(q)A_{\lambda
}^{a}(-q)\right) }\;
\nonumber \\
&=&\mathcal{N}\int \frac{d\beta e^{\beta }}{2\pi i\beta }DA_{\mu }e^{-\frac{1%
}{2g^{2}}\sum_{q}A_{\mu }^{a}(q)\mathcal{Q}_{\mu \nu }^{ab}A_{\nu
}^{b}(-q)}\;,  \label{r56}
\end{eqnarray}
with
\begin{equation}
\mathcal{Q}_{\mu \nu }^{ab}=\left( q^{2}\delta _{\mu \nu }+\left( \frac{1}{\alpha }-1\right) q_{\mu }q_{\nu }+\frac{\beta Ng^{2}}{N^{2}-1}\frac{1}{2V}%
\frac{1}{q^{2}}\delta _{\mu \nu }\right) \delta ^{ab}\;.
\label{r57}
\end{equation}
Thus
\begin{equation}
\mathcal{Z}_{\mathrm{quadr}}=\mathcal{N}\int \frac{d\beta e^{\beta
}}{2\pi i\beta }\left( \det \mathcal{Q}_{\mu \nu }^{ab}\right)
^{-\frac{1}{2}}\;. \label{r58}
\end{equation}
From
\begin{eqnarray}
\left( \det \mathcal{Q}_{\mu \nu }^{ab}\right) ^{-\frac{1}{2}} &=&e^{-\frac{1%
}{2}\log \det \mathcal{Q}_{\mu \nu }^{ab}}  \nonumber \\
&=&e^{-\frac{3}{2}\left( N^{2}-1\right) \sum_{q}\log \left( q^{2}+\frac{%
\beta Ng^{2}}{N^{2}-1}\frac{1}{2V}\frac{1}{q^{2}}\right) \;,}
\label{r59}
\end{eqnarray}
we get
\begin{equation}
\mathcal{Z}_{\mathrm{quadr}}=\mathcal{N}\int \frac{d\beta }{2\pi i}%
e^{f(\beta )}\;,  \label{r60}
\end{equation}
with
\begin{equation}
f(\beta )=\beta -\log \beta -\frac{3}{2}\left( N^{2}-1\right)
\sum_{q}\log \left( q^{2}+\frac{\beta
Ng^{2}}{N^{2}-1}\frac{1}{2V}\frac{1}{q^{2}}\right) \;.
\label{r61}
\end{equation}
Expression $\left( \ref{r60}\right) $ can be evaluated at the
saddle point, namely
\begin{equation}
\mathcal{Z}_{\mathrm{quadr}}\approx e^{f(\beta _{0})}\;,
\label{r62}
\end{equation}
where $\beta _{0}$ is determined by the minimum condition
\begin{equation}
f^{\prime }(\beta _{0})=0\;,  \label{r63}
\end{equation}
which yields
\begin{equation}
1-\frac{1}{\beta _{0}}-\frac{3}{4}\frac{Ng^{2}}{V}\sum_{q}\frac{1}{q^{4}+%
\frac{\beta _{0}Ng^{2}}{N^{2}-1}\frac{1}{2V}}=0\;.  \label{r64}
\end{equation}
Taking the thermodynamic limit, $V\rightarrow \infty $, and
setting
\begin{equation}
\gamma ^{4}=\frac{\beta _{0}Ng^{2}}{N^{2}-1}\frac{1}{2V}\;\;\;\mathrm{with}%
\;\;\;\;\;V\rightarrow \infty \;,  \label{r65}
\end{equation}
we get the following gap equation for the parameter $\gamma $%
\begin{equation}
\frac{3Ng^{2}}{4}\int \frac{d^{4}q}{\left( 2\pi \right) ^{4}}\frac{1}{%
q^{4}+\gamma ^{4}}=1\;,  \label{r66}
\end{equation}
where the term $1/\beta _{0}$ has been neglected in the
thermodynamic limit, according to eq.$\left( \ref{r65}\right) $.
The parameter $\gamma$ has the dimension of a mass. It is defined
by the gap equation $\left( \ref {r66}\right) $. In particular,
from
\begin{equation}
\int \frac{d^{4}q}{\left( 2\pi \right) ^{4}}\frac{1}{q^{4}+\gamma ^{4}}=%
\frac{\Omega _{4}}{\left( 2\pi \right) ^{4}}\int_{0}^{\Lambda }dq\frac{q^{3}%
}{q^{4}+\gamma ^{4}}=\frac{1}{16\pi ^{2}}\log \frac{\Lambda ^{2}}{\gamma ^{2}%
}\;,  \label{r67}
\end{equation}
where $\Omega _{4}=2\pi ^{2}$ stands for the four-dimensional
solid angle and $\Lambda $ is the ultraviolet cutoff, one gets
\begin{equation}
\gamma ^{2}=\Lambda ^{2}e^{-\frac{64\pi
^{2}}{3N}\frac{1}{g^{2}}}\;. \label{r68}
\end{equation}
To obtain the gluon propagator, we can now go back to expression
$\left( \ref {r56}\right) $ which, after substituting the saddle
point value $\beta =\beta _{0}$, becomes
\begin{equation}
\mathcal{Z}_{\mathrm{quadr}}=\mathcal{N}^{\prime }\mathcal{\;}\int
DA_{\mu }e^{-\frac{1}{2g^{2}}V\int \frac{d^{4}q}{\left( 2\pi
\right) ^{4}}A_{\mu }^{a}(q)\mathcal{Q}_{\mu \nu }^{ab}A_{\nu
}^{b}(-q)}\;,  \label{r69}
\end{equation}
with
\begin{equation}
\mathcal{Q}_{\mu \nu }^{ab}=\left( \left( q^{2}+\frac{\gamma ^{4}}{q^{2}}%
\right) \delta _{\mu \nu }+\left( \frac{1}{\alpha }-1\right)
q_{\mu }q_{\nu }\right) \delta ^{ab}\;.  \label{r70}
\end{equation}
The gluon propagator is obtained by evaluating the inverse of $\mathcal{Q}%
_{\mu \nu }^{ab}$ and taking the limit $\alpha \rightarrow 0$.
After a straightforward computation we get
\begin{equation}
\mathcal{D}_{\mu \nu }^{ab}(q)=\left\langle A_{\mu }^{a}(q)A_{\nu
}^{b}(-q)\right\rangle =\delta ^{ab}g^{2}\frac{q^{2}}{q^{4}+\gamma ^{4}}%
\left( \delta _{\mu \nu }-\frac{q_{\mu }q_{\nu }}{q^{2}}\right)
\;. \label{r71}
\end{equation}
One sees that for large $q$, $q^{2}\gg \gamma ^{2}$, one recovers
the usual perturbative behavior
\begin{equation}
\delta ^{ab}g^{2}\frac{1}{q^{2}}\left( \delta _{\mu \nu
}-\frac{q_{\mu }q_{\nu }}{q^{2}}\right) \;.  \label{r72}
\end{equation}
However, for small values of $q$, corresponding to the infrared
region, the behavior of the gluon propagator deeply differs from
the perturbative behavior. Notice in fact that $\mathcal{D}_{\mu
\nu }^{ab}(q)$ is suppressed in the infrared.

\begin{itemize}
\item  \textbf{Remark}

\textit{In the thermodynamic limit, $V\rightarrow \infty $, we
have neglected the term $1/\beta _{0}$ in eq.$\left(
\ref{r64}\right) $. As a consequence, the factor
$\mathcal{V}(C_{0})$ in eq.$\left( \ref{r1}\right) $ becomes
equivalent to the $\delta$-function $\delta \left( 1-\sigma
(0,A)\right) $. This means that the significant range of
integration in the path-integral turns out to coincide with the
region near the horizon $l_{1}$. }

\item  \textbf{Summary}

\textit{As a consequence of the restriction of the domain of
integration up to the first horizon $l_{1}$ in the path-integral,
the gluon propagator in the Landau gauge gets deeply modified in
the infrared, namely
\begin{equation}
\mathcal{D}_{\mu \nu }^{ab}(q)=\left\langle A_{\mu }^{a}(q)A_{\nu
}^{b}(-q)\right\rangle =\delta ^{ab}g^{2}\frac{q^{2}}{q^{4}+\gamma ^{4}}%
\left( \delta _{\mu \nu }-\frac{q_{\mu }q_{\nu }}{q^{2}}\right)
\;. \label{r74}
\end{equation}
The parameter $\gamma $, known as the Gribov mass, is defined by
the gap equation
\begin{equation}
\frac{3Ng^{2}}{4}\int \frac{d^{4}q}{\left( 2\pi \right) ^{4}}\frac{1}{%
q^{4}+\gamma ^{4}}=1\;.  \label{r75}
\end{equation}
}

\item  \textbf{Remark}

{\it Needless to say, the integral entering the equation $\left(
\ref{r75}\right)$ is divergent. Within the present approximation,
the gap equation $\left( \ref{r75}\right)$ has to be understood in
a regularized way by means of the introduction of a cutoff
$\Lambda$, as done in equations $\left( \ref{r67}\right)$, $\left(
\ref{r68}\right)$. In order to have a more precise meaning of this
equation and of the Gribov parameter $\gamma$, we should be able
to introduce a suitable set of counterterms allowing for a
renormalized version of the gap equation $\left(
\ref{r75}\right)$. In other words, we should have at our disposal
a local renormalizable effective theory implementing the
restriction to the region $C_{0}$. Without entering in details, we
mention that such a local formulation has been constructed by
Zwanziger \cite{z4}. Remarkably, the resulting effective theory
implementing the restriction to the first Gribov horizon $l_{1}$
turns out to be renormalizable \cite{z4}.  }

\end{itemize}

\subsection{The ghost propagator in the Landau gauge}

It remains now to discuss the infrared behavior of the ghost
propagator, which is obtained from expression \textit{$\left(
\ref{r28}\right) $ }upon contraction of the gauge fields, namely
\begin{equation}
\mathcal{G}(k)\approx \frac{1}{k^{2}}\frac{1}{\left( 1-\sigma
(k)\right) }\;, \label{r76}
\end{equation}
with
\begin{eqnarray}
\sigma (k) &=&\frac{N}{N^{2}-1}\frac{1}{k^{2}}\int
\frac{d^{4}q}{\left( 2\pi
\right) ^{4}}\frac{\left( k-q\right) _{\mu }k_{\nu }}{\left( k-q\right) ^{2}}%
\left\langle A_{\mu }^{a}(-q)A_{\nu }^{a}(q)\right\rangle \;  \nonumber \\
&=&Ng^{2}\frac{k_{\mu }k_{\nu }}{k^{2}}\int \frac{d^{4}q}{\left(
2\pi
\right) ^{4}}\frac{1}{\left( k-q\right) ^{2}}\frac{q^{2}}{q^{4}+\gamma ^{4}}%
\left( \delta _{\mu \nu }-\frac{q_{\mu }q_{\nu }}{q^{2}}\right)
\;.
\nonumber \\
&&  \label{r77}
\end{eqnarray}
Let us analyze the infrared behavior, $k\approx 0$, of $\left(
1-\sigma
(k)\right) $. Making use of the gap equation \textit{$\left( \ref{r75}%
\right) $} and of
\begin{equation}
\int \frac{d^{4}q}{\left( 2\pi \right) ^{4}}\frac{1}{q^{4}+\gamma ^{4}}%
\left( \delta _{\mu \nu }-\frac{q_{\mu }q_{\nu }}{q^{2}}\right) =\frac{3}{4}%
\delta _{\mu \nu }\int \frac{d^{4}q}{\left( 2\pi \right) ^{4}}\frac{1}{%
q^{4}+\gamma ^{4}}\;,  \label{r78}
\end{equation}
it follows that
\begin{equation}
Ng^{2}\frac{k_{\mu }k_{\nu }}{k^{2}}\int \frac{d^{4}q}{\left( 2\pi
\right) ^{4}}\frac{1}{q^{4}+\gamma ^{4}}\left( \delta _{\mu \nu
}-\frac{q_{\mu }q_{\nu }}{q^{2}}\right) =1\;.  \label{r79}
\end{equation}
Thus, for $\left( 1-\sigma (k)\right) $, we obtain
\begin{eqnarray}
\left( 1-\sigma (k)\right) &=&Ng^{2}\frac{k_{\mu }k_{\nu }}{k^{2}}\int \frac{%
d^{4}q}{\left( 2\pi \right) ^{4}}\left( 1-\frac{q^{2}}{\left( k-q\right) ^{2}%
}\right) \frac{1}{q^{4}+\gamma ^{4}}\left( \delta _{\mu \nu
}-\frac{q_{\mu
}q_{\nu }}{q^{2}}\right)  \nonumber \\
&=&Ng^{2}\frac{k_{\mu }k_{\nu }}{k^{2}}\int \frac{d^{4}q}{\left(
2\pi
\right) ^{4}}\frac{\left( k^{2}-2kq\right) }{\left( k-q\right) ^{2}}\frac{1}{%
q^{4}+\gamma ^{4}}\left( \delta _{\mu \nu }-\frac{q_{\mu }q_{\nu }}{q^{2}}%
\right)  \nonumber \\
&=&Ng^{2}\frac{k_{\mu }k_{\nu }}{k^{2}}\mathcal{P}_{\mu \nu
}(k)\;, \label{r80}
\end{eqnarray}
where
\begin{equation}
\mathcal{P}_{\mu \nu }(k)=\int \frac{d^{4}q}{\left( 2\pi \right) ^{4}}\frac{%
\left( k^{2}-2kq\right) }{\left( k-q\right) ^{2}}\frac{1}{q^{4}+\gamma ^{4}}%
\left( \delta _{\mu \nu }-\frac{q_{\mu }q_{\nu }}{q^{2}}\right)
\;. \label{r81}
\end{equation}
From this expression one sees that $\mathcal{P}_{\mu \nu }(k)$ is
convergent and non singular at $k=0$. In fact
\begin{equation}
\mathcal{P}_{\mu \nu }(0)=0\;,  \label{r82}
\end{equation}
from which it follows that, for $k\approx 0$,
\begin{equation}
\left. \mathcal{P}_{\mu \nu }(k)\right. _{k\rightarrow 0}\approx
k^{2}\int \frac{d^{4}q}{\left( 2\pi \right)
^{4}}\frac{1}{q^{2}}\frac{1}{q^{4}+\gamma ^{4}}\left( \delta _{\mu
\nu }-\frac{q_{\mu }q_{\nu }}{q^{2}}\right) \;. \label{r83}
\end{equation}
Since
\begin{eqnarray}
\int \frac{d^{4}q}{\left( 2\pi \right) ^{4}}\frac{1}{q^{2}}\frac{1}{%
q^{4}+\gamma ^{4}}\left( \delta _{\mu \nu }-\frac{q_{\mu }q_{\nu }}{q^{2}}%
\right) &=&\frac{3}{4}\delta _{\mu \nu }\int \frac{d^{4}q}{\left(
2\pi
\right) ^{4}}\frac{1}{q^{2}}\frac{1}{q^{4}+\gamma ^{4}}  \nonumber \\
&=&\delta _{\mu \nu }\frac{3}{4}\frac{\Omega _{4}}{\left( 2\pi \right) ^{4}}%
\int dq\frac{q}{q^{4}+\gamma ^{4}}  \nonumber \\
&=&\delta _{\mu \nu }\frac{3}{8}\frac{\Omega _{4}}{\left( 2\pi \right) ^{4}}%
\frac{\pi }{2}\frac{1}{\gamma ^{2}}  \nonumber \\
&=&\delta _{\mu \nu }\frac{3}{128\pi }\frac{1}{\gamma ^{2}}\;,
\label{r84}
\end{eqnarray}
we get
\begin{equation}
\left. \mathcal{P}_{\mu \nu }(k)\right. _{k\rightarrow 0}\approx
k^{2}\delta _{\mu \nu }\frac{3}{128\pi }\frac{1}{\gamma ^{2}}\;.
\label{r85}
\end{equation}
Therefore
\begin{equation}
\left( 1-\sigma (k)\right) _{k\rightarrow 0}\approx \frac{3Ng^{2}}{128\pi }%
\frac{1}{\gamma ^{2}}k^{2}\;,  \label{r86}
\end{equation}
and, for the infrared behavior of the ghost propagator
\begin{equation}
\left. \mathcal{G}(k)\right. _{k\rightarrow 0}\approx \frac{128\pi
\gamma ^{2}}{3Ng^{2}}\frac{1}{k^{4}}\;.  \label{r87}
\end{equation}
One sees thus that, while the gauge propagator is suppressed in
the infrared, the ghost propagator is enhanced at $k\approx 0$,
being indeed more singular than $1/k^{2}$.

\newpage

\section{Conclusions}

\begin{itemize}
\item  Following Gribov's suggestion, we have discussed the
implementation of the restriction of the domain of integration in
the path-integral up to the first horizon $l_{1}$. As a
consequence of this restriction, we have seen that, in the Landau
gauge, the gluon and ghost propagators get deeply modified. The
gluon propagator is suppressed in the infrared region, while the
ghost propagator is enhanced. These remarkable features might
signal that the Gribov copies could play a crucial role for a
better understanding of the behavior of Yang-Mills theories in the
infrared. Let us conclude this short excursion through Gribov's
work by mentioning a few important results which have been
obtained in the last two decades.

\item  {\bf General properties of the Gribov  region $C_{0}$}.

\noindent The Gribov region $C_{0}$ is defined as the set of the
gauge connections $\left\{ A_{\mu }\right\} $ which are
transverse, $\partial A=0$, and for which the Faddeev-Popov
operator is positive definite, $-\partial _{\mu }D_{\mu }>0$. The
boundary of $C_{0}$ is the first horizon $l_{1}$, where the first
vanishing eigenvalue of the operator $-\partial _{\mu }D_{\mu }$
appears. General properties of the region $C_{0}$ have been
established, namely
\begin{itemize}
{\it \item The region $C_{0}$ is convex and bounded in every
direction \cite{z1}. Essentially, this means that any point on
$l_1$ can be seen to have a finite distance to the origin of field
space. \item Every gauge orbit passes inside the Gribov horizon
$l_{1}$ \cite{z2}. This result provides a well defined support to
Gribov's proposal of restricting the domain of integration in the
path integral to the region $C_{0}$. \item The configuration
$A_{\mu }=0$ is contained in $C_{0}$. This means that the usual
perturbation theory lies within this region. }
\end{itemize}

\item  Nowadays, it is known that the Gribov region $C_{0}$ is not
free from Gribov copies, {\it i.e.} Gribov copies still exist
inside $C_{0}$ \cite{s,z2,z3,vb}\footnote{For instance, the
existence of additional copies inside the Gribov region $C_{0}$
can be inferred by means of Henyey's example \cite{hen}, as
discussed in \cite{vb}.}. To avoid the presence of these
additional copies, a further restriction to a smaller region,
known as the fundamental modular region $\Lambda$, should be
implemented. However, it is difficult to give an explicit
description of the region $\Lambda$. A review on the
implementation of the restriction to the modular region $\Lambda$
at the Hamiltonian level can be found in \cite{vb1}. 

\item In spite of the presence of copies inside the first horizon,
Gribov's suggestion of restricting the domain of integration in
the path-integral to the region $C_{0}$ captures nontrivial
nonperturbative aspects of Yang-Mills theories, as expressed by
the infrared suppression and the infrared enhancement of the gluon
and ghost propagators in the Landau gauge. Recently, it has been
argued in \cite{Zwanziger:2003cf} that the additional copies
existing inside $C_{0} $ have no influence on the expectation
values, so that averages calculated over $\Lambda $ or $C_{0}$ are
expected to give the same result. It is worth mentioning that this
behavior of the gluon and ghost propagators has received many
confirmations from lattice simulations \cite{l1,l2,l3,l4,l5,l6}.
Also, the suppression of the gluon propagator and the enhancement
of the ghost in the Landau gauge have been obtained within the
Schwinger-Dyson approach \cite{alk}.

\item  Finally, we remark that a local action for Yang-Mills
theories implementing the restriction of the domain of integration
to the interior of the first Gribov horizon has been obtained by
D. Zwanziger \cite{z4}. The restriction to the region $C_{0}$ is
achieved through the introduction of a nonlocal horizon function
in the Boltzmann weight defining the Yang-Mills measure. This
nonlocal term may be written in local form through the
introduction of suitable additional fields. Remarkably, the
resulting local action turns out to be multiplicatively
renormalizable to all orders, obeying the renormalization group
equations \cite{z4}.
\end{itemize}

\section*{Acknowledgments.}
S.P. Sorella thanks the Organizing Committee of the 13th Jorge
Andre Swieca Summer School for the kind invitation. The Conselho
Nacional de Desenvolvimento Cient\'{i}fico e Tecnol\'{o}gico
(CNPq-Brazil), the Faperj, Funda{\c c}{\~a}o de Amparo {\`a}
Pesquisa do Estado do Rio de Janeiro, the SR2-UERJ and the
Coordena{\c{c}}{\~{a}}o de Aperfei{\c{c}}oamento de Pessoal de
N{\'{i}}vel Superior (CAPES) are gratefully acknowledged for
financial support.
\newpage

\appendix
\section{Appendix A. Notations}

The pure Yang-Mills action in Euclidean space-time reads
\begin{equation}
S_{YM}=\frac{1}{4g^{2}}\int {d^{4}x}F_{\mu \nu }^{a}F_{\mu \nu }^{a}\;,
\label{ym}
\end{equation}
where $F_{\mu \nu }^{a}$ is the field strength
\begin{equation}
F_{\mu \nu }^{a}=\partial _{\mu }{A}_{\nu }^{a}-\partial _{\nu }{A}_{\mu
}^{a}+f^{abc}A_{\mu }^{b}{A}_{\nu }^{c}\;.
\end{equation}
The color index $a$ refers to the adjoint representation of a
semi-simple Lie group $G$ whose structure constants are given by
$f^{abc}$.  The generators $\left\{ \lambda ^{a}\right\} $ of the
gauge group $G$ are chosen to be anti-hermitian
\begin{equation}
\lambda ^{a}=-\lambda ^{a\dagger },
\end{equation}
with
\begin{equation}
\left[ \lambda ^{a},\lambda ^{b}\right] =f^{abc}\lambda ^{c}.
\end{equation}
Thus,
\begin{equation}
F_{\mu \nu }=\partial _{\mu }A_{\nu }-\partial _{\nu }{A}_{\mu }+\left[
A_{\mu },A_{\nu }\right] \;,  \label{f}
\end{equation}
where
\begin{equation}
A_{\mu }=A_{\mu }^{a}\lambda ^{a}\;,\;\;\;\;F_{\mu \nu }=F_{\mu \nu
}^{a}\lambda ^{a}\;.
\end{equation}
For the gauge transformations we have
\begin{eqnarray}
&A_{\mu }&\rightarrow \widetilde{A}_{\mu }={S}^{\dagger }\partial _{\mu }{S}%
+S^{\dagger }A_{\mu }{S}\;, \nonumber \\
&S& =e^{\alpha^a \lambda^a} \;, \label{gt}
\end{eqnarray}
from which it follows
\begin{equation}
\widetilde{F}_{\mu \nu }={S}^{\dagger }F_{\mu \nu }S\;.  \label{ft}
\end{equation}
At the infinitesimal level,
\begin{eqnarray}
S &=&1+\alpha \;,  \label{inf} \\
\alpha &=&\alpha ^{a}\lambda ^{a},  \nonumber
\end{eqnarray}
one has
\begin{eqnarray}
\delta A_{\mu } &=&\partial _{\mu }\alpha +\left[ A_{\mu },\alpha \right]
=D_{\mu }\alpha \;, \\
\delta F_{\mu \nu } &=&\left[ F_{\mu \nu },\alpha \right] \;.
\end{eqnarray}
In components, these transformations read
\begin{eqnarray}
\delta A_{\mu }^{a} &=&\partial _{\mu }\alpha ^{a}+f^{abc}A_{\mu }^{b}\alpha
^{c}=D_{\mu }^{ab}\alpha ^{b}, \\
\delta F_{\mu \nu }^{a} &=&f^{abc}F_{\mu \nu }^{b}\alpha ^{c}.
\end{eqnarray}
with the covariant derivative $D_\mu$ defined as
\begin{eqnarray}
D_{\mu } &=&\partial _{\mu }+[A_{\mu },\;] \;, \\
D_{\mu }^{ab} &=&\delta ^{ab}\partial _{\mu }-f^{abc}A_{\mu }^{c}
\;.
\end{eqnarray}

\newpage

\section{Appendix B. The Gribov pendulum}

As we have seen, eq.$\left( \ref{eq39}\right) $, the condition for
the existence of Gribov copies gives a differential equation
corresponding to a damped pendulum under the action of several
forces, see Fig.14. Let us discuss here its equations of motion%
\footnote{%
It is assumed that the pendulum has unit mass $m=1$ and unit
length $l=1$.}.

\vspace{1cm}


\begin{figure}[ht]
\begin{center}
        \scalebox{0.7}{\includegraphics{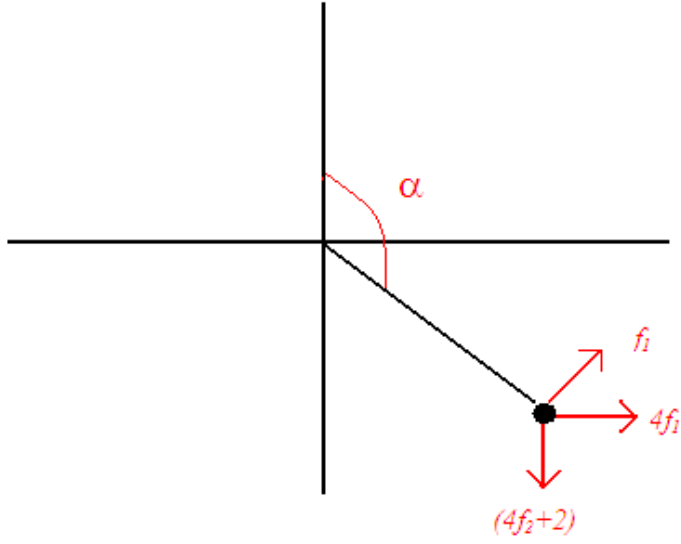}}
        \caption{The Gribov pendulum}
    \end{center}
\end{figure}

\vspace{2cm}

\noindent From Fig.14, we have
\begin{eqnarray}
\frac{d^{2}y}{d\tau ^{2}} &=&-(2+4f_{2})+4f_{1}\cos (\alpha -\frac{\pi }{2}%
)+T\cos (\pi -\alpha )\;,  \label{p1} \\
\frac{d^{2}x}{d\tau ^{2}} &=&4f_{1}-T\cos (\alpha -\frac{\pi }{2}%
)+4f_{1}\sin (\alpha -\frac{\pi }{2})\;,  \nonumber
\end{eqnarray}
where $T$ stands for the tension. From
\begin{eqnarray}
y &=&-\cos (\pi -\alpha )\;,  \label{p2} \\
x &=&\sin (\pi -\alpha )\;,  \nonumber
\end{eqnarray}
we have
\begin{eqnarray}
-\frac{d^{2}\cos (\pi -\alpha )}{d\tau ^{2}} &=&-(2+4f_{2})+4f_{1}\cos
(\alpha -\frac{\pi }{2})+T\cos (\pi -\alpha )  \nonumber \;,  \\
\frac{d^{2}\sin (\pi -\alpha )\;}{d\tau ^{2}} &=&4f_{1}-T\cos (\alpha -\frac{%
\pi }{2})+4f_{1}\sin (\alpha -\frac{\pi }{2}) \label{p3} \;.
\end{eqnarray}
Recalling that
\begin{eqnarray}
\cos (\pi -\alpha ) &=&-\cos \alpha \;,  \label{p4} \\
\cos (\alpha -\frac{\pi }{2}) &=&\sin \alpha \;,  \nonumber \\
\sin (\pi -\alpha ) &=&\sin \alpha \;,  \nonumber \\
\sin (\alpha -\frac{\pi }{2}) &=&-\cos \alpha \;,  \nonumber
\end{eqnarray}
it follows
\begin{eqnarray}
\frac{d^{2}\cos \alpha }{d\tau ^{2}} &=&-(2+4f_{2})+4f_{1}\sin \alpha -T\cos
\alpha \;,  \label{p5} \\
\frac{d^{2}\sin \alpha \;}{d\tau ^{2}} &=&4f_{1}-T\sin \alpha -4f_{1}\cos
\alpha \;,  \nonumber
\end{eqnarray}
\textit{i.e.}
\begin{eqnarray}
\sin \alpha \frac{d^{2}\cos \alpha }{d\tau ^{2}} &=&-(2+4f_{2})\sin \alpha
+4f_{1}\sin ^{2}\alpha -T\sin \alpha \cos \alpha \;,  \label{p6} \\
\cos \alpha \frac{d^{2}\sin \alpha \;}{d\tau ^{2}} &=&4f_{1}\cos \alpha
-T\sin \alpha \cos \alpha -4f_{1}\cos ^{2}\alpha \;.  \nonumber
\end{eqnarray}
Eliminating the tension $T$, one gets
\begin{equation}
\sin \alpha \frac{d^{2}\cos \alpha }{d\tau ^{2}}-\cos \alpha \frac{d^{2}\sin
\alpha \;}{d\tau ^{2}}=-(2+4f_{2})\sin \alpha +4f_{1}\sin ^{2}\alpha
-4f_{1}\cos \alpha +4f_{1}\cos ^{2}\alpha \;,  \label{p7}
\end{equation}
which becomes
\begin{equation}
-\sin \alpha \frac{d\left( \alpha ^{\prime }\sin \alpha \right) }{d\tau }%
-\cos \alpha \frac{d\left( \alpha ^{\prime }\cos \alpha \right) \;}{d\tau }%
=-(2+4f_{2})\sin \alpha -4f_{1}\cos \alpha +4f_{1}\;.  \label{p8}
\end{equation}
Thus
\begin{equation}
\alpha ^{\prime \prime }-(2+4f_{2})\sin \alpha +4f_{1}(1-\cos \alpha )=0\;.
\label{p10}
\end{equation}
Finally, adding the damping term $\alpha ^{\prime }$, one gets
\begin{equation}
\alpha ^{\prime \prime }+\alpha ^{\prime }-(2+4f_{2})\sin \alpha
+4f_{1}(1-\cos \alpha )=0\;,  \label{p11}
\end{equation}
which is precisely the Gribov condition $\left( \ref{eq39}\right) $.

\newpage

\section{Appendix C. Brief introduction to homotopy and winding number}

We shall give here a short introduction to the homotopy and to the
winding number, following the references by S. Coleman \cite{col}
and by P. Goddard and P. Mansfield \cite{gm}.

\subsection{Homotopy}

We shall be interested in the study of continuous mappings
$\left\{ \phi \right\} $ between the $n-$dimensional hyper-sphere
$S^{n}$ and the coset space $\mathcal{M}=G/H$, where $G$ is a Lie
group and $H$ a subgroup

\begin{equation}
\phi (x):S^{n}\rightarrow \mathcal{M}\;,\;\;\;x\in S^{n}  \label{c1}
\end{equation}

\begin{itemize}
\item  \textbf{Definition:} \textit{Two continuous maps $\phi $, $g$ are
said to be homotopic if there exists a map $F(x,t)$, with $0\leq t\leq 1$%
\begin{equation}
F(x,t):S^{n}\times \left[ 0,1\right] \rightarrow \mathcal{M\;},\;  \label{c2}
\end{equation}
which interpolates continuously between them, namely
\begin{equation}
F(x,0)=\phi \;,\;\;\;{{\mathrm{and\ \,\,\,\,\,\,}}F(x,1)=g\;\;.}  \label{c3}
\end{equation}
}
\end{itemize}

\noindent The existence of a homotopy between $\phi $ and $g$ will be
denoted by $\phi \sim $ $g$. The set of maps $\left\{ \phi :S^{n}\rightarrow
\mathcal{M}\right\} $ can be divided into disjoint classes of mutually
homotopic maps, the homotopy classes, denoted by $\pi _{n}(\mathcal{M})$.

\subsection{The winding number}

In the case in which $\mathcal{M}=S^{n}$, \textit{i.e.}
\begin{equation}
\phi :S^{n}\rightarrow S^{n}\;,  \label{c4}
\end{equation}
it can be shown that the equivalence homotopy classes are labelled
by the \textit{winding number}: two maps, $\phi $,
$g:S^{n}\rightarrow S^{n}$, can
be continuously deformed into one another if and only if $\phi (x)$ and $%
g(x)\;$cover $S^{n}$ the same number of times as $x$ covers it once. Thus $%
\pi _{n}(S^{n})=\mathcal{Z}$, the set of all integers.

\begin{itemize}
\item  \textbf{Example 1}. $g:S^{1}\rightarrow S^{1}$

\textit{Let us consider the mapping $g(\theta ):S^{1}\rightarrow S^{1}$,
where $\theta \in \left[ 0,2\pi \right] $ and $S^{1}$ is the unit circle.
For the identity map we have
\begin{equation}
g^{(1)}(\theta )=e^{i\theta }=\cos \theta +i\sin \theta \;.  \label{c5}
\end{equation}
}

\textit{As $\theta \in \left[ 0,2\pi \right] $ covers $S^{1}$, $%
g^{(1)}(\theta )$ covers $S^{1}$ once. The map
\begin{equation}
g^{(2)}(\theta )=e^{i2\theta }=\cos 2\theta +i\sin 2\theta \;,  \label{c6}
\end{equation}
covers $S^{1}$ twice as $\theta $ covers $S^{1}$. In general,
\begin{equation}
g^{(\nu )}(\theta )=e^{i\nu \theta }=\cos \nu \theta +i\sin \nu \theta \;,
\label{c7}
\end{equation}
with $\nu $ integer, covers $S^{1}$ $\nu -$times as $\theta $ covers $S^{1}$
once. The integer $\nu $ is called the winding number. It measures the
number of times we wind around $S^{1}$ as we go once around the circle in
two-space. }

\item  \textit{\textbf{The following result holds}: }

\textit{Every mapping from $S^{1}\rightarrow S^{1}$ is homotopic to one of
the mappings $g^{(\nu )}(\theta )$, with $\nu $ integer. From this result it
follows that we can associate a winding number with every continuous mapping
from $S^{1}\rightarrow S^{1}$. }
\end{itemize}

\noindent The winding number can be represented by the integral
\begin{equation}
\nu =-\frac{i}{2\pi }\int_{0}^{2\pi }d\theta \left( g^{(\nu )}(\theta
)\right) ^{-1}\frac{d}{d\theta }g^{(\nu )}(\theta )\;.  \label{c8}
\end{equation}
Moreover, it can be proven that the quantity $\int_{0}^{2\pi }d\theta
g^{-1}\partial _{\theta }g$ is invariant under continuous deformations,
\textit{i.e.} it has a topological meaning. Indeed, denoting by $\delta
g=i\left( \delta \lambda \right) g$ a general infinitesimal deformation,
where $\delta \lambda $ is an infinitesimal real function on the circle $%
S^{1}$, it follows
\begin{eqnarray}
gg^{-1} &=&1 \;\Longrightarrow \left( \delta g\right)
g^{-1}=-g\left( \delta
g^{-1}\right) \;,  \label{c9} \\
\delta g^{-1} &=&-g^{-1}\left( \delta g\right)
g^{-1}\;=-ig^{-1}\delta \lambda \;. \nonumber
\end{eqnarray}
Thus
\begin{eqnarray}
\delta \int_{0}^{2\pi }d\theta g^{-1}\partial _{\theta }g &=&\int_{0}^{2\pi
}d\theta \left( -ig^{-1}\delta \lambda \partial _{\theta }g+ig^{-1}\left(
\partial _{\theta }\delta \lambda \right) g+ig^{-1}\delta \lambda \partial
_{\theta }g\right)  \nonumber \\
&=&i\int_{0}^{2\pi }d\theta \partial _{\theta }\delta \lambda =0\;.
\label{c10}
\end{eqnarray}
Also, if
\begin{equation}
g(\theta )=g^{(\nu _{1})}(\theta )g^{(\nu _{2})}(\theta )\;,  \label{c11}
\end{equation}
it follows that
\begin{equation}
\nu =\nu _{1}+\nu _{2}\;.  \label{c12}
\end{equation}

\begin{itemize}
\item  \textbf{Example 2}. $g:S^{3}\rightarrow SU(2)$

\textit{Let us discuss now the homotopy between the three-dimensional
hyper-sphere $S^{3}$ and the group $SU(2)$%
\begin{equation}
g:S^{3}\rightarrow SU(2)\;.  \label{c13}
\end{equation}
The unit hyper-sphere $S^{3}$ can be parametrized by local coordinates $%
x_{\mu },\;\mu =1,2,3,4,\;x_{\mu }x_{\mu }=1$. We may also choose three
angles $\left\{ \theta _{1},\;\theta _{2},\;\theta _{3}\right\} $ to
parametrize $S^{3}$. The group $SU(2)$ is the group of unitary unimodular
two by two matrices. Any such matrix can be written as
\begin{equation}
g=aI+i\overrightarrow{b}\cdot \overrightarrow{\sigma }\;,  \label{c144}
\end{equation}
}

\textit{where $a,\overrightarrow{b}$ are real parameters and $I$, $%
\overrightarrow{\sigma }$are the unit and the Pauli matrices, respectively.
From
\begin{equation}
\sigma _{1}=\left(
\begin{array}{ll}
0 & 1 \\
1 & 0
\end{array}
\right) ,\;\;\;\;\sigma _{2}=\left(
\begin{array}{ll}
0 & i \\
-i & 0
\end{array}
\right) ,\;\;\;\;\sigma _{3}=\left(
\begin{array}{ll}
1 & 0 \\
0 & -1
\end{array}
\right) \;.  \label{c15}
\end{equation}
it turns out
\begin{equation}
g=\left(
\begin{array}{ll}
a+ib_{3} & b_{2}+ib_{1} \\
-b_{2}+ib_{1} & a-ib_{3}
\end{array}
\right) \;,  \label{c16}
\end{equation}
so that the condition $\det g=1$ gives
\begin{equation}
\det g=1\Longrightarrow a^{2}+\overrightarrow{b}\cdot \overrightarrow{b}=1\;.
\label{c177}
\end{equation}
}

\textit{Thus $SU(2)$ has the topology of the sphere $S^{3}$.
Therefore, the mapping $\left( \ref{c13}\right) $ becomes a
mapping between two hyper-spheres $S^{3}$
\begin{equation}
g:S^{3}\rightarrow S^{3}\;,  \label{c18}
\end{equation}
}

\textit{the homotopy classes being classified by the winding number $\pi
_{3}(S^{3})=\mathcal{Z}$. Examples of the mapping $g:S^{3}\rightarrow S^{3}$
are given by
\begin{eqnarray}
g^{(0)}(x) &=&1\;,\;\;\mathrm{trivial\;mapping}\;\nu =0\;,  \label{c19} \\
g^{(1)}(x) &=&x_{4}I+i\overrightarrow{x}\cdot \overrightarrow{\sigma }%
\;\;,\;\;\;\mathrm{identity\;mapping}\;\nu =1\;,  \nonumber \\
g^{(\nu )}(x) &=&\left( g^{(1)}(x)\right) ^{\nu },\;\;\nu \mathrm{{\ }%
\mathrm{integer}{,\ }\mathrm{\;winding}\;\nu \;.}  \nonumber
\end{eqnarray}
}

\item  \textit{\textbf{The following results hold}: }

\item  \textit{\textbf{Result 1. }Every mapping from $S^{3}$ to $S^{3}$ is
homotopic to one of the mapping $g^{(\nu )}(x)$ of equation $\left( \ref{c19}%
\right) $. }

\item  \textit{\textbf{Result 2 (R. Bott)}. Let $G$ be a simple Lie group.
Any continuous mapping from $S^{3}$ to $G$ can be continuously deformed into
a mapping of $S^{3}$ into an $SU(2)$ subgroup of $G$. Thus, everything that
can be established for $SU(2)$ is true for an arbitrary simple Lie group, in
particular for $SU(N)$. }
\end{itemize}

\noindent In the case of the mapping $g:S^{3}\rightarrow S^{3}$, the
expression $\left( \ref{c8}\right) $ for the winding number generalizes to
\begin{equation}
\nu =\frac{1}{24\pi ^{2}}Tr\int_{S^{3}}dS_{\mu }\varepsilon _{\mu \nu \rho
\sigma }\left( g^{-1}\frac{\partial }{\partial x_{\nu }}g\right) \left(
g^{-1}\frac{\partial }{\partial x_{\rho }}g\right) \left( g^{-1}\frac{%
\partial }{\partial x_{\sigma }}g\right) \;,  \label{c22}
\end{equation}
where $dS_{\mu }$ stands for the surface element of $S^{3}$. This expression
can also be rewritten as
\begin{equation}
\nu =\frac{1}{24\pi ^{2}}Tr\int_{S^{3}}d\theta _{1}d\theta _{2}d\theta
_{3}\varepsilon _{ijk}\left( g^{-1}\frac{\partial }{\partial \theta _{i}}%
g\right) \left( g^{-1}\frac{\partial }{\partial \theta _{j}}g\right) \left(
g^{-1}\frac{\partial }{\partial \theta _{k}}g\right) \;.  \label{c23}
\end{equation}

\subsection{Application: Instantons in Euclidean Yang-Mills theories}

As an application of the homotopy and of the winding number, let
us discuss here the instanton solution of Euclidean Yang-Mills
theories. Instantons are classical solutions of the equations of
motion of pure Euclidean Yang-Mills theories which have finite
action. Let $\left\{ T^{a}\right\} $ be the anti-hermitian
generators of a Lie group $G$,
\begin{equation}
\left[ T^{a},T^{b}\right] =f^{abc}T^{c}\;.  \label{c24}
\end{equation}
Following Coleman \cite{col}, the Cartan inner product is defined by
\begin{equation}
\left( T^{a},T^{b}\right) =\delta ^{ab}\;.  \label{c25}
\end{equation}
For instance, in the case of $SU(2)$, which we shall take as gauge group $G$%
, we have
\begin{eqnarray}
T^{a} &=&-\frac{i}{2}\sigma ^{a}\;,\;\;\;\;Tr\left( T^{a}T^{b}\right) =-%
\frac{1}{2}\delta ^{ab}\;,  \label{c26} \\
\left[ T^{a},T^{b}\right] &=&-\frac{1}{4}\left[ \sigma ^{a},\sigma
^{b}\right] =-\frac{i}{2}\varepsilon ^{abc}\sigma ^{c}=\varepsilon
^{abc}T^{c}\;.  \nonumber
\end{eqnarray}
Thus
\begin{equation}
\left( T^{a},T^{b}\right) =-2Tr\left( T^{a}T^{b}\right) \;.  \label{c27}
\end{equation}
Let us start with the Yang-Mills Euclidean action
\begin{equation}
S_{YM}=\frac{1}{4g^{2}}\int {d^{4}x}\left( F_{\mu \nu },F_{\mu \nu }\right)
\;,  \label{c28}
\end{equation}
with
\begin{equation}
F_{\mu \nu }(A)=\partial _{\mu }A_{\nu }-\partial _{\nu }{A}_{\mu }+\left[
A_{\mu },A_{\nu }\right] \;.  \label{c29}
\end{equation}
The classical equations of motion are
\begin{equation}
D_{\mu }F_{\mu \nu }=\partial _{\mu }F_{\mu \nu }+\left[ A_{\mu },F_{\mu \nu
}\right] =0\;.  \label{c30}
\end{equation}
In order to have finite action, and recalling that ${d^{4}x=d\Omega }%
_{4}r^{3}dr$, one requires that $F_{\mu \nu }$ falls off faster than $%
1/r^{3}$ when $r\rightarrow \infty $, namely
\begin{equation}
F_{\mu \nu }\sim 1/r^{3}\;\;\;\mathrm{for}\;\;\;r\rightarrow \infty \;.
\label{c31}
\end{equation}
This condition implies that, when $r\rightarrow \infty $%
\begin{equation}
A_{\mu }=g^{-1}\partial _{\mu }g+O(1/r^{2})\;\;\;\;\mathrm{for\;\;\;}%
r\rightarrow \infty \;.  \label{c32}
\end{equation}
Notice that, for a pure gauge configuration,
$A_{\mu}=g^{-1}\partial _{\mu }g$, one has
\[
F_{\mu \nu }(g^{-1}\partial g)=0 \;.
\]
In fact
\begin{eqnarray}
F_{\mu \nu }(g^{-1}\partial g) &=&\partial _{\mu }(g^{-1}\partial _{\nu
}g)-\partial _{\nu }(g^{-1}\partial _{\mu }g)+\left[ (g^{-1}\partial _{\mu
}g),(g^{-1}\partial _{\nu }g)\right]  \nonumber \\
&=&\left( \partial _{\mu }g^{-1}\right) \partial _{\nu }g+g^{-1}\partial
_{\mu }\partial _{\nu }g-\left( \partial _{\nu }g^{-1}\right) \partial _{\mu
}g-g^{-1}\partial _{\nu }\partial _{\mu }g  \nonumber \\
&&+g^{-1}\left( \partial _{\mu }g\right) g^{-1}\partial _{\nu
}g-g^{-1}\left( \partial _{\nu }g\right) g^{-1}\partial _{\mu }g  \nonumber
\\
&=&\left( \partial _{\mu }g^{-1}\right) \partial _{\nu }g-\left( \partial
_{\nu }g^{-1}\right) \partial _{\mu }g-\left( \partial _{\mu }g^{-1}\right)
\partial _{\nu }g+\left( \partial _{\nu }g^{-1}\right) \partial _{\mu }g
\nonumber \\
&=&0  \label{c33}
\end{eqnarray}
The boundary of the four-dimensional Euclidean space-time at infinity, $%
r\rightarrow \infty $, is given by the three hyper-sphere
$S_{\infty }^{3}$.
The behavior of the gauge field $A_{\mu }$ at infinity, eq.$\left( \ref{c32}%
\right) $, allows to define a map between the hyper-sphere
$S_{\infty }^{3}$ and $SU(2)$;
\begin{equation}
g(x):S_{\infty }^{3}\rightarrow SU(2)\;.  \label{c34}
\end{equation}
Since $SU(2)$ has the topology of $S^{3}$, the mapping $\left( \ref{c34}%
\right) $ can be characterized by the winding number $\nu $ corresponding to
the homotopy $\pi _{3}(S^{3})=\mathcal{Z}$. This means that the classical
solutions of the equations of motion in pure Yang-Mills with finite action
can be classified by the winding number $\nu $.

\noindent In order to find classical solutions of the equations of motion,
it is useful to consider the identity
\begin{equation}
\int {d^{4}x}\left( F_{\mu \nu }\mp ^{*}F_{\mu \nu },F_{\mu \nu }\mp
^{*}F_{\mu \nu }\right) \geq 0\;,  \label{c35}
\end{equation}
where $^{*}F_{\mu \nu }$ is the dual of $F_{\mu \nu }$%
\begin{eqnarray}
^{\ast }F_{\mu \nu } &=&\frac{1}{2}\varepsilon _{\mu \nu \rho \sigma
}F_{\rho \sigma }\;,  \label{c36} \\
\varepsilon _{\mu \nu \rho \sigma }\varepsilon _{\mu \nu \lambda \delta }
&=&2\left( \delta _{\rho \lambda }\delta _{\sigma \delta }-\delta _{\rho
\delta }\delta _{\sigma \lambda }\right) \;.  \nonumber
\end{eqnarray}
From eq.$\left( \ref{c35}\right) $ we have
\begin{equation}
\int {d^{4}x}\left( F_{\mu \nu }\mp ^{*}F_{\mu \nu },F_{\mu \nu }\mp
^{*}F_{\mu \nu }\right) =\int {d^{4}x}\left( \left( F_{\mu \nu },F_{\mu \nu
}\right) +\left( ^{*}F_{\mu \nu },^{*}F_{\mu \nu }\right) \mp 2\left( F_{\mu
\nu },^{*}F_{\mu \nu }\right) \right)  \label{c37}
\end{equation}
Since $^{*}\left( ^{*}F_{\mu \nu }\right) =F_{\mu \nu }$, it follows
\begin{equation}
\frac{1}{4g^{2}}\int {d^{4}x}\left( F_{\mu \nu },F_{\mu \nu }\right) \geq
\frac{1}{4g^{2}}\left| \int {d^{4}x}\left( F_{\mu \nu },^{*}F_{\mu \nu
}\right) \right|  \label{c38}
\end{equation}
The bound $\left( \ref{c38}\right) $ is saturated when
\begin{equation}
F_{\mu \nu }=\pm ^{*}F_{\mu \nu }=\pm \frac{1}{2}\varepsilon _{\mu \nu \rho
\sigma }F_{\rho \sigma }\;.  \label{c39}
\end{equation}
This condition is a first order differential equation. The solutions to the
self-dual equation
\begin{equation}
F_{\mu \nu }=^{*}F_{\mu \nu }=\frac{1}{2}\varepsilon _{\mu \nu \rho \sigma
}F_{\rho \sigma }\;,\;  \label{c40}
\end{equation}
are called \textbf{instantons} (anti-instantons are solutions of $F_{\mu \nu
}=-^{*}F_{\mu \nu }$). Thus, for an instanton, we have the equality
\begin{equation}
\frac{1}{4g^{2}}\int {d^{4}x}\left( F_{\mu \nu },F_{\mu \nu }\right) =\frac{1%
}{4g^{2}}\left| \int {d^{4}x}\left( F_{\mu \nu },^{*}F_{\mu \nu }\right)
\right| \;.  \label{c41}
\end{equation}

\begin{itemize}
\item  It is useful to observe that, from the self-dual condition, $F_{\mu
\nu }=^{*}F_{\mu \nu }$, one has
\begin{equation}
D_{\mu }F_{\mu \nu }=\frac{1}{2}\varepsilon _{\mu \nu \rho \sigma }D_{\mu
}F_{\rho \sigma }=0\;,  \label{c42}
\end{equation}
due to the Bianchi identity. This means that instantons are
solutions of the equations of motion of pure Euclidean Yang-Mills
theories.

\item  Another important property of the instantons is that they give
vanishing contribution to the energy-momentum tensor $\Theta _{\mu \nu }$,
as it is apparent from
\begin{equation}
\Theta _{\mu \nu }=\frac{1}{2g^{2}}\left( F_{\mu \rho }^{a}+^{*}F_{\mu \rho
}^{a}\right) \left( F_{\nu \rho }^{a}-^{*}F_{\nu \rho }^{a}\right) \;,
\label{c43}
\end{equation}
\end{itemize}

\noindent It is useful to show now that the quantity $\int {d^{4}x}\left(
F_{\mu \nu },^{*}F_{\mu \nu }\right) =\int {d^{4}x}\left( F_{\mu \nu
}^{a}{}^{*}F_{\mu \nu }^{a}\right) $\ is directly related to the winding
number $\nu $. In order to establish the relationship between $\int {d^{4}x}%
\left( F_{\mu \nu },^{*}F_{\mu \nu }\right) $ and $\nu $, let us first prove
the identity
\begin{eqnarray}
Tr\left( F_{\mu \nu }{}^{*}F_{\mu \nu }\right) &=&\partial _{\mu }\mathcal{G}%
_{\mu }\;,  \label{c44} \\
\mathcal{G}_{\mu } &=&Tr\varepsilon _{\mu \nu \rho \sigma }\left( A_{\nu
}F_{\rho \sigma }-\frac{2}{3}A_{\nu }A_{\rho }A_{\sigma }\right) \;.
\nonumber
\end{eqnarray}
Making use of the Bianchi identity, $\varepsilon _{\mu \nu \rho
\sigma }D_{\mu }F_{\rho \sigma }=0$, we have
\begin{eqnarray}
\partial _{\mu }\mathcal{G}_{\mu } &=&Tr\varepsilon _{\mu \nu \rho \sigma
}\partial _{\mu }\left( A_{\nu }F_{\rho \sigma }-\frac{2}{3}A_{\nu }A_{\rho
}A_{\sigma }\right)  \label{c45} \\
&=&Tr\varepsilon _{\mu \nu \rho \sigma }\left( \left( \partial _{\mu }A_{\nu
}\right) F_{\rho \sigma }+A_{\nu }\partial _{\mu }F_{\rho \sigma }-2\left(
\partial _{\mu }A_{\nu }\right) A_{\rho }A_{\sigma }\right)  \nonumber \\
&=&Tr\varepsilon _{\mu \nu \rho \sigma }\left( \frac{1}{2}\left( \partial
_{\mu }A_{\nu }-\partial _{\nu }A_{\mu }\right) F_{\rho \sigma }-A_{\nu
}\left[ A_{\mu },F_{\rho \sigma }\right] -\left( \partial _{\mu }A_{\nu
}-\partial _{\nu }A_{\mu }\right) A_{\rho }A_{\sigma }\right) \;  \nonumber
\\
&=&Tr\varepsilon _{\mu \nu \rho \sigma }\left( \frac{1}{2}F_{\mu \nu
}F_{\rho \sigma }-\frac{1}{2}\left[ A_{\mu },A_{\nu }\right] F_{\rho \sigma
}-A_{\nu }A_{\mu }F_{\rho \sigma }+A_{\nu }F_{\rho \sigma }A_{\mu }-F_{\mu
\nu }A_{\rho }A_{\sigma }\right)  \nonumber \\
&&+Tr\varepsilon _{\mu \nu \rho \sigma }\left( \left[ A_{\mu },A_{\nu
}\right] A_{\rho }A_{\sigma }\right)  \nonumber \\
&=&Tr\varepsilon _{\mu \nu \rho \sigma }\left( \frac{1}{2}F_{\mu \nu
}F_{\rho \sigma }-A_{\mu }A_{\nu }F_{\rho \sigma }+A_{\mu }A_{\nu }F_{\rho
\sigma }+A_{\mu }A_{\nu }F_{\rho \sigma }-A_{\rho }A_{\sigma }F_{\mu \nu
}\right)  \nonumber \\
&&+Tr\varepsilon _{\mu \nu \rho \sigma }\left( \left[ A_{\mu },A_{\nu
}\right] A_{\rho }A_{\sigma }\right)  \nonumber \\
&=&Tr\varepsilon _{\mu \nu \rho \sigma }\left( \frac{1}{2}F_{\mu \nu
}F_{\rho \sigma }+\left[ A_{\mu },A_{\nu }\right] A_{\rho }A_{\sigma
}\right) \;.  \nonumber
\end{eqnarray}
The term $Tr\varepsilon _{\mu \nu \rho \sigma }\left[ A_{\mu },A_{\nu
}\right] A_{\rho }A_{\sigma }$ vanishes due to the Jacoby identity

\begin{eqnarray}
Tr\varepsilon _{\mu \nu \rho \sigma }\left[ A_{\mu },A_{\nu }\right] A_{\rho
}A_{\sigma } &=&\frac{1}{2}Tr\varepsilon _{\mu \nu \rho \sigma }\left[
A_{\mu },A_{\nu }\right] \left[ A_{\rho },A_{\sigma }\right]  \nonumber \\
&=&\frac{1}{2}\varepsilon _{\mu \nu \rho \sigma }A_{\mu }^{a}A_{\nu
}^{b}A_{\rho }^{c}A_{\sigma }^{d}f^{abm}f^{cdn}TrT^{m}T^{n}  \nonumber \\
&=&-\frac{1}{4}\varepsilon _{\mu \nu \rho \sigma }A_{\mu }^{a}A_{\nu
}^{b}A_{\rho }^{c}A_{\sigma }^{d}f^{abm}f^{cdm}  \nonumber \\
&=&0\;.  \label{c46}
\end{eqnarray}
Thus, we conclude that
\begin{equation}
\partial _{\mu }\mathcal{G}_{\mu }=Tr\varepsilon _{\mu \nu \rho \sigma }%
\frac{1}{2}F_{\mu \nu }F_{\rho \sigma }=Tr\left( F_{\mu \nu }{}^{*}F_{\mu
\nu }\right) \;.  \label{c47}
\end{equation}
Moreover, making use of the Stokes theorem we obtain
\begin{equation}
Tr\int d^{4}x\left( F_{\mu \nu }{}^{*}F_{\mu \nu }\right) =\int
d^{4}x\partial _{\mu }\mathcal{G}_{\mu }=Tr\int_{S_{\infty }^{3}}dS_{\mu
}\varepsilon _{\mu \nu \rho \sigma }\left( A_{\nu }F_{\rho \sigma }-\frac{2}{%
3}A_{\nu }A_{\rho }A_{\sigma }\right) \;.  \label{c48}
\end{equation}
For a classical solution of the equations of motion with finite
action, we have that on the hyper-sphere at infinity $S_{\infty
}^{3}$

\begin{eqnarray}
A_{\mu } &=&g^{-1}\partial _{\mu }g\;\;\;\;\mathrm{on}\;\;S_{\infty }^{3}%
\mathrm{\;}  \label{c49} \\
F_{\mu \nu } &=&0\;\;\;\mathrm{on}\;\;S_{\infty }^{3}\;.  \nonumber
\end{eqnarray}
Therefore, for an instanton solution

\begin{eqnarray}
Tr\int d^{4}x\left( F_{\mu \nu }{}^{*}F_{\mu \nu }\right)
&=&Tr\int_{S_{\infty }^{3}}dS_{\mu }\varepsilon _{\mu \nu \rho \sigma
}\left( A_{\nu }F_{\rho \sigma }-\frac{2}{3}A_{\nu }A_{\rho }A_{\sigma
}\right)  \label{c50} \\
&=&-\frac{2}{3}Tr\int_{S_{\infty }^{3}}dS_{\mu }\varepsilon _{\mu \nu \rho
\sigma }\left( A_{\nu }A_{\rho }A_{\sigma }\right)  \nonumber \\
&=&-\frac{2}{3}Tr\int_{S_{\infty }^{3}}dS_{\mu }\varepsilon _{\mu \nu \rho
\sigma }\left( g^{-1}\partial _{\nu }g\right) \left( g^{-1}\partial _{\rho
}g\right) \left( g^{-1}\partial _{\sigma }g\right)  \nonumber \\
&=&-16\pi ^{2}\nu \;,  \nonumber
\end{eqnarray}
namely
\begin{equation}
\nu =-\frac{1}{16\pi ^{2}}Tr\int d^{4}x\left( F_{\mu \nu }{}^{*}F_{\mu \nu
}\right) \;,  \label{c51}
\end{equation}
or
\begin{equation}
\nu =\frac{1}{32\pi ^{2}}\int d^{4}x\left( F_{\mu \nu }{},^{*}F_{\mu \nu
}\right) =\frac{1}{64\pi ^{2}}\;\int d^{4}x\varepsilon _{\mu \nu \rho \sigma
}F_{\mu \nu }^{a}F_{\rho \sigma }^{a}\;.  \label{c52}
\end{equation}
The expression for $\nu $, eq.$\left( \ref{c52}\right) $, is called the
Pontryagin index, whilst the integrand is the Pontryagin density. Thus, for
an instanton with winding number $\nu $%
\begin{equation}
S_{YM}=\frac{1}{4g^{2}}\int {d^{4}x}F_{\mu \nu }^{a}F_{\mu \nu }^{a}=\frac{%
8\pi ^{2}}{g^{2}}\left| \nu \right| \;.\;  \label{c53}
\end{equation}
In the case of $SU(2)$, the explicit solution for the instanton with $\nu =1$
has been given by Belavin, Polyakov, Schwartz, Tyupkin, and reads

\begin{eqnarray}
A_{\mu } &=&f(r^{2})\left( g^{(1)}\right) ^{-1}\partial _{\mu }g^{(1)}\;,
\label{c54} \\
g^{(1)} &=&\frac{x_{4}+i\overrightarrow{x}\cdot \overrightarrow{\sigma }}{r}%
\;,\;\;\;\;r=\sqrt{x_{\mu }x_{\mu }}\;,  \nonumber \\
f(r^{2}) &=&\frac{r^{2}}{r^{2}+\rho ^{2}}\;,  \nonumber
\end{eqnarray}
where $\rho $ is an arbitrary constant, called the size of the instanton.

$\;$\pagebreak

\section{Appendix D. Polar coordinates}

Let us remind here some useful relationships in polar coordinates:






\begin{eqnarray}
x &=&r\sin \theta \cos \varphi \;,  \nonumber \\
y &=&r\sin \theta \sin \varphi \;,  \nonumber \\
z &=&r\cos \theta \;.  \label{1d}
\end{eqnarray}

\noindent Thus, for the orthonormal basis $\left( \overrightarrow{e}_{r},\;%
\overrightarrow{e}_{\theta },\;\overrightarrow{e}_{\varphi }\right) $ we
have
\begin{eqnarray}
\overrightarrow{e}_{r} &=&\sin \theta \cos \varphi \overrightarrow{e}%
_{x}+\sin \theta \sin \varphi \overrightarrow{e}_{y}+\cos \theta
\overrightarrow{e}_{z}\;,  \label{2d} \\
\overrightarrow{e}_{\theta } &=&\cos \theta \cos \varphi \overrightarrow{e}%
_{x}+\cos \theta \sin \varphi \overrightarrow{e}_{y}-\sin \theta
\overrightarrow{e}_{z}\;,  \nonumber \\
\overrightarrow{e}_{\varphi } &=&\overrightarrow{e}_{r}\times
\overrightarrow{e}_{\theta }=-\sin \varphi \overrightarrow{e}_{x}+\cos
\varphi \overrightarrow{e}_{y}\;.  \nonumber
\end{eqnarray}
Let $\overrightarrow{v}$ be a vector. We have
\begin{equation}
\overrightarrow{v}=v_{x}\overrightarrow{e}_{x}+v_{y}\overrightarrow{e}%
_{y}+v_{z}\overrightarrow{e}_{z}\;=v_{r}\overrightarrow{e}_{r}+v_{\theta }%
\overrightarrow{e}_{\theta }+v_{\varphi }\overrightarrow{e}_{\varphi }\;,\;
\label{3d}
\end{equation}
with
\begin{eqnarray}
v_{r} &=&\overrightarrow{v}\cdot \overrightarrow{e}_{r}=v_{x}\sin \theta
\cos \varphi +v_{y}\sin \theta \sin \varphi +v_{z}\cos \theta \;,  \nonumber
\\
v_{\theta } &=&\overrightarrow{v}\cdot \overrightarrow{e}_{\theta
}=v_{x}\cos \theta \cos \varphi +v_{y}\cos \theta \sin \varphi -v_{z}\sin
\theta \;,  \nonumber \\
v_{\varphi } &=&\overrightarrow{v}\cdot \overrightarrow{e}_{\varphi
}=-v_{x}\sin \varphi +v_{y}\cos \varphi \;.  \label{4d}
\end{eqnarray}
Also,
\begin{eqnarray}
\overrightarrow{\nabla }f &=&\frac{\partial f}{\partial r}\overrightarrow{e}%
_{r}+\frac{1}{r}\frac{\partial f}{\partial \theta }\overrightarrow{e}%
_{\theta }+\frac{1}{r\sin \theta }\frac{\partial f}{\partial \varphi }%
\overrightarrow{e}_{\varphi }\;,  \nonumber \\
\overrightarrow{\nabla }^{2}f &=&\frac{1}{r^{2}}\frac{\partial }{\partial r}%
\left( r^{2}\frac{\partial f}{\partial r}\right) +\frac{1}{r^{2}\sin \theta }%
\frac{\partial }{\partial \theta }\left( \sin \theta \frac{\partial f}{%
\partial \theta }\right) +\frac{1}{r^{2}\sin ^{2}\theta }\frac{\partial ^{2}f%
}{\partial \varphi ^{2}}\;,  \nonumber \\
\overrightarrow{\nabla }\cdot \overrightarrow{v} &=&\frac{1}{r^{2}}\frac{%
\partial \left( r^{2}v_{r}\right) }{\partial r}+\frac{1}{r\sin \theta }\frac{%
\partial \left( \sin \theta v_{\theta }\right) }{\partial \theta }+\frac{1}{%
r\sin \theta }\frac{\partial v_{\varphi }}{\partial \varphi }\;.  \label{5d}
\end{eqnarray}

\end{document}